\documentclass[aps,showpacs,twocolumn,dvips]{revtex4}
\usepackage{feynmp}
\usepackage{amssymb}
\usepackage{epsfig}
\usepackage{graphicx}
\usepackage{subfigure}
\usepackage{mathrsfs}
\usepackage{hyperref}
\usepackage{cleveref}
\usepackage{cases}
\usepackage{bbm}
\usepackage{xcolor}
\usepackage{tabularx}
\usepackage{array}
\usepackage{multirow}
\usepackage{tikz}
\usepackage{calc}
\usepackage{pifont}%

\usepackage{lmodern}
\usepackage{textcomp}

\begin{document}

%%%%%%%%%%%%%%%%%%%%%%%%%%%%%%%%%%%%%%%%%%%%%%%%%%%%%%%%%%%%%%%%%%%%%%%%%%%%%%%%%%%%%%%
%%%%%%%%%%%%%%%%%%%%%%%%%%%%%%%%%%%%%%%%%%%%%%%%%%%%%%%%%%%%%%%%%%%%%%%%%%%%%%%%%%%%%%%

\title{Incommensurate magnetic states induced by ordering competition in $\mathrm{Ba_{1-x}Na_xFe_2As_2}$}

\date{\today}

\author{Jing Wang}
\altaffiliation{E-mail address: jing$\textunderscore$wang@tju.edu.cn}
\affiliation{Department of Physics, Tianjin University, Tianjin 300072, P.R. China}

\begin{abstract}
Quantum criticality nearby a certain magnetic phase transition beneath the
superconducting dome of $\mathrm{Ba_{1-x}Na_xFe_2As_2}$ is
attentively studied by virtue of a phenomenological theory in conjunction with
renormalization group approach. We report that
ordering competition between magnetic and superconducting fluctuations
is capable of coaxing incommensurate (IC) magnetic states to experience distinct fates
depending upon their spin configurations. The $C_2$-symmetry IC magnetic stripe with
perpendicular magnetic helix dominates over other $C_2$-symmetry magnetic competitors
and hints at a potential candidate for the unknown $C_2$-symmetry magnetic state.
Amongst $C_4$-symmetry IC magnetic phases, IC charge spin density wave
is substantiated to be superior,  
shedding light on the significant intertwining of charge and spin degrees of freedom.
Meanwhile, ferocious fluctuations render
a sharp fall of superfluid density alongside dip
of critical temperature as well as
intriguing behavior of London penetration depth.
\end{abstract}


\maketitle

%%%%%%%%%%%%%%%%%%%%%%%%%%%%%%%%%%%%%%%%%%%%%%%%%%%%%%%%%%%%%%%%%%%%%%%%%%%%%%%%%%%%%
%%%%%%%%%%%%%%%%%%%%%%%%%%%%%%%%%%%%%%%%%%%%%%%%%%%%%%%%%%%%%%%%%%%%%%%%%%%%%%%%%%%%%

%\vspace{0.65cm}

\section{Introduction}\label{Sec_intro}

The last dozen years have witnessed considerably intense research devoted to
iron pnictides of $\mathrm{BaFe_2As_2}$
family~\cite{Hinkov2010NPhys,Goldman2010PRB,Stewart2011RMP,
Fisher2011RPP,Basov2011NPhys,Chubukov2012ARCMP,
Osborn2014NatureComm,Hardy2015NComm,Allred2015PRB,
Wang2016PRB,Andersen2017NComm,Hardy2018PRL-BaNaFeAs,
Prozorov2019PRB-BaKFeAs}, whose phase diagrams
are ubiquitously borne out of both superconducting (SC) and
diverse kinds of magnetic orders mediated by %separated and
quantum phase transitions (QPTs)~\cite{Vojta2003RPP}. Notwithstanding magnetism
is an antagonistic state versus superconductivity, they compete and
collaborate other than coexist with each other~\cite{Basov2011NPhys,Chubukov2012ARCMP,Hirschfeld2004.13134}.
This accordingly poses a substantial challenge as to what the connection is between
magnetic and SC states, providing a crucial ingredient to
glue Cooper pairing~\cite{Basov2011NPhys,Chubukov2012ARCMP}.
In the light of abundant magnetic states in $\mathrm{BaFe_2As_2}$~\cite{Goldman2010PRB,Osborn2014NatureComm,
Wang2016PRB,Hardy2015NComm,Allred2015PRB,Andersen2017NComm,Hardy2018PRL-BaNaFeAs,
Prozorov2019PRB-BaKFeAs}, one of the most imperative and realistic quests of
understanding this very compound, prior to exploring the ultimate SC nature,
is how to unambiguously identify concrete configurations
of magnetic states around QPTs in that different states are associated with
distinguished fluctuations which play a pivotal role in establishing its phase diagram.

\begin{figure}
\centering
\epsfig{file=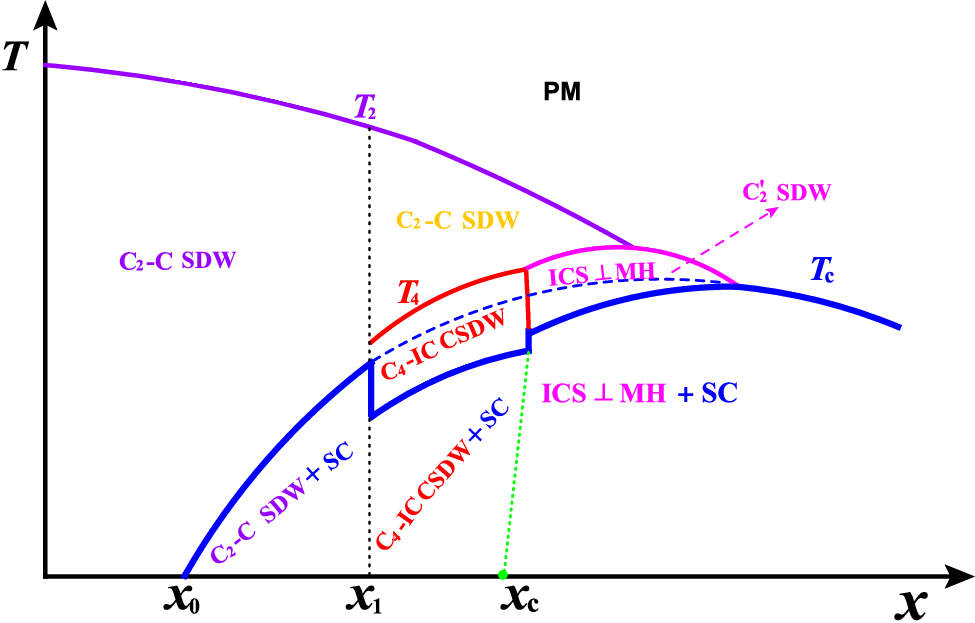,height=6.3cm,width=8.5cm}
\vspace{0.05cm}
\caption{(Color online) Schematic $x-T$ phase diagram of
$\mathrm{Ba_{1-x}Na_xFe_2As_2}$ in the vicinity of a vital
magnetic quantum critical point (QCP) located at $x_c$. PM, SDW,
and SC are shortened notations for paramagnetism, spin density wave (SDW), and superconductivity
with $T_{2,4}$ and $T_c$ denoting the critical temperatures from PM
to $C_{2,4}$ SDW and non-SC to SC, respectively.  Ordering competition under the ferocious
quantum fluctuations bears out that $C_2$ ICS $\perp$ MH state
is a good candidate for the cryptic $C_2$ magnetic state ($C'_2$ SDW) and
the leading $C_4$ SDW close by the QCP is preferable
to be an IC CSDW, which are manifestly substantiated and supported by the combination of
Table~\ref{table_criteria-fates} and Fig.~\ref{Fig_stable}.
Instead, other potential candidates, including $C_2$ ICS, DPMH, and MH SDW states
as well as $C_4$ IC SVC and SWC states, cannot survive in the neighboring regime
of the QCP as addressed in Appendix~\ref{Appendix_stab-magnetic}
(the abbreviations of states hereby are consistent
with those in Table~\ref{table_criteria-fates}'s).}\label{Fig_schematic_phase_diagram}
\end{figure}

Instead of global scenario, the focus of this paper is on finding specific magnetic states that reside close
to magnetic QPTs in the phase diagram of $\mathrm{Ba_{1-x}Na_xFe_2As_2}$~\cite{Goldman2010PRB,
Osborn2014NatureComm,Wang2016PRB,Andersen2017NComm,Hardy2018PRL-BaNaFeAs}.
This compound provides a versatile platform to investigate ordering-competition
impacts on stabilities of magnetic states and relations with SC state.
On one hand, it hosts a rather rich phase diagram with typical doping-tuned
magnetic QPTs compared to other $\mathrm{BaFe_2As_2}$ systems. It is of
unique interest to asseverate there exists an elusive $C_2$-symmetry ($C_2$)
magnetic phase in $\mathrm{Na}$-doped system
reported recently by Wang \emph{et al.}~\cite{Wang2016PRB}, which is hitherto
enigmatic and remains an open topic. On the other, three commensurate plus kinds
of IC magnetic states might all be possible candidates inhabiting in its phase diagram~\cite{Osborn2014NatureComm,Hardy2015NComm,Fernandes2016PRB,
Schmalian2016PRB,Wang2017PRB,Andersen2018PRX}. To be specific, the commensurate magnetic states involve the
stripe spin density wave (SDW), charge spin density wave (CSDW),
and spin vortex crystal (SVC)~\cite{Fernandes2016PRB,
Schmalian2016PRB,Yu1706,Schmalian2018PRB}. In addition, the IC magnetic states cover
four different $C_2$-symmetry ($C_2$) IC cases consisting of
$C_2$ IC stripe (ICS), $C_2$ magnetic helix (MH), $C_2$ IC
magnetic stripe with perpendicular magnetic helix (ICS $\perp$ MH),
and $C_2$ double parallel magnetic helix (DPMH), as well as
three distinct $C_4$-symmetry ($C_4$) IC situations involving $C_4$ IC CSDW,
$C_4$ IC SVC, and $C_4$ IC spin-whirl crystal (SWC)~\cite{Andersen2018PRX}.
Due to their own peculiarities, these distinct states
conventionally bring forward various outcomes. Questions are naturally raised:
which one is the prime $C_4$-symmetry ($C_4$) magnetic order in the shadow
of some QPT and what is the optimal state characterizing the mystic $C_2$ magnetic state? 
We respond by taking advantage of a phenomenological
theory together with the Wilsonian renormalization
group (RG)~\cite{Wilson1975RMP}. The answers
are of notable help to deeply understand the phase diagram and even offer instructive insights
into pairing mechanism. Fig.~\ref{Fig_schematic_phase_diagram}
schematically illustrates our central results driven by ordering competition.

The rest of paper is organized as follows. In
Sec.~\ref{Sec_model}, we establish the phenomenological effective theory
and provide the coupled RG equations after performing
the one-loop momentum-shell RG analysis. Next, within
Sec.~\ref{Sec_IC-states} we endeavor to select the most favorable
IC SDW states among all potential candidates under the influence of
strong quantum fluctuations induced by the QCP. Sec.~\ref{Sec_rho-lambda}
is accompanied by an investigation of unusual physical implications including
both superfluid density and London penetration depth caused by
the ferocious ordering competition near the QCP. Finally,
we briefly summarize the primary conclusions in Sec.~\ref{Sec_summary}.

%\vspace{0.35cm}

\section{Effective theory and RG analysis}\label{Sec_model}

\subsection{Effective theory}\label{Sec_S_eff}

The fermi surfaces of $\mathrm{BaFe_2As_2}$
compounds under a three-band model consist of
one hole pocket at the center of Brillouin zone $\mathbf{Q}_\Gamma = (0,0)$
and two electron pockets centered at two fixed momenta
$\mathbf{Q}_X=(\pi,0)$ and
$\mathbf{Q}_Y = (0,\pi)$~\cite{Knolle2010PRB,Fernandes2013PRL,
Chowdhury2013PRL,Chubukov2012ARCMP}.
From microscopical considerations, both magnetic and SC states are
rooted in interactions among excited quasiparticles from these Fermi pockets~\cite{Fernandes2013PRL,Chowdhury2013PRL,Fernandes2016PRB,
Schmalian2016PRB,Chubukov2008PRB,Knolle2010PRB,Chubukov2012ARCMP}. Concretely,
a magnetic state is composed of
two basic magnetic order parameters $\mathbf{M}_X$
and $\mathbf{M}_Y$, which are designated by $\mathbf{M}_j =
\sum_{\mathbf{k}}c^\dagger_{\Gamma,\mathbf{k}
\alpha}\vec{\sigma}_{\alpha\beta} c_{j,\mathbf{k} +
\mathbf{Q}_j\beta}$ with $j = X,Y$~\cite{Fang2008PRB,Chubukov2008PRB,
Fernandes_Schmalian2010PRL,Fernandes2013PRL,Chowdhury2013PRL}.
To involve IC magnetic states, ordering vectors are
afterwards distributed as $\mathbf{Q}_X=(\pi-\delta,0)$
and $\mathbf{Q}_Y=(0,\pi-\delta)$ with $\delta$ being a small correction
for generic wavevectors.
This indicates that the magnetic order
parameters are regarded as a complex quantity
$M_{\mathbf{Q}_{X,Y}}\neq M^*_{\mathbf{Q}_{X,Y}}\equiv M_{-\mathbf{Q_{X,Y}}}$,
which is in striking contrast to the commensurate case with $\delta=0$ and
$M^*_{Q_{X,Y}}=M_{Q_{X,Y}}$~\cite{Fernandes2016PRB,
Schmalian2016PRB,Andersen2018PRX}.

We begin with the extended Landau-Ginzuburg free energy after integrating out the fermionic ingredients~\cite{Schulz1990PRL,Fernandes2016PRB,
Schmalian2016PRB,Anderson1704,Andersen2018PRX}
\begin{eqnarray}
f&=&\alpha(|\mathbf{M}_X|^2+|\mathbf{M}_Y|^2)+\frac{\beta_2}{2}
(|\mathbf{M}_X|^2+|\mathbf{M}_Y|^2)^2\nonumber\\
&&+\frac{\beta_1-\beta_2}{2}(|\mathbf{M}^2_X|^2+|\mathbf{M}^2_Y|^2)
+(g_1-\beta_2)|\mathbf{M}_X|^2|\mathbf{M}_Y|^2\nonumber\\
&&+\frac{g_2}{2}(|\mathbf{M}_X\cdot\mathbf{M}_Y|^2
+|\mathbf{M}_X\cdot\mathbf{M}^*_Y|^2),\label{Eq_free-energy}
\end{eqnarray}
with $\alpha$, $\beta_{1,2}$, and $g_{1,2}$
being fundamental structure parameters.
It deserves to be pointed out that the QCP at $x_1$ in Fig.~\ref{Fig_schematic_phase_diagram}
associated with commensurate states was studied previously~\cite{Fernandes2016PRB,Schmalian2016PRB,
Wang2017PRB}. In order to determine the unknown $C_2$ and $C_4$ IC SDWs,
we hereafter concentrate on the magnetic QPT denoted by $x_c$
in Fig.~\ref{Fig_schematic_phase_diagram}.

After designating
$\mathbf{M}_X\equiv M_X\cos\theta \mathbf{n}_X$ and
$\mathbf{M}_Y\equiv M_Y\sin\theta \mathbf{n}_Y$, where
$\theta\in(0,\pi/2)$ and $|\mathbf{n}_{X,Y}|^2=1$ specify
the spin configurations of magnetic states, we go beyond mean-field level
and construct a phenomenological effective field theory~\cite{Fernandes2013PRL,
Fernandes2016PRB}, which captures main information of ordering competition
including both $C_{2,4}$-symmetric IC magnetic and SC
fluctuations~\cite{Halperin1974PRL,Kleinert2003NPB,Wang2014PRD,Wang2017PRB}.
To this end, the phenomenological effective action~\cite{Fernandes2013PRL,
Fernandes2016PRB} can be casted as
%\begin{eqnarray}
%S=\int\mathcal{L}=\int\mathcal{L}_{\mathrm{SDW}}+\int\mathcal{L}_{\mathrm{SC}}
%+\int\mathcal{L}_{\mathrm{SDW-SC}},\label{Eq_L}
%\end{eqnarray}
\begin{eqnarray}
S=\!\int \!d^d\mathcal{L}=\!\!\int \!d^d\mathcal{L}_{\mathrm{SDW}}+\!\int \!d^d\mathcal{L}_{\mathrm{SC}}
+\!\int \!d^d\mathcal{L}_{\mathrm{SDW-SC}},\label{Eq_L}
\end{eqnarray}
where $\mathcal{L}_{\mathrm{SDW}}$, $\mathcal{L}_{\mathrm{SC}}$, and
$\mathcal{L}_{\mathrm{SDW-SC}}$ correspond to SDW, SC orders, and their interplay,
respectively.

At first, we examine $\mathcal{L}_{\mathrm{SDW}}$.
An angle $\theta\in[0,\pi/2]$ is employed to specify the direction of
magnetic order parameter $\mathbf{M}$ in the spin space. Accordingly,
the order parameter can be divided into two components
$\mathbf{M}_X\equiv M_X\cos\theta \mathbf{n}_X$ and
$\mathbf{M}_Y\equiv M_Y\sin\theta \mathbf{n}_Y$ by projecting $\mathbf{M}$
onto the spin vectors $\mathbf{n}_X$ and $\mathbf{n}_Y$, which characterize the spin configurations
of magnetic states with $|\mathbf{n}_{X,Y}|^2=1$ and
whose concrete values depending upon the types of candidate
states~\cite{Andersen2018PRX}. Inserting them into
the free energy density~(\ref{Eq_free-energy}) by adding
the dynamical terms of magnetic order parameters then gives rise
to~\cite{Schulz1990PRL,Fernandes2013PRL,
Fernandes2016PRB,Wang2017PRB}
\begin{widetext}
\vspace{-0.5cm}
\begin{eqnarray}
\mathcal{L}_{\mathrm{SDW}}
&=&\Bigl[|\mathbf{n}_X\cos\theta|^2\frac{1}{2}(\partial_\mu M_X)^2
+\alpha(|\mathbf{n}_X|^2\cos^2\theta)M^2_X\Bigr]
+\Bigl[|\mathbf{n}_Y\sin\theta|^2\frac{1}{2}(\partial_\mu M_Y)^2
+\alpha(|\mathbf{n}_Y|^2\sin^2\theta)M^2_Y\Bigr]\nonumber\\
&&+\frac{\beta_1-\beta_2}{2}(|\mathbf{n}^2_X|^2\cos^4\theta M^4_X
+|\mathbf{n}^2_Y|^2\sin^4\theta M^4_Y)+
\frac{\beta_2}{2}(|\mathbf{n}_X|^4\cos^4\theta M^4_X +|\mathbf{n}_Y|^4\sin^4\theta M^4_Y)\nonumber\\
&&+g_1|\mathbf{n}_X|^2 |\mathbf{n}_Y|^2\cos^2\theta\sin^2\theta  M^2_XM^2_Y +\frac{g_2}{2}\cos^2\theta\sin^2\theta
(|\mathbf{n}_X\cdot \mathbf{n}_Y|^2+|\mathbf{n}_X\cdot \mathbf{n}^*_Y|^2)M^2_XM^2_Y.\label{Eq_SDW}
\end{eqnarray}
\end{widetext}

We next consider $\mathcal{L}_{\mathrm{SC}}$.
In order to obtain SC fluctuations in the ordered state, we bring out the
the following contribution by employing the condition
$\partial_\mu A_\mu=0$~\cite{Halperin1974PRL}
\begin{eqnarray}
\mathcal{L}_{\mathrm{SC}}
&=&\partial_\mu\Delta^\dagger\partial_\mu \Delta+a_s\Delta^2(k)
+\frac{u_s}{2}\Delta^4(k)+\frac{\alpha_A}{2}A^2\nonumber\\
&&-\frac{1}{4}\left(\partial_\mu A_\nu-\partial_\nu A_\mu\right)^2
+\lambda_{\Delta A}|\Delta|^2A^2.\label{Eq_SC}
\end{eqnarray}
As the system enters the SC ordered state around the SDW QCP,
we need to expand the SC order parameter by introducing two new gapless fields
\begin{eqnarray}
\Delta=V_0+\frac{(h+i\eta)}{\sqrt{2}},\hspace{0.02cm}
\langle h \rangle=\langle \eta \rangle=0,
V_0\equiv\langle \Delta\rangle=\sqrt{\frac{-a_s}{u_s}},\label{Eq_SC-2}
\end{eqnarray}
which help us to extract the potential fluctuation of SC order parameter~\cite{Kleinert2003NPB},
to make the $\mathbf{A}$ massive after absorbing the gapless Goldstone particles.
Combing Eq.~(\ref{Eq_SC}) and Eq.~(\ref{Eq_SC-2}), after discarding the constant terms and
choosing some transformation to make $\eta=0$ due to the local gauge
invariance~\cite{Kleinert2003NPB}, we obtain
\begin{eqnarray}
\mathcal{L}_{\mathrm{SC}}
&=&\frac{1}{2}(\partial_\mu h)^2-a_sh^2+\frac{u_s}{8}h^4
+\frac{\sqrt{-2a_su_s}}{2}h^3\nonumber\\
&&-\frac{1}{4}\left(\partial_\mu A_\nu-\partial_\nu A_\mu\right)^2
+\frac{\alpha_A}{2}A^2\nonumber\\
&&+\lambda_{\Delta A}\sqrt{\frac{-2a_s}{u_s}}hA^2
+\frac{\lambda_{\Delta A}}{2}h^2A^2,
\end{eqnarray}
where the ``mass" of field $A$ is defined as $\alpha_A\equiv\lambda_{\Delta A}\frac{-2a_s}{u_s}$.

Finally, we introduce $\mathcal{L}_{\mathrm{SC}}$.
The interplay between SC and SDW order parameters can be written as~\cite{Wang2017PRB},
\begin{eqnarray}
\mathcal{L}_{\mathrm{SDW-SC}}
&=&\lambda(|\mathbf{M}_X|^2+|\mathbf{M}_Y|^2)\Delta^2+
\kappa(|\mathbf{M}_X\cdot\mathbf{M}_Y|\nonumber\\
&&+|\mathbf{M}_X\cdot\mathbf{M}^*_Y|)\Delta^2.\label{Eq_SDW-SC}
\end{eqnarray}
Based on the information of $\mathcal{L}_{\mathrm{SDW}}$ and
$\mathcal{L}_{\mathrm{SC}}$, we are left with our
effective theory
\begin{widetext}
\vspace{-0.5cm}
\begin{eqnarray}
\mathcal{L}_{\mathrm{eff}}
&=&\left[\frac{1}{2}(\partial_\mu M_X/\mathcal{C})^2
+\alpha_XM^2_X+\frac{\beta_X}{2}M^4_X\right]
+\left[\frac{1}{2}(\partial_\mu M_Y/\mathcal{S})^2
+\alpha_YM^2_Y+\frac{\beta_Y}{2}M^4_Y\right]
+\left[-\frac{1}{4}\left(\partial_\mu A_\nu-\partial_\nu A_\mu\right)^2
+\frac{\alpha_A}{2}A^2\right]\nonumber\\
&&+\left[\frac{1}{2}(\partial_\mu h)^2+a_hh^2+\frac{\beta_h}{2}h^4
+\gamma_hh^3\right]+\alpha_{XY}M_XM_Y+\gamma_{XYh}M_XM_Yh
+\gamma_{X^2h}M^2_Xh+\gamma_{Y^2h}M^2_Yh+\gamma_{hA^2}hA^2\nonumber \\ \nonumber \\
&&+\lambda_{XY}M^2_XM^2_Y+\lambda_{Xh}M^2_Xh^2
+\lambda_{Yh}M^2_Yh^2+\lambda_{XYh}M_XM_Yh^2+\lambda_{hA}h^2A^2,\label{Eq_L-eff}
\end{eqnarray}
\end{widetext}
with $\mathcal{C}\equiv1/|\mathbf{n}_X\cos\theta|^2$
and $\mathcal{S}\equiv1/|\mathbf{n}_Y\sin\theta|^2$.
$M_{X,Y}$ point to magnetic fluctuations and $h, A$
are auxiliary fields to absorb SC fluctuations.
We here dub factors in (\ref{Eq_L-eff}) such as $\alpha_X$ etc.
the effective parameters to prevent them from being confused with fundamental parameters
appearing in Eq.~(\ref{Eq_free-energy}). Two series of parameters are bridged by virtue
of following relationships,
\begin{widetext}
\vspace{-0.39cm}
\begin{eqnarray}
\alpha_h&\equiv&(-a_s),\beta_h\equiv\frac{u_s}{4},\gamma_h\equiv\frac{\sqrt{-2a_su_s}}{2},
\alpha_A\equiv\frac{-2\lambda_{\Delta A}a_s}{u_s},
\gamma_{hA^2}\equiv\lambda_{\Delta A}\sqrt{\frac{-2a_s}{u_s}},
\lambda_{hA}\equiv\frac{\lambda_{\Delta A}}{2},\label{Eq_bridge-1}\\
\alpha_{X}&\equiv&\left(a-\frac{\lambda a_s}{u_s}\right)(|\mathbf{n}_X|^2\cos^2\theta),
\beta_{X}\equiv\beta_2\Bigl(|\mathbf{n}_X|^4\cos^4\theta\Bigr)
+(\beta_1-\beta_2)\Bigl(|\mathbf{n}^2_X|^2\cos^4\theta\Bigr),\label{Eq_bridge-2}\\
\alpha_{Y}&\equiv&\left(a-\frac{\lambda a_s}{u_s}\right)(|\mathbf{n}_Y|^2\sin^2\theta),
\beta_{Y}\equiv\beta_2\Bigl(|\mathbf{n}_Y|^4\sin^4\theta\Bigr)
+(\beta_1-\beta_2)\Bigl(|\mathbf{n}^2_Y|^2\sin^4\theta \Bigr),\label{Eq_bridge-3}\\
\alpha_{XY}&\equiv&\frac{-a_s\kappa}{u_s}(|\cos\theta\sin\theta\mathbf{n}_X\cdot \mathbf{n}_Y|
+|\cos\theta\sin\theta\mathbf{n}_X\cdot \mathbf{n}^*_Y|),\label{Eq_bridge-4}\\
\gamma_{XYh}&=&\kappa\sqrt{\frac{-2a_s}{u_s}}(|\cos\theta\sin\theta\mathbf{n}_X\cdot \mathbf{n}_Y|+|\cos\theta\sin\theta\mathbf{n}_X\cdot \mathbf{n}^*_Y|),\label{Eq_bridge-5}\\
\gamma_{X^2h}&\equiv&\lambda\sqrt{\frac{-2a_s}{u_s}}
\Bigl(|\mathbf{n}_X|^2\cos^2\theta\Bigr),
\gamma_{Y^2h}\equiv\lambda\sqrt{\frac{-2a_s}{u_s}}
\Bigl(|\mathbf{n}_Y|^2\sin^2\theta\Bigr),\label{Eq_bridge-6}\\
\lambda_{XY}&\equiv&g_1
\cos^2\theta\sin^2\theta \Bigl(|\mathbf{n}_X|^2 |\mathbf{n}_Y|^2\Bigr)
+\frac{g_2}{2}\cos^2\theta\sin^2\theta\Bigl(|\mathbf{n}_X\cdot \mathbf{n}_Y|^2
+|\mathbf{n}_X\cdot \mathbf{n}^*_Y|^2\Bigr),\label{Eq_bridge-7}\\
\lambda_{Xh}&\equiv&\frac{\lambda}{2}\Bigl(|\mathbf{n}_X|^2\cos^2\theta\Bigr),
\lambda_{Yh}\equiv\frac{\lambda}{2}
\Bigl(|\mathbf{n}_Y|^2\sin^2\theta\Bigr),\label{Eq_bridge-8}\\
\lambda_{XYh}&\equiv&\frac{\kappa}{2}(|\cos\theta\sin\theta\mathbf{n}_X\cdot \mathbf{n}_Y|+|\cos\theta\sin\theta\mathbf{n}_X\cdot \mathbf{n}^*_Y|),\label{Eq_bridge-9}
\end{eqnarray}
\end{widetext}
where $\kappa$ and $\lambda_{\Delta A}$ cannot
be represented by original parameters $a$, $a_s$, $u_s$, $\lambda$, $\beta_1$, $\beta_2$,
$g_1$, and $g_2$ appearing in the free energy~(\ref{Eq_free-energy}),
and therefore comes up with two supplementary fundamental parameters.
It is necessary to point out these effective
parameters are intermediate auxiliary variables
but instead the fundamental parameters play a
central role in pinning down the specific SDW sates.

\subsection{RG analysis}

As aforementioned in Sec.~\ref{Sec_S_eff},
the concrete spin configuration state would be essentially determined by the
fundamental parameters. In order to examine the stabilities
of all potential states, we need to construct the energy-dependent
coupled RG equations of the fundamental parameters.
To this end, we compute one-loop corrections to all effective parameters
in Eq.~(\ref{Eq_L-eff}) and derive the corresponding RG evolutions within
Wilsonian RG framework~\cite{Wilson1975RMP,
Wang2014PRD,Wang2017PRB} via integrating out the fast fields in the
momentum shell $e^{-l}\Lambda<k<\Lambda$ with the running scale $l>0$.
Since the fundamental parameters defined in Eq.~(\ref{Eq_free-energy})
dictate the physical properties, it heralds undeviatingly that a pillar
of task consists in refining their flow equations.
To this end, we resort to the strategy in Refs.~\cite{Wang2014PRD,Wang2017PRB}.
Combining RG flows of effective parameters and connections with
fundamental parameters~(\ref{Eq_bridge-1})-(\ref{Eq_bridge-9}) yields
a set of coupled RG equations
\begin{eqnarray}
\frac{d \mathcal{X}_i}{d l} = \sum_j\mathcal{F}_{ij}\mathcal{X}_j,\label{Eq_flow_X_i}
\end{eqnarray}
with $\mathcal{X}_{i/j}$ serving as the fundamental parameters~\cite{notice} and $\mathcal{F}_{ij}$
standing for evolution coefficients as a function of $\mathcal{X}_{i/j}$.
This necessitates bearing in mind that the coupled RG evolutions
hinge heavily upon the spin configurations of magnetic
fluctuations, namely the relationships between $|\mathbf{n}^2_X|^2$, $|\mathbf{n}_X|^4$,
$|\mathbf{n}^2_Y|^2$, and $|\mathbf{n}_Y|^4$, which give rise to seven independent classes
of RG evolutions. The details of Eq.~(\ref{Eq_flow_X_i}) are stored completely
in Appendix~\ref{Appendix-coupled-RG}.

\begin{table*}[htbp]
\caption{Collections of low-energy fates for IC magnetic states in
$\mathrm{Ba_{1-x}Na_xFe_2As_2}$. The first line enumerates seven distinguished
types of IC magnetic states as well as the second and third lines provide their related spin
configurations and schematic illustrations, respectively.
In addition, the fourth line shows stable constraints
as functions of fundamental interaction parameters~\cite{Andersen2018PRX}
and the last line presents the corresponding low-energy stabilities.
Herein, \ding{52} and \ding{56} stand for a stable state (i.e., the prevailing candidate
by the side of the magnetic QCP) and
an unstable state, respectively.}\label{table_criteria-fates}
\vspace{0.5cm}
\hspace{-0.9cm}
\begin{tabular}{p{1.6cm}<{\centering}p{2.55cm}<{\centering}p{2.55cm}<{\centering}
p{1.80cm}<{\centering}
p{2.55cm}<{\centering}p{2.55cm}<{\centering}p{1.9cm}<{\centering}p{1.76cm}<{\centering}}
\hline
\hline
\rule{0pt}{15pt}IC magnetic states & $\mathbb{ICS}$ & $\mathbb{MH}$ & $\mathbb{ICS}$ $\perp$ $\mathbb{MH}$
& $\mathbb{DPMH}$ & $\mathbb{IC}$ $\mathbb{CSDW}$ & $\mathbb{IC}$ $\mathbb{SVC}$ & $\mathbb{SWC}$  \\
\rule{0pt}{15pt}Spin configurations
& $\mathbf{n}_X=(0,0,1), \mathbf{n}_Y=(0,0,0)$
& $\mathbf{n}_X=\frac{1}{\sqrt{2}}(i,0,1), \mathbf{n}_Y=(0,0,0)$
& $\mathbf{n}_X=(0,0,1),\mathbf{n}_Y=\frac{1}{\sqrt{2}}(i,1,0), $
& $\mathbf{n}_X=\mathbf{n}_Y=\frac{1}{\sqrt{2}}(i,0,1), $
& $\mathbf{n}_X=(0,0,1), \mathbf{n}_Y=(0,0,1)$
& $\mathbf{n}_X=(0,0,1), \mathbf{n}_Y=(0,1,0)$
& $\mathbf{n}_X=\frac{1}{\sqrt{2}}(i,0,1),\mathbf{n}_Y=\frac{1}{\sqrt{2}}(0,i,1)$    \\
\rule{0pt}{15pt}Schematic illustrations
& Fig.~\ref{Fig_spin-configuration}(a)
& Fig.~\ref{Fig_spin-configuration}(d)
& Fig.~\ref{Fig_spin-configuration}(f)
& Fig.~\ref{Fig_spin-configuration}(g)
& Fig.~\ref{Fig_spin-configuration}(b)
& Fig.~\ref{Fig_spin-configuration}(c)
& Fig.~\ref{Fig_spin-configuration}(h)   \\
\rule{0pt}{15pt}Stable constraints
& $\beta_1-\beta_2<0$ with $\frac{g_2}{|\beta_1-\beta_2|}>0$,
$\frac{g_1-\beta_2}{|\beta_1-\beta_2|}>-1$
or $\frac{g_2}{|\beta_1-\beta_2|}<0$,
$\frac{g_1-\beta_2-0.9g_2}{|\beta_1-\beta_2|}>-1$
& $\beta_1-\beta_2>0$ with
$\frac{g_2}{|\beta_1-\beta_2|}>0$,
$\frac{g_1-\beta_2}{|\beta_1-\beta_2|}>0$
or $\frac{g_2}{|\beta_1-\beta_2|}<0$,
$\frac{g_1-\beta_2-0.9g_2}{|\beta_1-\beta_2|}>-1$
& $\beta_1-\beta_2>0$, $\frac{g_2}{|\beta_1-\beta_2|}>2$,
$\frac{g_1-\beta_2}{|\beta_1-\beta_2|}<0$
& $\beta_1-\beta_2>0$,
$\frac{g_2}{|\beta_1-\beta_2|}<0$,
$\frac{g_1-\beta_2-0.9g_2}{|\beta_1-\beta_2|}<-1$
& $\beta_1-\beta_2<0$, $\frac{g_2}{|\beta_1-\beta_2|}<0$,
$\frac{g_1-\beta_2-0.9g_2}{|\beta_1-\beta_2|}<-1$
or $\beta_1-\beta_2>0$, $\frac{g_2}{|\beta_1-\beta_2|}<-1$,
$\frac{g_1-\beta_2-0.9g_2}{|\beta_1-\beta_2|}<-1$
& $\beta_1-\beta_2<0$,
$\frac{g_2}{|\beta_1-\beta_2|}>0$,
$\frac{g_1-\beta_2}{|\beta_1-\beta_2|}<-1$
& $\beta_1-\beta_2>0$, $0<\frac{g_2}{|\beta_1-\beta_2|}$,
$\frac{g_2}{|\beta_1-\beta_2|}<2$,
$\frac{g_1-\beta_2}{|\beta_1-\beta_2|}<0$ \\
\rule{0pt}{15pt}Fates of magnetic states
& \ding{56}
& \ding{56}
& \ding{52}
& \ding{56}
& \ding{52}
& \ding{56}
& \ding{56}   \\
\hline
\hline
\end{tabular}
\end{table*}

\vspace{0.2cm}
\section{Stabilities of incommensurate magnetic states}\label{Sec_IC-states}

\begin{figure}
\hspace{-0.0cm}
\includegraphics[width=1.0in]{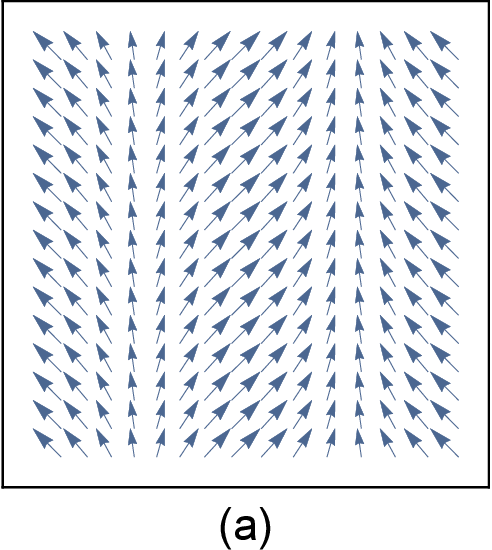}\hspace{0.2cm}
\includegraphics[width=1.0in]{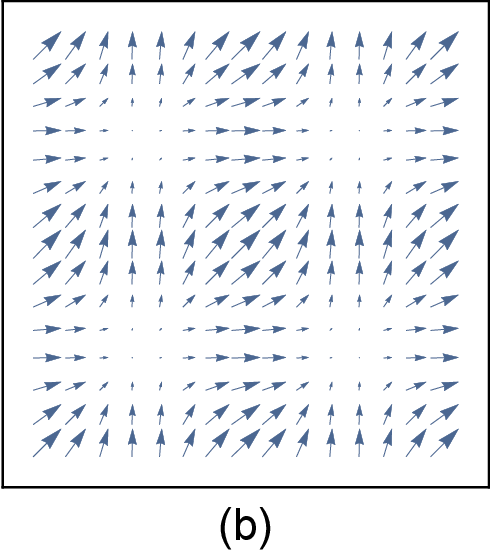}\hspace{0.2cm}
\includegraphics[width=1.0in]{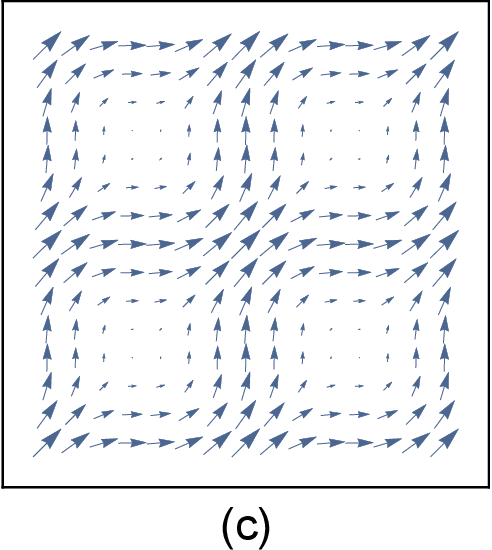}
\vspace{0.15cm}
\\
\hspace{-0.0cm}
\includegraphics[width=1.0in]{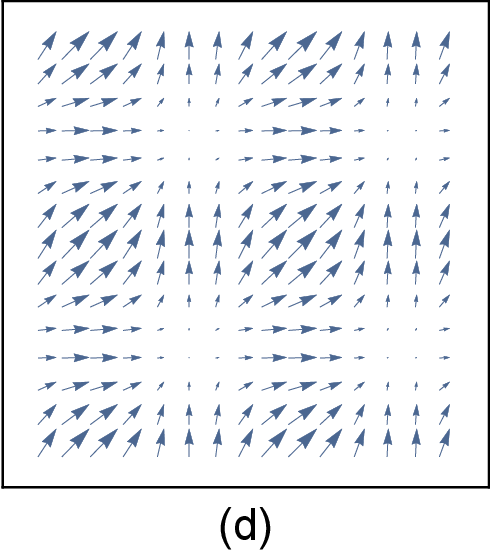}\hspace{0.2cm}
\includegraphics[width=1.0in]{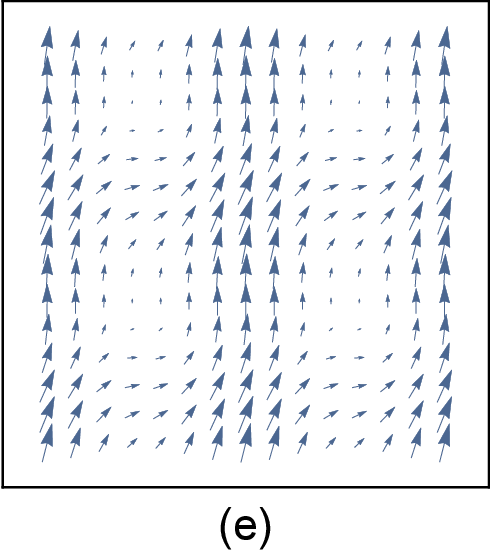}\hspace{0.2cm}
\includegraphics[width=1.0in]{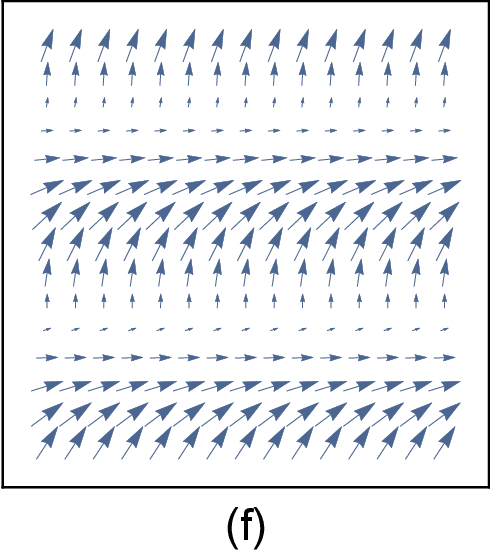}
\vspace{0.15cm}
\\
\hspace{-0.0cm}
\includegraphics[width=1.0in]{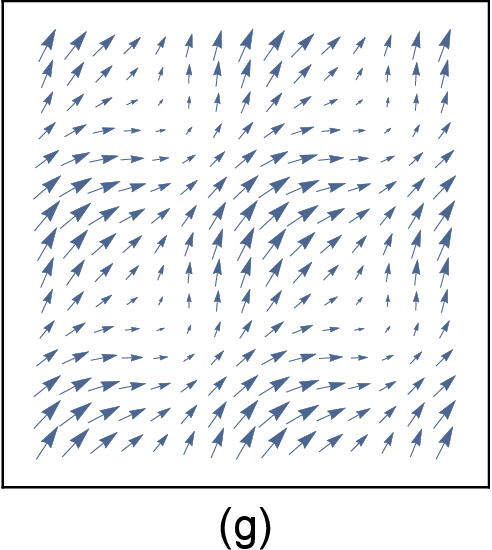}\hspace{0.2cm}
\includegraphics[width=1.0in]{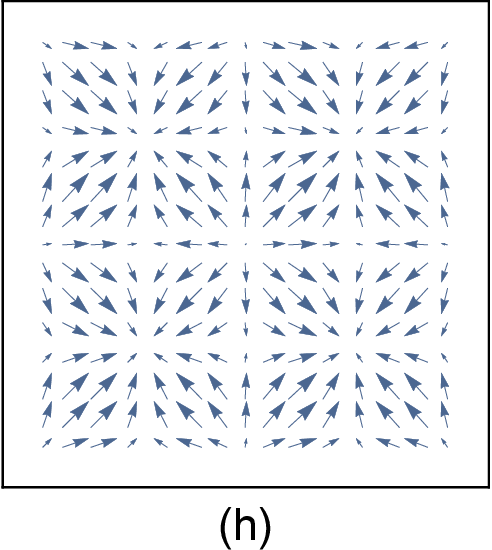}\hspace{0.2cm}
\includegraphics[width=1.0in]{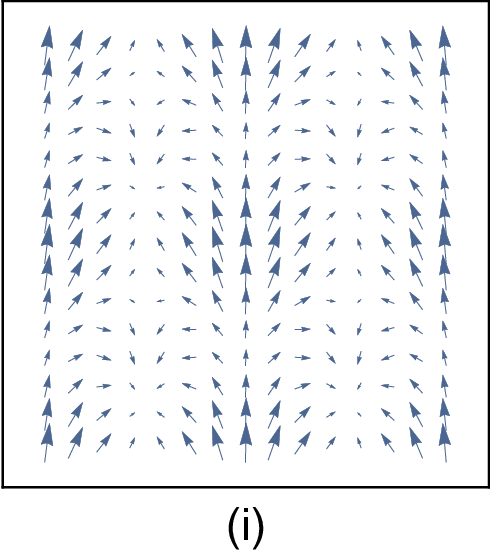}
\vspace{0.05cm}
\caption{(Color online) Schematic profile illustrations for nine distinct types of IC
spin configurations (at the starting point of RG flows)~\cite{Andersen2018PRX}:
(a) $C_2$ ICS , (b) $C_4$ IC CSDW, (c) $C_4$ IC SVC, (d) $C_2$ IC MH,
(e) $C_2$ ICS $\parallel$ MH, (f) $C_2$ ICS $\perp$ MH, (g) $C_2$
IC DPMH, (h) $C_4$ IC SWC, and (i) $C_2$ IC SWC.}\label{Fig_spin-configuration}
\end{figure}

With the help of energy-dependent flows of fundamental parameters, we
are now in a suitable situation to % go beyond the mean-field analysis and
study the stabilities of IC magnetic states triggered by some magnetic QCP.
As to $\mathrm{BaFe_2As_2}$ compounds, many experimental efforts~\cite{Goldman2010PRB,Osborn2014NatureComm,Wang2016PRB,Hardy2015NComm,
Allred2015PRB,Andersen2017NComm,Hardy2018PRL-BaNaFeAs,
Prozorov2019PRB-BaKFeAs}
corroborate that magnetism occupies major space of phase diagram
in terms of various states with distinguished symmetries
and spin configurations. In particular, compound $\mathrm{Ba_{1-x}Na_xFe_2As_2}$~\cite{Osborn2014NatureComm,Wang2016PRB,
Andersen2017NComm,Hardy2018PRL-BaNaFeAs}
harbors a complicated but fascinating phase diagram sketched in
Fig.~\ref{Fig_schematic_phase_diagram}, indicating a string of magnetic states for both
$C_2$ and $C_4$ symmetries are allowed with proper variations
of temperature and doping. Besides three commensurate
states, i.e., stripe spin density wave (SSDW), charge spin density wave (CSDW),
and spin vortex crystal (SVC)~\cite{Fernandes2016PRB,
Schmalian2016PRB,Yu1706,Schmalian2018PRB}, Christensen \emph{et al.}~\cite{Andersen2018PRX}
recently advocated that potential IC magnetic states are clustered into
nine inequivalent breeds. Moreover, seven of them can be realized
with confined parameters of mean-field free energy in the phase
diagram~\cite{Andersen2018PRX}, which cover four kinds of $C_2$ IC
cases involving $C_2$ IC stripe (ICS), $C_2$ magnetic helix (MH), $C_2$
IC magnetic stripe with perpendicular magnetic helix (ICS $\perp$ MH), and
$C_2$ double parallel magnetic helix (DPMH), as well as three distinct
$C_4$ IC situations consisting of $C_4$ IC CSDW,
$C_4$ IC SVC, and $C_4$ IC spin-whirl crystal (SWC).
Their spin configurations as well as their stability constraints and final fates are
catalogued point-to-point in Table~\ref{table_criteria-fates}.
In order to roughly capture the structural information of
distinct types of SDW states, Fig.~\ref{Fig_spin-configuration} presents
the relevant schematic illustrations of related spin configurations
for potential IC magnetic states.

\subsection{Setup and Strategy}\label{Sec_strategy}

Despite being an underlying antagonist against SC state,
magnetism is assumed to be of intimate relevance to
superconductivity as they are closely adjacent to each other or
even coexist near the magnetic QPT.
To be concrete, we concentrate on a particular point in Fig.~\ref{Fig_schematic_phase_diagram},
namely the QCP at $T=0$ that separates $C_2$ and $C_4$ IC magnetic states
labeled by $x_c$. Generally, the related magnetic
fluctuations compete so furiously that are always
responsible for physics in the shadow of QPT
including quantum critical regime with higher
temperatures~\cite{Vojta2003RPP,Fernandes2013PRL,Chowdhury2013PRL}.
Considering that individualities of diverse states, in spite of hosting common
magnetic generalities, have different consequences, we thereafter
contemplate the magnetic states on both sides of this QPT.

As it concerns the issue on intricate relationship between magnetism and
superconductivity, a hallmark of fathoming overall phase diagram is
tantamount to pinpointing the specific construction of each magnetic state.
As a corollary, it is appropriate that one investigates
how the ordering competition affects the magnetic state at the edge of the QCP by means
of RG flows~(\ref{Eq_flow_X_i}) in collaboration with the stable magnetic criteria
itemized in the second line of Table~\ref{table_criteria-fates}.

\begin{figure}
\centering
\includegraphics[width=2.5in]{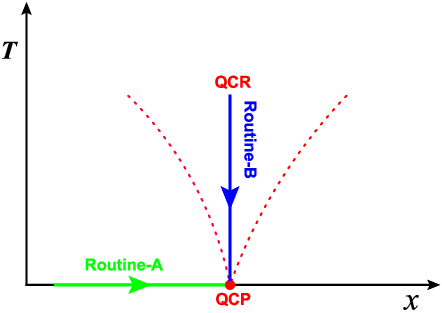}\\
\vspace{-0.15cm}
\caption{(Color online) Schematic illustrations for two distinct routines,
Routine-A and Routine-B, to access the quantum critical point (QCP)
in the $T-x$ plane with $T$ and $x$ corresponding to temperature and
non-thermal parameter, respectively.
Hereby, QCP and QCR denote the quantum critical point and quantum critical regime~\cite{Vojta2003RPP}.}\label{Fig1_routines}
\end{figure}

\begin{figure}
\centering
\includegraphics[width=3.5in]{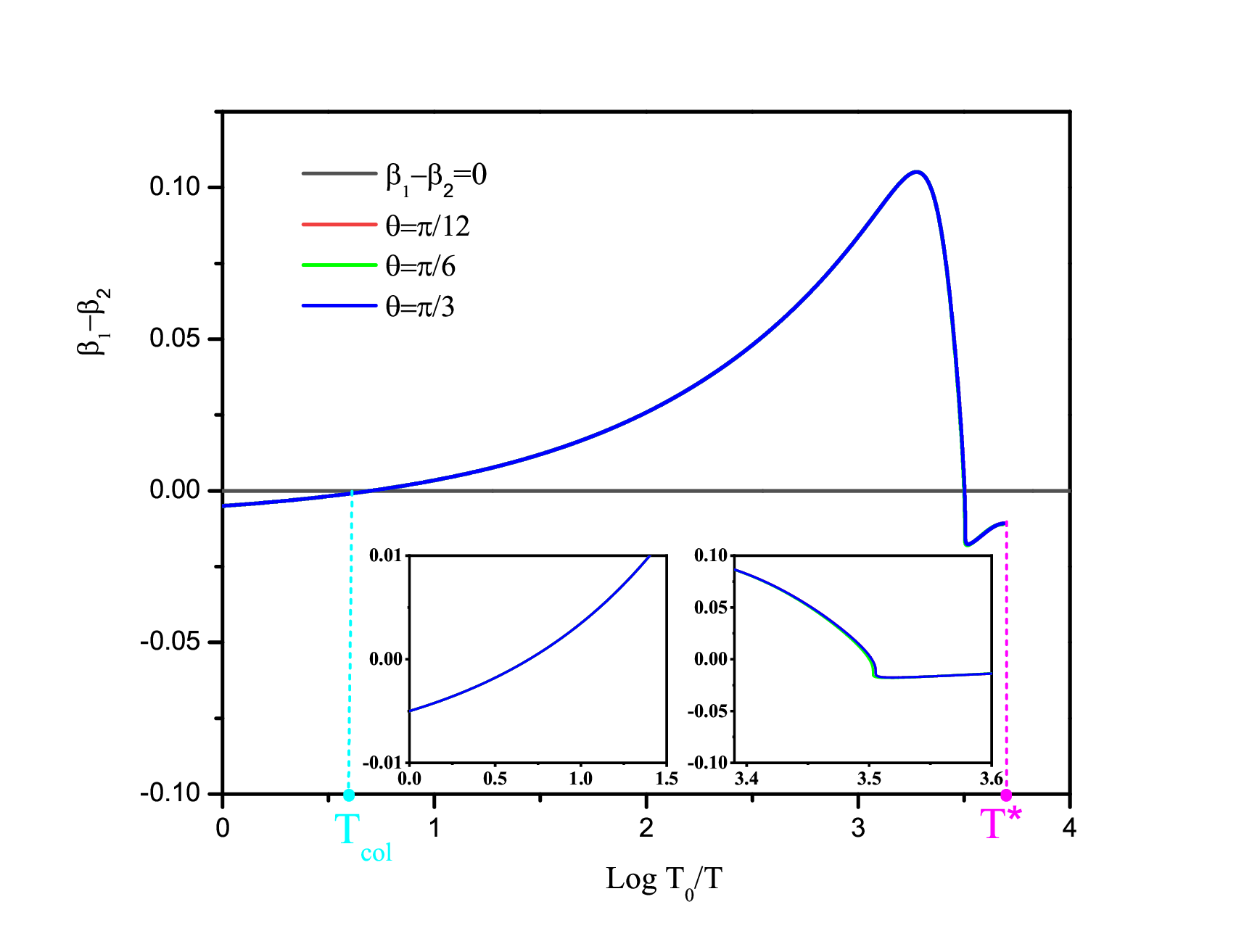}
\vspace{-0.65cm}
\caption{(Color online) Temperature-dependent stable constraints of
the $C_2$-symmetry ICS state. Hereby, the angle $\theta$ is designated
in Sec.~\ref{Sec_S_eff} to specify the direction of magnetic order
in the spin space and the initial values of related
fundamental parameters are chosen as
($g_1=0.01$, $g_2=0.01$, $u_s=0.05$, $\lambda=0.01$,
$\beta_1=0.005$, $\beta_2=0.01$) for $g_2>0$ (the $g_2<0$ case
exhibits the similar results and hence are not shown) with three
representative $\theta=\pi/12,\pi/6,\pi/3$ to satisfy the
ICS's stable constraint (the qualitative results are insensitive to
initial values of parameters). Apparently, the sign of $\beta_1-\beta_2$
is changed once temperature is slightly lowered at $T_{\mathrm{col}}$
(hereby $T_{\mathrm{col}}$ codifies the very temperature
at which the ICS stable constrain is jeopardized and collapsed),
indicating the violation of the stability constraint which does not
make it a good candidate to survive under the strong
ordering competition. In addition, the $T^\ast$ is associated
with the underlying fixed point and its related comments
are addressed in Sec.~\ref{Sec_comments}.}\label{Fig_C2-IC-stripe}
\end{figure}

To proceed, we briefly address our strategy to judge which magnetic candidate
is the most stable/favorable state. In principle, there exist two distinct routines
to access the QCP as schematically illustrated in Fig.~\ref{Fig1_routines}.
Although the essential physics should be captured by either routine-A or routine-B,
we are supposed to witness the physical behaviors along routine-B
to examine the stabilities of all potential states.
In order to be relevant with the schematic phase diagram,
we adopt $T=T_0e^{-l}$ with $T_0$ the initial temperature to
measure the evolution variable~\cite{Wang2014PRD,Wang2017PRB,
Fernandes2012PRB,Chubukov2012NP,She2015PRB,
Balents2014PRX,Lee2017PRX,Huh2008PRB,
Xu2008PRB,Foster2008PRB,Chubukov2016PRX,Metlitski2015PRB}.
Subsequently, several procedures are followed to investigate whether a
certain SDW state is a suitable candidate. At first, one
needs to tune the initial values of fundamental interaction parameters
to satisfy the corresponding stable constraint and hence make
sure that the starting point ($T_0$) is located at such SDW state
in the quantum critical regime (QCR) of Fig.~\ref{Fig1_routines}.
While we assume the constraint condition is developed
in a regime away from the QCP, this SDW state is always stable owing to the
absence of quantum fluctuations. In comparison, as approaching the QCP,
we have to carefully check whether the stable constraints are still
satisfied as the quantum fluctuations become more and more important
and play a dominant role in selecting the potential states. To this end,
one presents the energy-dependent behaviors of these
restrictions after extracting the information from the coupled RG equations
of all fundamental interaction parameters~(\ref{Eq_flow_X_i}).
With these in hand, it is suitable to determine the stability of such
SDW for accessing the QCP. It would be a preferable state once the
constraint is well preserved by lowering the energy scale (approaching the QCP).
Otherwise, the state is easily melted by ferocious fluctuations
and henceforth not a good candidate.

Accordingly, parallelling the
similar steps above, we can examine the stabilities of all candidate states
one by one on an equal footing and finally select the most favorable SDW
states nearby the QCP, which are schematically summarized in
Fig.~\ref{Fig_schematic_phase_diagram} and analyzed in the forthcoming sections.

\begin{figure*}
\hspace{-1.3cm}
\includegraphics[width=2.0in]{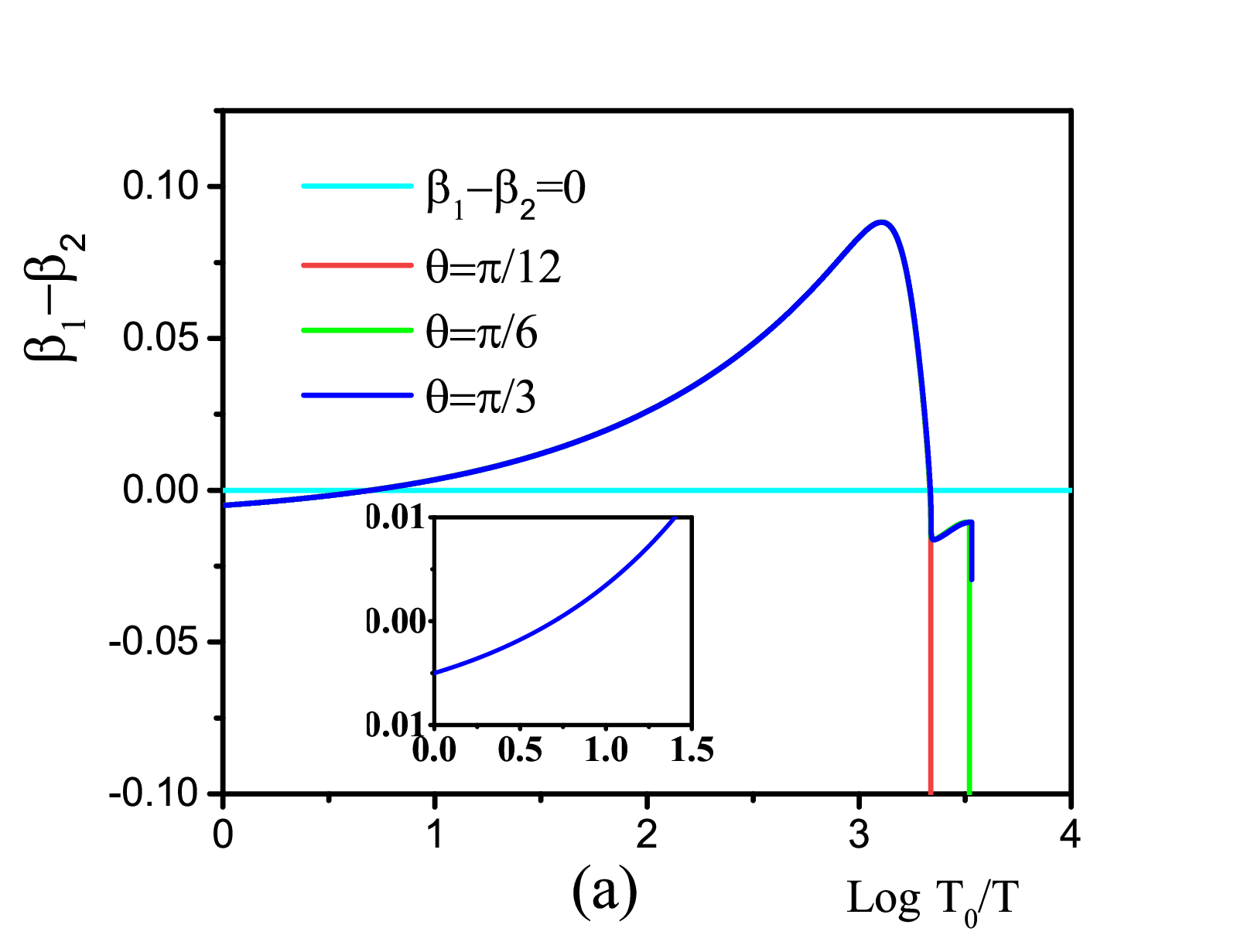}\hspace{-0.8cm}
\includegraphics[width=2.0in]{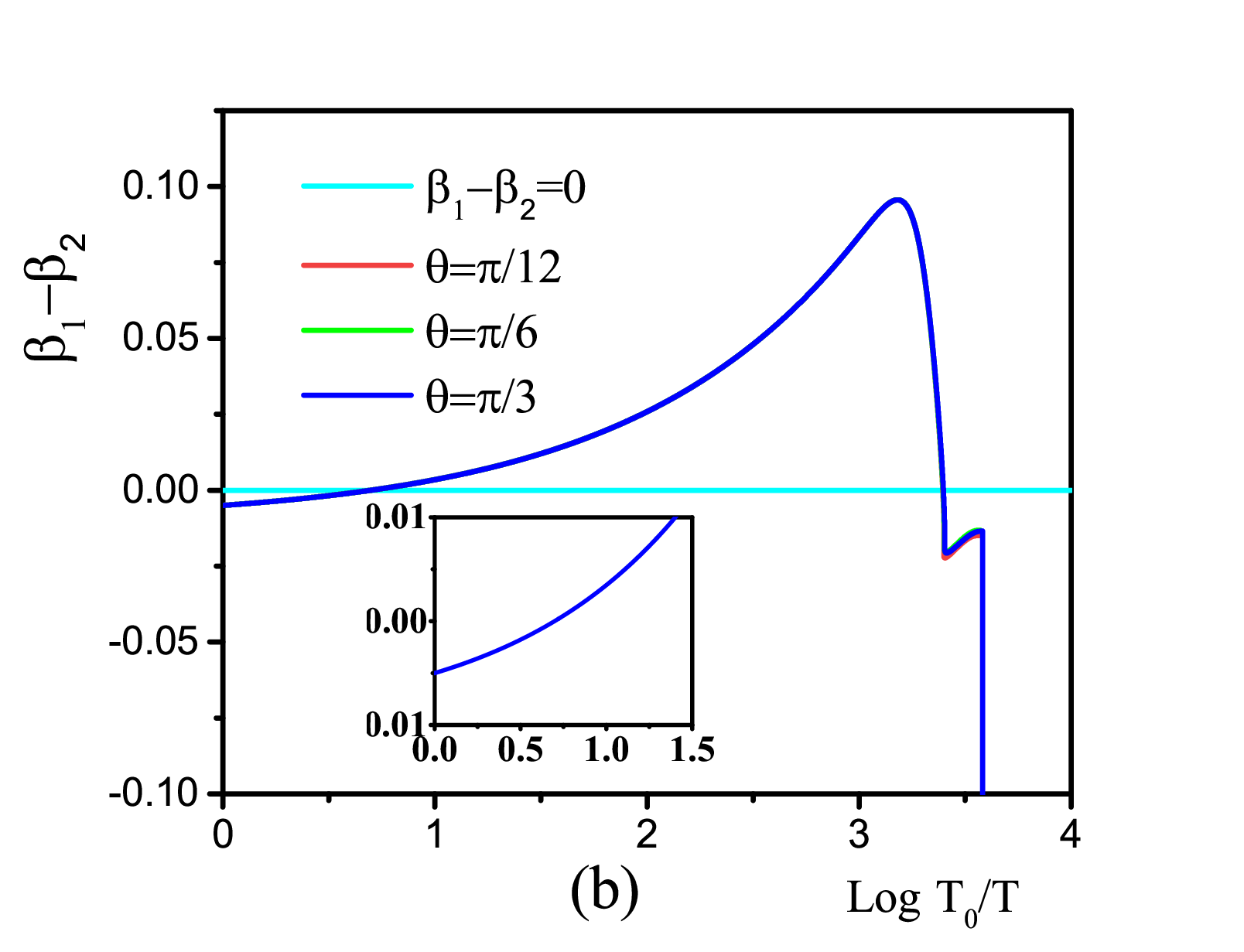}\hspace{-0.8cm}
\includegraphics[width=2.0in]{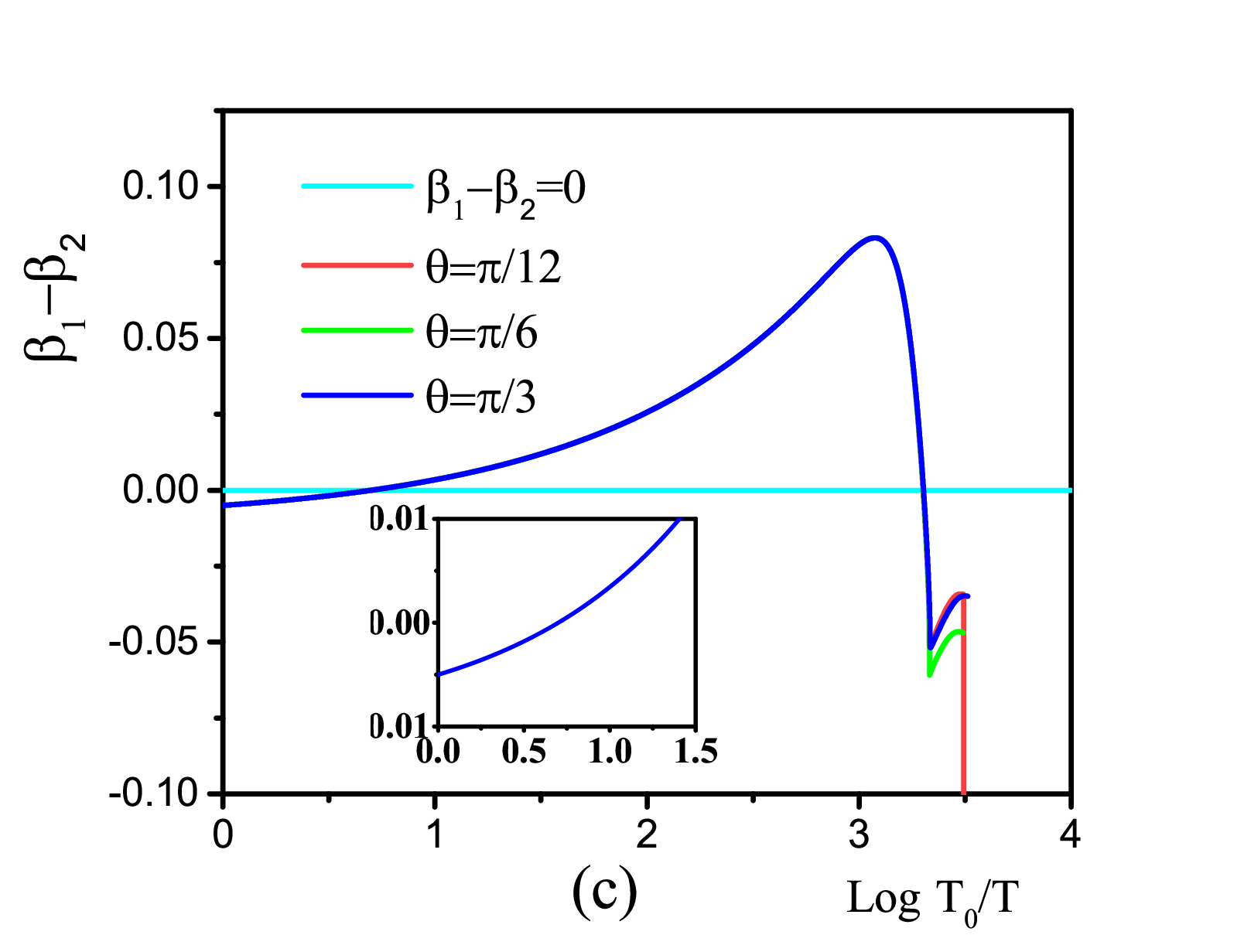}\\
\vspace{-0.1cm}
\hspace{-0.6cm}
\includegraphics[width=2.0in]{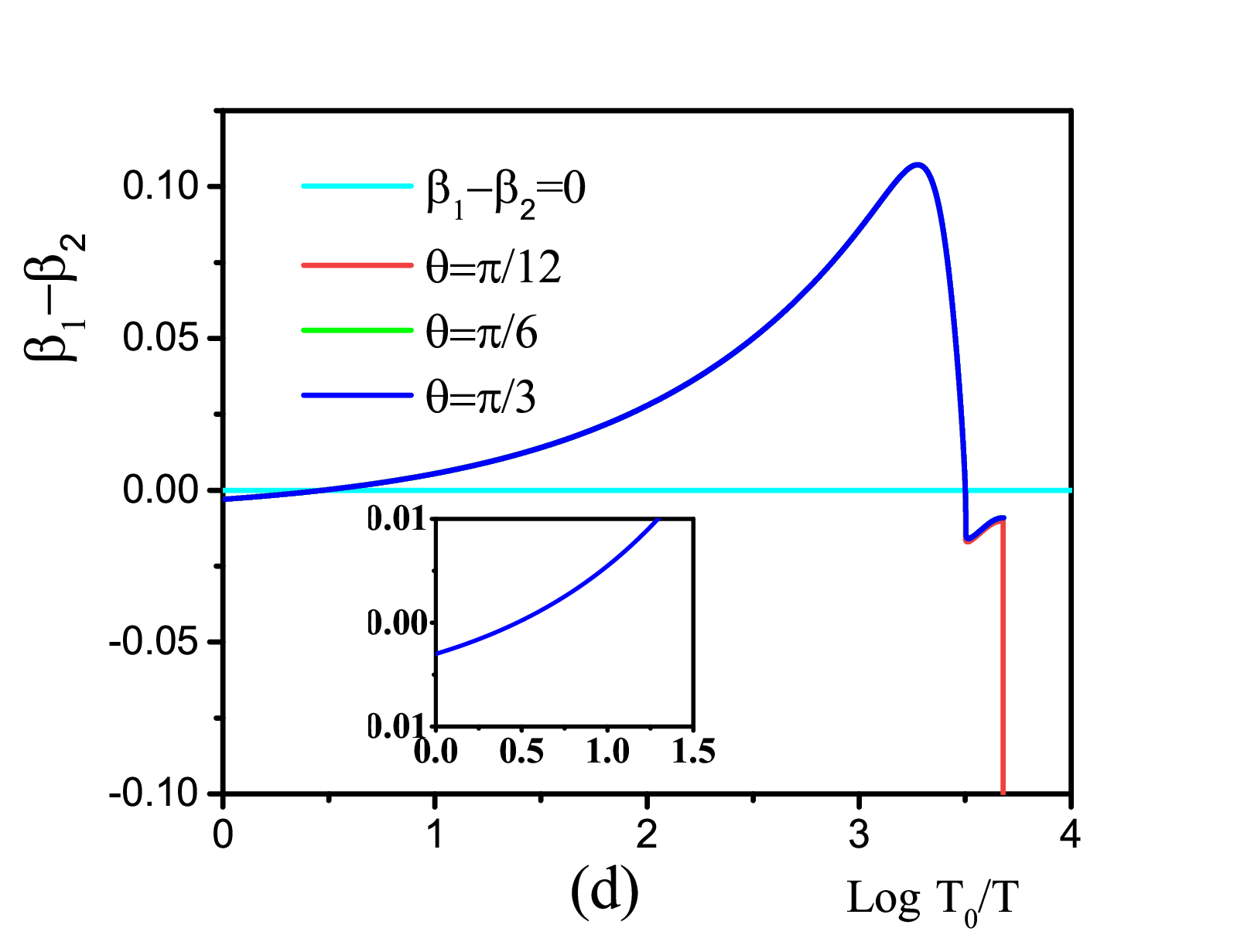}\hspace{-0.86cm}
\includegraphics[width=2.0in]{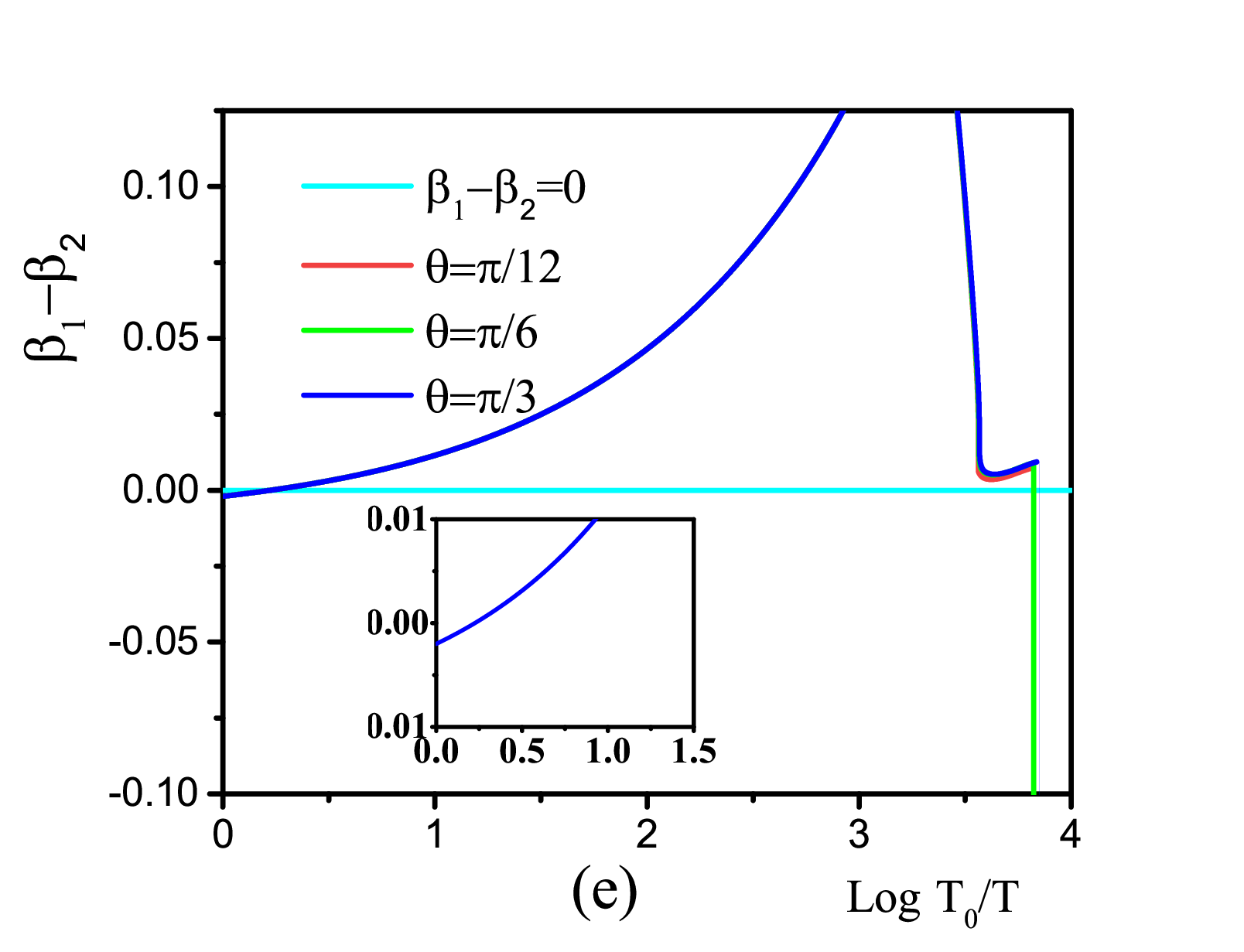}\hspace{-0.86cm}
\includegraphics[width=2.0in]{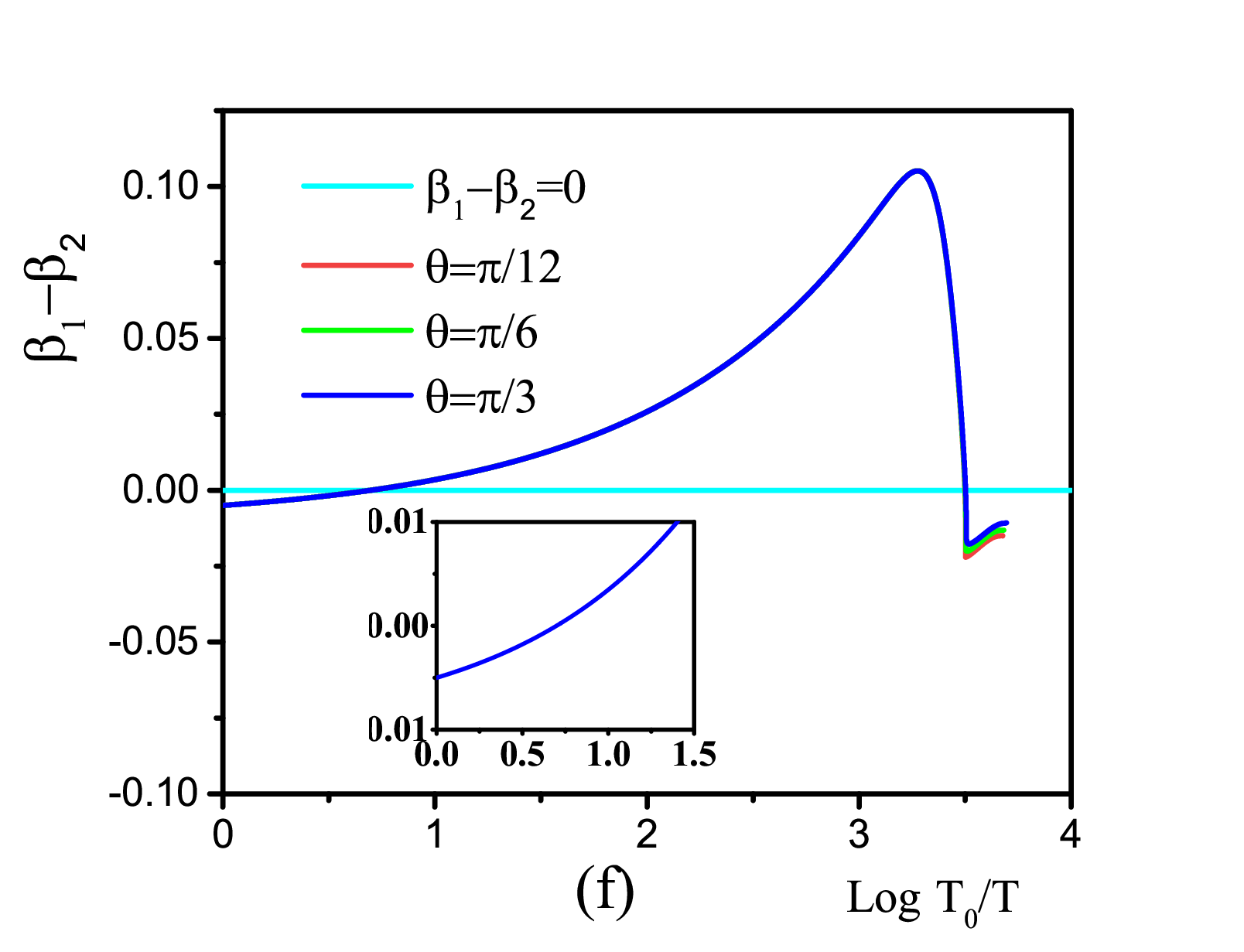}\hspace{-0.96cm}
\includegraphics[width=2.0in]{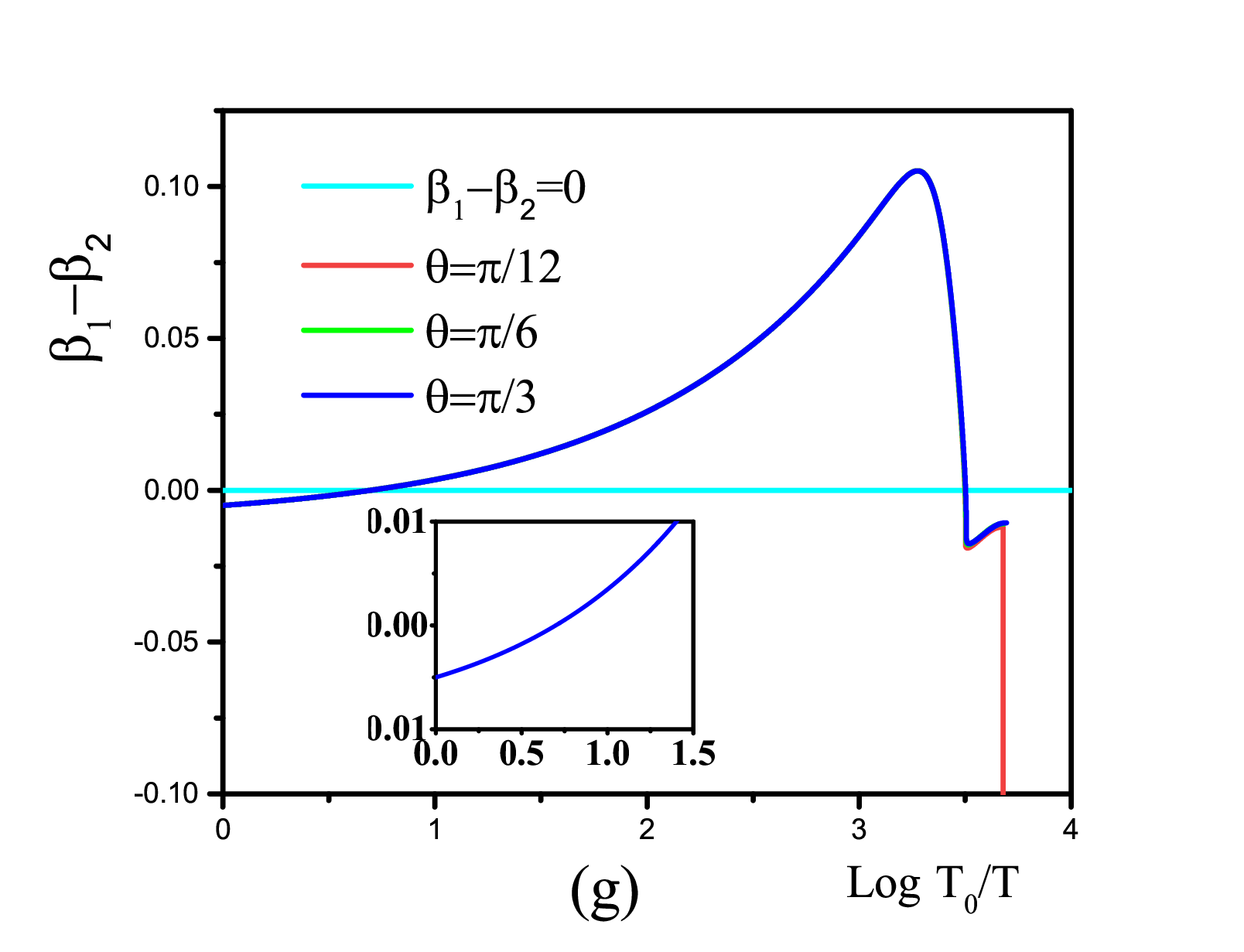}\\
\vspace{-0.15cm}
\caption{(Color online) Stability of tendency of temperature-dependence
$\beta_1-\beta_2$ for $C_2$ ICS magnetic state against adjusting the
initial value of Fig.~\ref{Fig_C2-IC-stripe}:
(a) $a_s$, (b) $u_s$, (c) $\lambda$, (d) $\beta_2$,
(e) $\beta_1$, (f) $\lambda_{\Delta A}$, and (g) $\kappa$
(all cases satisfy the stable constraints of such a state).
The angle $\theta$ serves as the direction of magnetic order
in the spin space and $\theta=\pi/12,\pi/6,\pi/3$
are selected as three representative values. Insets: the sign-change region of $\beta_1-\beta_2$.}\label{Fig_C2-ICS-against-initial-parameters}
\end{figure*}

\subsection{Fates of magnetic states}

By employing the strategy in Sec.~\ref{Sec_strategy},
we within this subsection endeavor to inspect the fates
of all magnetic candidate states one by one via performing
the RG analysis with respect to the coupled RG evolutions
that involve the fluctuations and ordering competitions
as well as their interplay.

\subsubsection{Warm-up}\label{Sec_warm-up}

Let us consider the $C_2$ IC stripe (ICS)
magnetic state for an instance and show how to determine whether
it is a good candidate (stable/favorable SDW state) for warm-up.
The configurations of spin vectors
for such state read $\mathbf{n}_X=(0,0,1)$ and
$\mathbf{n}_Y=(0,0,0)$~\cite{Andersen2018PRX}. As shown in Table~\ref{table_criteria-fates},
its stable constraints can be either $(\beta_1-\beta_2)<0$,
$g_2/|\beta_1-\beta_2|>0$, $(g_1-\beta_2)/|\beta_1-\beta_2|>-1$
or $(\beta_1-\beta_2)<0$, $g_2/|\beta_1-\beta_2|<0$,
$(g_1-\beta_2-0.9g_2)/|\beta_1-\beta_2|>-1$~\cite{Andersen2018PRX}.
If this state is favorable around the QCP in the real materials,
it must be adequately stable against
the quantum fluctuations when approaching the QCP.
Accordingly, we can initially choose a higher temperature $T_0$ away
from the QCP as our starting point, at which the initial interaction parameters
are supposed to satisfy the stable constraints of $C_2$ ICS state.
Next, we take some representatively initial values of fundamental parameters that obey
the stable constraint at $T_0$ and then perform numerical RG analysis of
related RG equations by approaching the QCP along Routine-B as
schematically shown in Fig.~\ref{Fig1_routines} (namely, by
lowering the temperature).

Due to the differences of spin configurations and fluctuations,
we recall that all candidate states in Table~\ref{table_criteria-fates}
possess their own RG equations for the fundamental parameters collected
in Appendix~\ref{Appendix-coupled-RG}, which dictate the fates of
stable constraints of certain states. It is therefore the related RG
equations that are in charge of the stability of
ICS state when approaching the QCP. After extracting the energy-dependent
information from such RG equations, the corresponding
numerical results in Fig.~\ref{Fig_C2-IC-stripe} display the
temperature dependence of flows for the associated
fundamental parameters.

In order to compare the robustness with other candidate states,
it is helpful to denominate the very temperature as
$T_{\mathrm{col}}$ at which the candidate state's stable constrain
is collapsed. From Fig.~\ref{Fig_C2-IC-stripe}, we can infer
that the sign change of $\beta_1-\beta_2$ is occurred explicitly once
temperature is slightly lowered at $T=T_{\mathrm{col}}$
owing to the effects of ordering competition,
hinting at the destruction of the stable constraint.
In addition, the basic results are
insensitive to the specific values of $\theta$, which
are generally rooted in the symmetry of a candidate
state~\cite{Andersen2018PRX}. As for the ICS state that does not satisfy the
$C_4$ symmetry, the magnetic components are inequivalent in two directions
indicating $\theta\neq\pi/4$, and hence, without loss of generality,
three representative values $\theta=\pi/12,\pi/6,\pi/3$
are chosen to perform the numerical calculations.
In principle, there exists another critical point that describes the fixed point of parameters
can be accessed at $l=l^\star$ or a related $T=T^\ast$ beyond
which the parameters are divergent or
unphysical as labeled in Fig.~\ref{Fig_C2-IC-stripe}.
As presented in Appendix~\ref{Appendix-coupled-RG}, different types of candidate
SDW states exhibit distinct quantum fluctuations, which give rise to
their own sets of RG equations. Henceforth, it is of particular necessity to address that the state with
$T<T_{\mathrm{col}}$ is no longer the ICS state but an uncertain state, which
possesses unknown but distinct RG equations compared to those of the ICS state.
Accordingly, the evolutions of parameters obtained by
obeying RG equations of ICS state are unphysical at $T<T_{\mathrm{col}}$
in Fig.~\ref{Fig_C2-IC-stripe}. This implies that one can neglect the behaviors
of parameters at $T<T_{\mathrm{loc}}$ in that
whether a candidate state is robust can be determined as $T$
approaches $T_{\mathrm{loc}}$ from  $T>T_{\mathrm{loc}}$. For convenience,
the curves with $T<T_{\mathrm{col}}$ are preserved for comparison
with the numerical results of other states.

For completeness,  it is worth inspecting whether the
fate of an SDW state is robust against the initial
fundamental parameters. To this end, we regard the initial
condition in Fig.~\ref{Fig_C2-IC-stripe} as a reference point
and tune an initial parameter of this point but
keep all others invariant to form distinct representative groups of
initial conditions, all of which are required to
meet the stable criteria of $C_2$ ICS state. The numerical
results in Fig.~\ref{Fig_C2-ICS-against-initial-parameters} share the similar
tendency of $\beta_1-\beta_2$ to its counterpart
in Fig.~\ref{Fig_C2-IC-stripe}, evincing the robustness of stability
against the variation of initial condition. As for all other types of
candidates states, the basic results are analogous and thus not shown hereby.

To wrap up, in the spirt of strategy addressed in Sec.~\ref{Sec_strategy},
we can infer that $C_2$ ICS is not
a stable state against the quantum fluctuations in the low-energy regime
and hence not a good candidate for IC magnetic state nearby the QCP.

\begin{figure}
\centering
\includegraphics[width=3.8in]{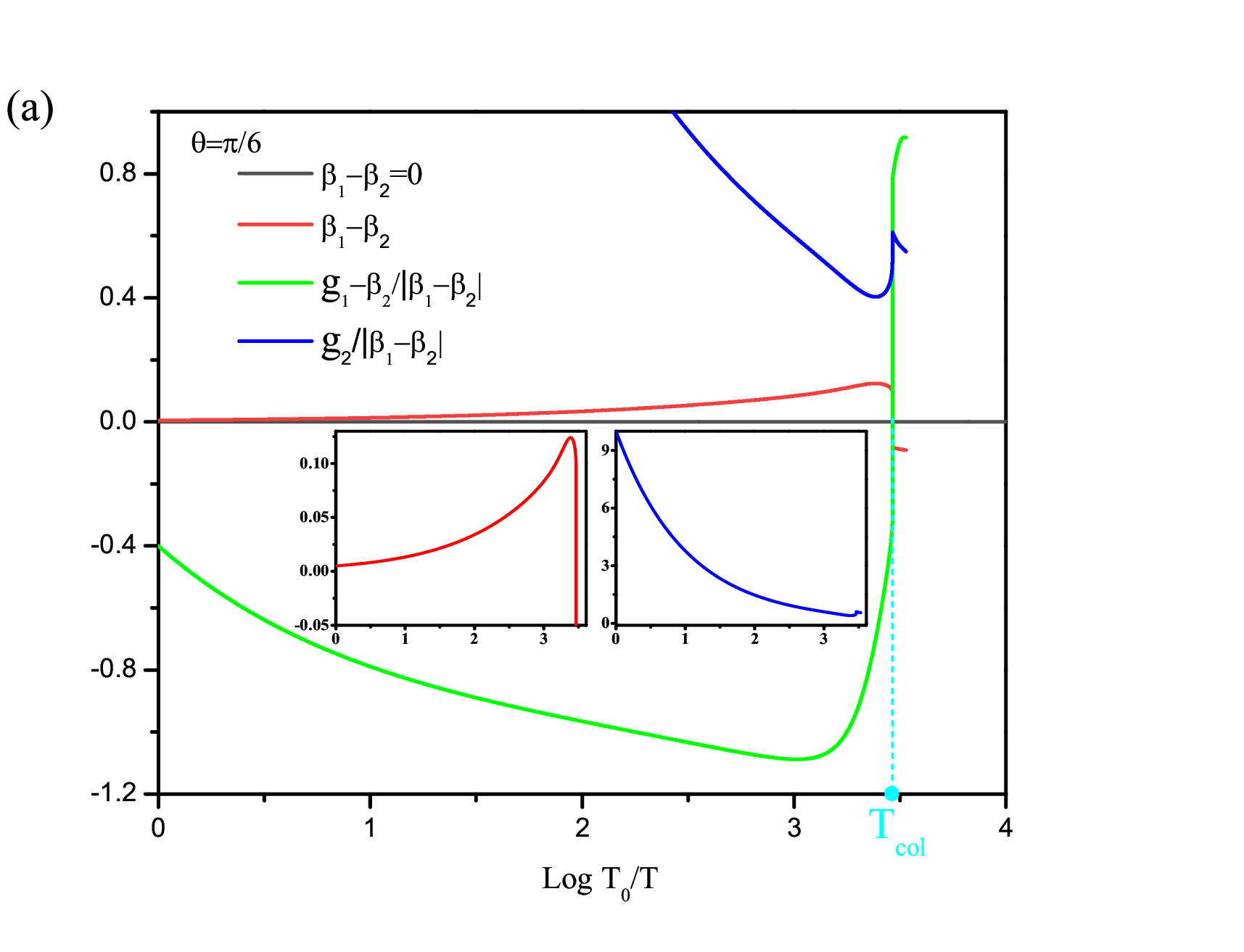}\\
\vspace{-0.85cm}
\includegraphics[width=3.8in]{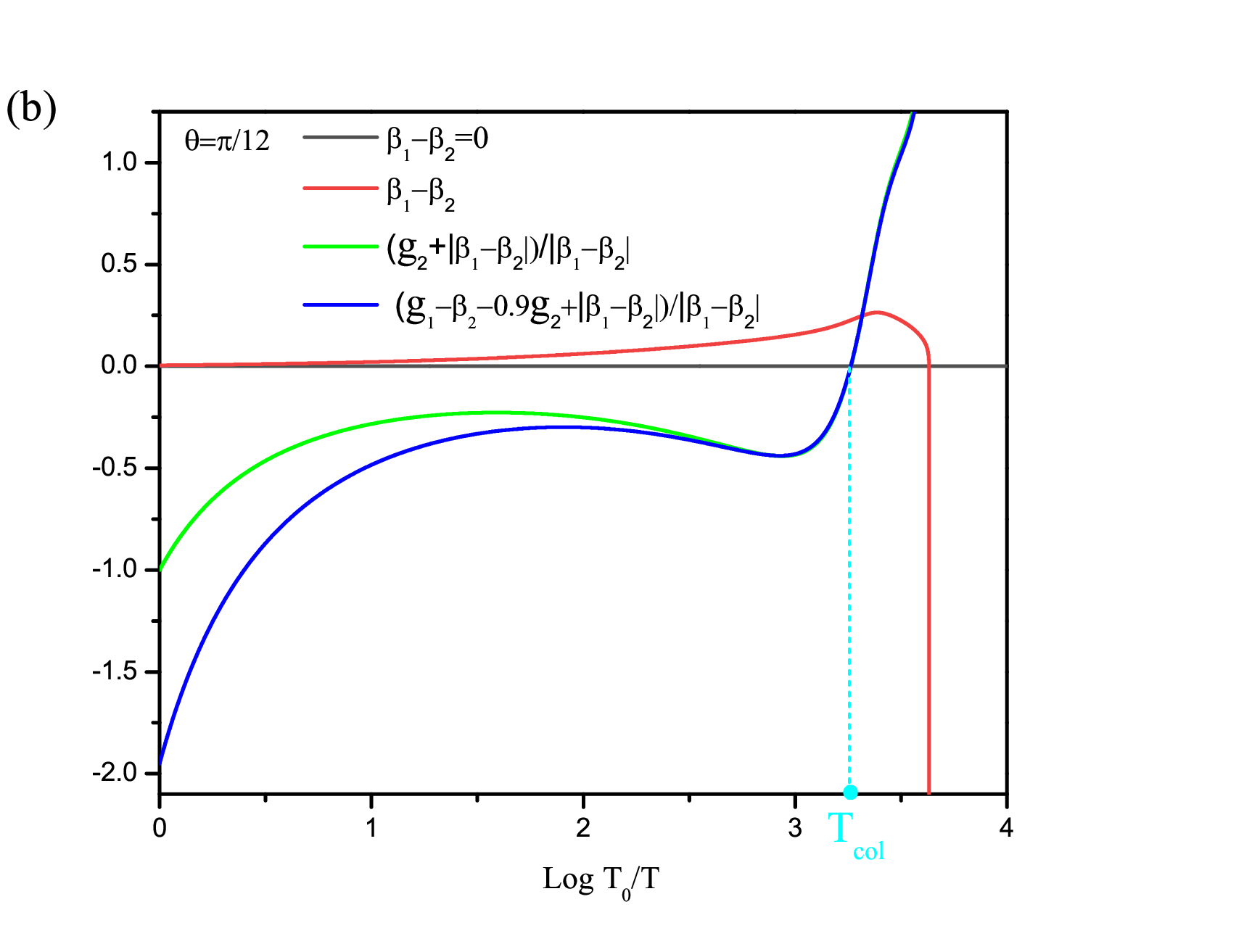}
\vspace{-0.75cm}
\caption{(Color online) Temperature-dependent stable constraints of (a)
$C_2$ ICS $\perp$ MH and (b) $C_4$ IC CSDW %case-2
listed in Table~\ref{table_criteria-fates}.
Hereby, the angle $\theta$ is designated
in Sec.~\ref{Sec_S_eff} to specify the direction of magnetic order
in the spin space and two representative values
$\pi/6$ and $\pi/12$ are selected for cases (a) and (b), respectively.
The initial values of related fundamental parameters are chosen as
($g_1=0.01$, $g_2=0.01$, $u_s=0.05$, $\lambda=0.01$,
$\beta_1=0.01$, $\beta_2=0.005$) for ICS $\perp$ MH and
($g_1=-0.015$, $g_2=-0.01$, $u_s=0.05$, $\lambda=0.01$,
$\beta_1=0.01$, $\beta_2=0.005$) for IC CSDW states
(the basic results are insensitive to initial values of parameters).
$T_{\mathrm{col}}$ labels the very temperature at which the corresponding
stable constrains are jeopardized and collapsed.
Insets: enlarged regions for $\beta_1-\beta_2$ and $g_2/|\beta_1-\beta_2|$.}\label{Fig_stable}
\end{figure}

\subsubsection{Stable states}\label{Sec_stable-states}

After paralleling the procedures for $C_2$ ICS state in Sec.~\ref{Sec_warm-up},
we check the stabilities of all candidate states collected in Table~\ref{table_criteria-fates}.
This not only bears witness to the crucial % substantiates
role of ordering competition but also sheds light on
fates of all types of IC magnetic states.

To be concrete, Fig.~\ref{Fig_stable} exhibits the
temperature (energy) dependence of correlated fundamental
parameters, which carry the low-energy characteristics for
both $C_2$ ICS $\perp$ MH and $C_4$ IC CSDW.
At the outset, we find that stable constraints
for $C_2$ ICS $\perp$ MH shown in Fig.~\ref{Fig_stable}(a)
are well protected with a decrease of temperature.
They are sabotaged by extremely
strong fluctuations only until the magnetic QCP is sufficiently
accessed at the collapsed temperature
$T_{\mathrm{col}}\sim10^{-4}T_0$ (taking $T_0=100$\,K for instance,
$T_\mathrm{col}\sim10^{-2}$\,K). This evidently signals that
$C_2$ ICS $\perp$ MH is of particular robustness withstanding
ordering competition. In reminiscence of the unknown
$C_2$ magnetic state, which is located at a little deviation from the magnetic QCP
portrayed in Fig.~\ref{Fig_schematic_phase_diagram},
we are aware that $C_2$ ICS $\perp$ MH is therefore deemed to be a reasonable
candidate for this mysterious $C_2$ state that differs substantially from
conventional $C_2$ stripe state. In addition,
Fig.~\ref{Fig_stable}(b) proposes firmly robust temperature-dependent
constraints for $C_4$ IC CSDW.
Moreover, the basic results bearing the similarities to
the $C_2$ ICS presented in Sec.~\ref{Sec_warm-up}
are insusceptible to the variances of
starting parameters as long as they satisfy
the restricted conditions listed in
Table~\ref{table_criteria-fates}
(the related further discussions will be briefly
delivered in the forthcoming
subsection~\ref{Sec_comments}).

On the basis of these, we then come to a conclusion that IC CSDW, like its
commensurate counterpart~\cite{Wang2017PRB},
behaves dominantly compared to other types of IC $C_4$ magnetic states.
This $C_4$ magnetic state is hence the most applicable
choice on the left side of magnetic QCP $x_c$ in Fig.~\ref{Fig_schematic_phase_diagram},
which compete, coexist, and cooperate with SC state.
Furthermore, apart from the two applicable states including $C_2$ ICS $\perp$ MH
and $C_4$ IC CSDW, ordering competition surrounded by magnetic QCP is not in
favour of all other types of IC magnetic states listed in Table~\ref{table_criteria-fates}.
In terminological language, given these states are prone to easily feel
plus efficiently receive the fluctuation corrections even far away from a magnetic QCP,
they are fairly sensitive and fragile to ordering competition, resulting in
undeviating breakdown themselves as temperature is reduced.
This broadly suggests that one is unable to solely fix the configuration
of $C_2$ IC SDW above $C_4$ IC CSDW and $C_2$ ICS $\perp$ MH as displayed
in Fig.~\ref{Fig_schematic_phase_diagram}, which may either be $C_2$ ICS,
$C_2$ DPMH, or $C_2$ MH.  % or their intermixed state.
Details of verifying stabilities
of IC magnetic states are provided in
Appendix~\ref{Appendix_stab-magnetic}. Last but not the least important,
we deliver that, as for the region close enough to the QCP
with $T<T_{\mathrm{col}}$, ordering competition
is so ferociously that no magnetic state
can exist alone but instead there might be a coexistence
of multiple IC magnetic states.

\subsection{Relevant comments and explanations}\label{Sec_comments}

Before going further, we stop to address three relevant issues
with comments and explanations.

To begin with, we highlight the major concerns between Ref.~\cite{Andersen2018PRX}
and this work are different and then explain the reason for adjusting the initial
values not very largely. Concisely, the authors of Ref.~\cite{Andersen2018PRX} focus on
how many possible SDW states can be generated and where do they reside in
the parameter space via tuning a series of energy-independent parameters. The potential
states are separated by several boundaries that are developed by the related parameters and
not directly associated with the QCP. In comparison, the phase diagram in Fig.~\ref{Fig_schematic_phase_diagram}
with a QCP is constructed by the temperature and doping, which indicates
that the boundaries of these two situations are
not the same thing. Additionally, our target is to
examine and determine which are the most favorable states among all candidates
neighboring the QCP. Following the strategy in Sec.~\ref{Sec_strategy},
we confine the initial parameters
to satisfy the related stable constraints of a candidate state
and judge whether such a state
is suitable to exist nearby QCP with the help of
the corresponding RG equations. In order to make sure the starting point is 100\% of the
candidate state and avoid the possible influence of other states as
different states are associated with different RG equations, it is more
suitable to choose the initial parameters a little away from the very
boundary of Ref.~\cite{Andersen2018PRX} but near the
QCP in Fig.~\ref{Fig_schematic_phase_diagram}.
This may be ascribed to a shortage of our strategy in that
the RG equations of parameters are based upon
the quantum fluctuations around the QCP and we can
only deal with the candidate states one by one but
cannot tackle two or more mixed states simultaneously.

Afterward, we move on to deliver several comments on the underlying
fixed points (FPs) of parameters in the lowest-energy limit.
For convenience, let us suppose that the FPs can be accessed at
$T=T^\ast$, beyond which the parameters are divergent or
unphysical as labeled in Fig.~\ref{Fig_C2-IC-stripe} for an example.
From Fig.~\ref{Fig_C2-IC-stripe} or Fig.~\ref{Fig_stable}
(whose $T^\ast$ can be designated analogously to Fig.~\ref{Fig_C2-IC-stripe}'s
and have not been shown for brevity),
we can infer that the FPs can be accessed either at
a much or a little lower energy scale for
an unstable ($T^\ast\ll T_{\mathrm{col}}$) or a stable
($T^\ast< T_{\mathrm{col}}$) candidate state,
implying the parameters do not satisfy the restricted
conditions within $T^\ast<T<T_{\mathrm{col}}$.
As a consequence, one can already
judge whether some candidate state survives and
which are the most favorable states among
potential candidates around the QCP before the FPs
are exactly approached. Indeed, the FPs may be
instructive to other interesting behaviors
which are out scope of our main target and
worth systematically studying in future.

Furthermore, it is necessary to present some words
on the IC parameter $\delta$ in Sec.~\ref{Sec_S_eff}, which
does not directly appear in the effective action but is indirectly
reflected by imposing the order parameters
$M_{\mathbf{Q}_{X,Y}}\neq M^*_{\mathbf{Q}_{X,Y}}$ described in Sec.~\ref{Sec_S_eff}.
There exists a little distinction from Puga \emph{et al.}'s pioneering work on
the sine-Gordon model~\cite{Beck1982PRB},
in which the parameter $\delta$ is explicit in their effective theory.
Henceforth, one can regard such parameter as an
interaction parameter and examine the transition
between a commensurate and an IC state via tracking the evolution of
parameter $\delta$. However, all potential states hereby are restricted
to IC states and the focus is put on the stability of certain IC state
without involving the transition in Ref.~\cite{Beck1982PRB}.

\begin{figure}
\centering
\hspace{0.8cm}
\includegraphics[width=3.9in]{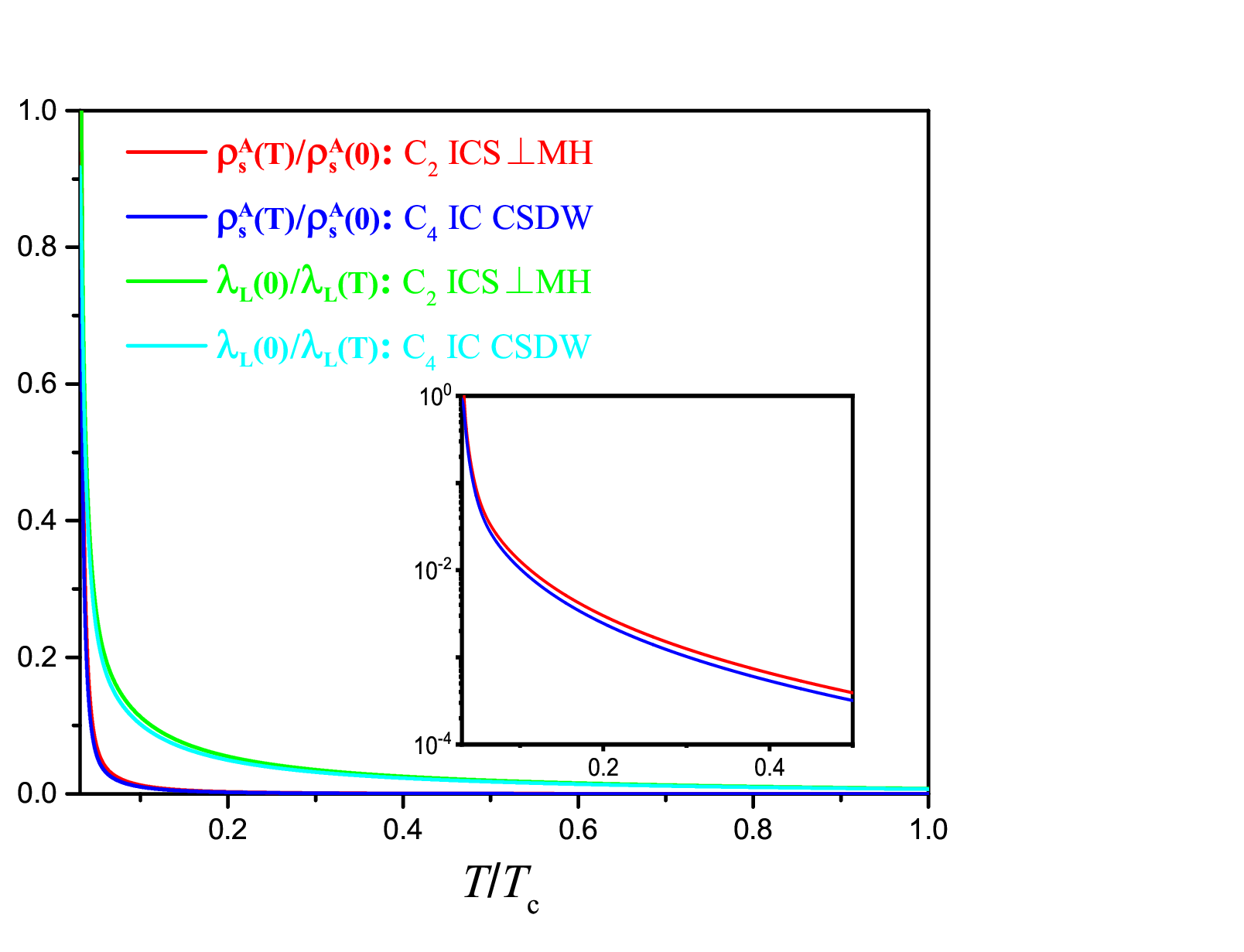}
\vspace{-0.5cm}
\caption{(Color online) Superfluid density and London penetration depth
as a function of temperature at $\theta=6/\pi$ affected
by $C_2$ ICS $\perp$ MH and $C_4$ IC CSDW states neighboring
the magnetic QCP. Hereby, $T_c$ designates the related critical temperature
without the ordering competition and the angle $\theta$
characterizes the direction of magnetic order in the
spin space, which is assigned a representative
value ($\pi/6$) for numerical evaluation (the essential features are
insusceptible to beginning values of interaction parameters and
concrete directions of such magnetic order).
Inset: enlarged regions for $\rho_s$ displaying
difference between the two cases.}\label{Fig7_rho_lambda}
\end{figure}

\vspace{0.35cm}

\section{Superfluid density and London penetration depth}\label{Sec_rho-lambda}

Generally, the quantum critical region accompanied by a certain QCP is
a fertile ground for generating unusual physical behaviors caused by the
strong fluctuations, which are of qualitative distinction from
the scopes out of control by the QCP.
According to Sec.~\ref{Sec_stable-states}, the most favorable SDW states
for the left and right sides of the QCP correspond to
the $C_4$ IC CSDW and $C_2$ ICS $\perp$ MH states, respectively.
These different sorts of magnetic states would be responsible for
distinct fates of physical implications around the QCP.
In order to make the logic self-consistent,
we follow this clue and endeavor to evaluate
the temperature dependence of superfluid density and
London penetration depth around both sides of the QCP by taking into account
the quantum fluctuations of these two preferable states.

As magnetic states steadily compete and coexist with a SC order,
it is of great temptation to examine how the superfluid density ($\rho_s$) and
London penetration depth ($\lambda_L$) are influenced in the presence of ordering competition,
which has two particularly important implications. In principle,
$\rho_s(T)$ can be evaluated as $\rho_s(T) = \rho^A_s(T) -
\rho_n(T)$, where $\rho^A_s(T)\propto \alpha_A(T)$ stems from
the mass of vector field $\mathbf{A}$ that obey RG equations due to Anderson-Higgs
mechanism~\cite{Halperin1974PRL} and $\rho_n(T)$ grasps
the density of thermally excited normal (non-SC) fermionic
quasiparticles (QPs), respectively. Approaching the QCP, ordering competition is
dominant and thus the normal QPs effects can be neglected
implying $\rho_s(T)\sim \rho^A_s(T)$.

Fig.~\ref{Fig7_rho_lambda} clearly shows
that $\rho_s(T)$ is notably
suppressed by the ordering competition~\cite{Wang2017PRB,Liu2012PRB,Koshelev2020}.
Because critical temperature $T_c$ is nominated by $\rho_s(T_c) = \rho^A_s(T)
- \rho_n(T_c) = 0$, one can infer that it would be intensively
reduced in the absence of $\rho_n(T)$. As explicitly delineated
in the inset of Fig.~\ref{Fig7_rho_lambda},
it is worth declaring that the drop of $T_c$ caused by the
$C_4$ IC CSDW is a little more than its $C_2$ ICS $\perp$ MH's counterpart,
which is also apparently exposed in Fig.~\ref{Fig_schematic_phase_diagram}.
Albeit a slight splitting, principal tendencies are qualitatively compatible
with recent experiments~\cite{Hardy2015NComm,Wang2016PRB,Hardy2018PRL-BaNaFeAs}.
As for the unconventional derivation of temperature-dependent $\rho_s$ from
usual $s$-wave gap symmetry's, there are two underlying reasons.
A major concern is the possible alteration of the pairing gap symmetry driven by
so ferocious fluctuations around the QCP. In addition, the fundamental interaction
parameters can also display anomalous energy-dependence behaviors as
accessing the QCP, which  enter into the analytical expression of $\rho_s$
and hence can indirectly influence the tendency of superfluid.

For qualitative discussions, we single out the $s$-wave gap symmetry as a toy and tentative
substitute. In this respect, the London penetration depth is expressed as
$\lambda_L(0)/\lambda_L(T)=\sqrt{\rho_s(T)}$~\cite{Helfand1966PR}.
As a consequence, $\lambda_L(0)/\lambda_L(T)$ shares an
analogous temperature-dependent trajectory with $\rho_s$ under the impact of ordering
competition as depicted in Fig.~\ref{Fig7_rho_lambda}. Although $\mathrm{BaFe_2As_2}$ system
possesses a more intricate gap structure~\cite{Chubukov2012ARCMP},
this primitive result might uncover parts of central ingredients that are in charge
of $\lambda_L$'s property. For completeness, we adopt the method in
Sec.~\ref{Sec_stable-states} and check that the basic conclusions concerning
the superfluid density in Fig.~\ref{Fig7_rho_lambda}
are robust under the variation of couplings between SDW and SC, which
are embodied by the initial fundamental parameters.

As a consequence, the behaviors of physical obversables
indirectly corroborate $C_4$ IC CSDW and $C_2$ ICS $\perp$ MH states are
favorable SDW states compared to the other candidates in
Table~\ref{table_criteria-fates}. This implies that the
phenomenological theory can qualitatively
capture the key information around the QCP. In addition,
the primary conclusions concerning the preferable SDW
states neighboring the putative QCP are relatively
stable and self-consistent.

\vspace{0.35cm}

\section{Summary}\label{Sec_summary}

To recapitulate,
we study and discern the probable IC magnetic states induced by subtle ordering competition
in the proximity of certain QPT below the SC dome of $\mathrm{Ba_{1-x}Na_xFe_2As_2}$.
Specifically, we find that $C_2$ ICS $\perp$ MH survives to be a good candidate for the
obscure $C_2$ magnetic state and IC $C_4$ CSDW points to the reasonable IC state
in the vicinity of the magnetic QPT. In addition, we address that
superfluid density in tandem with critical temperature and London penetration depth
manifest critical behaviors attesting to ordering competition around the QCP.

Fig.~\ref{Fig_schematic_phase_diagram} schematically presents our primary conclusions, whose
overall structure is borrowed from the experimental results
in Ref.~\cite{Wang2016PRB}. However, it is worth pointing out that
the spin configurations of magnetic states for both sides of
the QCP are unclear and hence only labeled by SDW state in Wang \emph{et al.}'s
work~\cite{Wang2016PRB}. In sharp contrast, we explicitly determine that the most
favorable candidates for the left and right sides correspond
to the $C_4$ IC CSDW and $C_2$ ICS $\perp$ MH states by virtue
of one-loop RG analysis. In addition, we theoretically address that
the $C_4$ IC CSDW state is more harmful to the superconductivity.
The conclusions are qualitatively concomitant with recent experiments~\cite{Hardy2015NComm,Wang2016PRB,Hardy2018PRL-BaNaFeAs}.
In this sense, we offer a relatively operable
strategy to select out the most favorable states
around the QCP, with which one can in
principle examine whether
some magnetic state is a preferable
state against the influence of quantum fluctuations.
We expect our results are profitable to further understand the phase diagram of
$\mathrm{Ba_{1-x}Na_xFe_2As_2}$ and explore the correspondence
between SC and magnetic states in the
iron-based superconductors.

\section*{ACKNOWLEDGEMENTS}

J.W. is partially supported by the National Natural
Science Foundation of China under Grant No. 11504360 and
highly grateful to Ya-Jie Zhou for stimulating discussions on
the June 11, 2020. The author also acknowledges
the Referee A from PRL for the constructive
comments and suggestions.

\vspace{0.2cm}
\appendix

\section{Coupled RG equations of fundamental interaction parameters}\label{Appendix-coupled-RG}

After performing one-loop analysis of
effective theory~\cite{Wilson1975RMP,Wang2014PRD,Wang2017PRB} via integrating out the fields in the
momentum shell $e^{-l}\Lambda<k<\Lambda$ with $l>0$ the running scale,
we can derive flows of effective parameters in Eq.~(\ref{Eq_L-eff}). Combining these equations and connections~(\ref{Eq_bridge-1})-(\ref{Eq_bridge-9})
~\cite{Wang2014PRD,Wang2017PRB}, the coupled RG equations for
fundamental parameters can be derived.

Before going further, it is necessary to highlight that
the fundamental parameters $g_{1,2}$ only appear in Eq.~(\ref{Eq_bridge-7}).
This implies that they do not evolve independently. In this sense,
it is hereafter convenient to introduce a parameter
\begin{eqnarray}
\hat{g}
&\equiv&g_1\cos^2\theta\sin^2\theta (|\mathbf{n}_X|^2 |\mathbf{n}_Y|^2)
\nonumber\\
&&+\frac{g_2}{2}\cos^2\theta\sin^2\theta(|\mathbf{n}_X\cdot \mathbf{n}_Y|^2
+|\mathbf{n}_X\cdot \mathbf{n}^*_Y|^2),
\end{eqnarray}
to describe the information of $g_{1,2}$.

After long but straightforward calculations~\cite{Wang2014PRD,Wang2017PRB},
we eventually obtain the coupled RG equations of all fundamental interaction parameters around
the magnetic QCP, which include $\alpha,\beta_{1,2},\hat{g}$ and
$\kappa$ specifying the characters of spin configurations as well as
$a_s, u_s, \lambda_{\Delta A}$ stemming from SC fluctuations.
These coupled RG evolutions are closely dependent upon the spin configurations of magnetic
fluctuations, namely the relationships between $|\mathbf{n}^2_X|^2$, $|\mathbf{n}_X|^4$,
$|\mathbf{n}^2_Y|^2$, $|\mathbf{n}_Y|^4$, which are divided into two main sorts of situations.

For type-I case,  at which $|\mathbf{n}^2_X|^2\neq|\mathbf{n}_X|^4$
and $|\mathbf{n}^2_Y|^2=|\mathbf{n}_Y|^4$
or $|\mathbf{n}^2_X|^2=|\mathbf{n}_X|^4$ and $|\mathbf{n}^2_Y|^2\neq|\mathbf{n}_Y|^4$,
both $\beta_1$ and $\beta_2$ flow independently and thus the coupled evolutions
are written as
\begin{widetext}
\begin{small}
\begin{eqnarray}
\frac{da_s}{dl}
&=&2a_s-\frac{1}{4\pi^2}\Bigl\{\frac{9a_su_s(1+4a_s)}{2}+
\frac{2\mathcal{S}\mathcal{E}_1^2a_s\lambda^2}{u_s}
[1-4\mathcal{S}\mathcal{E}_1(a-\frac{\lambda a_s}{u_s})]
+\frac{2\mathcal{C}\mathcal{D}_1^2a_s\lambda^2}{u_s}
[1-4\mathcal{C}\mathcal{D}_1(a-\frac{\lambda a_s}{u_s})]\nonumber\\
&&+\frac{32a_s\lambda^2_{\Delta A}}{3u_s}
(1+\frac{4\lambda_{\Delta A}a_s}{u_s})+\frac{(\mathcal{S}\mathcal{E}_1
+\mathcal{C}\mathcal{D}_1)\lambda}{2}
+\frac{3u_s(1+2a_s)}{4}-(a-\frac{\lambda a_s}{u_s})
(\mathcal{E}_1^2\mathcal{S}^2
+\mathcal{D}_1^2\mathcal{C}^2)\lambda\nonumber\\
&&+\lambda_{\Delta A}(1+\frac{2\lambda_{\Delta A}a_s}{u_s})
+\frac{\mathcal{C}\mathcal{S}\mathcal{F}^2a_s\kappa^2}{4u_s}
[1-2(\mathcal{C}\mathcal{D}_1+\mathcal{S}\mathcal{E}_1
)(a-\frac{\lambda a_s}{u_s})]\Bigr\},\label{RG-eqs-type-I-as}\\
\frac{da}{dl}
&=&2(a-\frac{\lambda a_s}{u_s})+
\frac{1}{4\pi^2}\Bigl\{\frac{\lambda}{2}
+\frac{\mathcal{S}\hat{g}}{\mathcal{D}_1}
+3\mathcal{C}[\beta_2\mathcal{D}_1
+(\beta_1-\beta_2)\frac{\mathcal{D}_2}{\mathcal{D}_1}]
-\frac{2\mathcal{S}^2\mathcal{E}_1\hat{g}}{\mathcal{D}_1}(a-\frac{\lambda a_s}{u_s})
\nonumber\\
&&-6\mathcal{C}^2(a-\frac{\lambda a_s}{u_s})[\beta_2\mathcal{D}_1^2
+(\beta_1-\beta_2)\mathcal{D}_2]
+a_s\lambda+\frac{4\mathcal{C}\mathcal{D}_1a_s\lambda^2}{u_s}
[1-2(\mathcal{C}\mathcal{D}_1(a-\frac{\lambda a_s}{u_s})-a_s)]\nonumber\\
&&-\frac{\mathcal{S}\mathcal{F}^2\kappa^2}{8\mathcal{D}_1}
[1-2(\mathcal{S}\mathcal{E}_1(a-\frac{\lambda a_s}{u_s})-a_s)]\Bigr\}
+\left(\frac{\lambda}{u_s}\frac{da_s}{dl}
+\frac{a_s}{u_s}\frac{d\lambda}{dl}-\frac{a_s\lambda}
{u^2_s}\frac{du_s}{dl}\right),\label{RG-eqs-type-I-a}\\
\frac{du_s}{dl}
&=&u_s+\frac{1}{2\pi^2}\Bigl\{-18a_su_s^2(1+6a_s)
-\frac{16\mathcal{C}^3\mathcal{D}_1^3a_s\lambda^3}{3u_s}
[1-6\mathcal{C}(a-\frac{\lambda a_s}{u_s})\mathcal{D}_1]
-\frac{16\mathcal{S}^3\mathcal{E}_1^3a_s\lambda^3}{3u_s}
[1-6\mathcal{S}(a-\frac{\lambda a_s}{u_s})\mathcal{E}_1]\nonumber\\
&&+8\lambda^2(\mathcal{S}^3\mathcal{E}_1^3
+\mathcal{C}^3\mathcal{D}_1^3)
(a-\frac{\lambda a_s}{u_s})-18a_su_s^2-\frac{9u_s^2}{2}
-2\lambda^2(\mathcal{S}^2\mathcal{E}_1^2
+\mathcal{C}^2\mathcal{D}_1^2)-\frac{32\lambda^2_{\Delta A}}{3}(\frac{4\lambda_{\Delta A}a_s}{u_s}+1)\nonumber\\
&&
-\frac{11072a_s\lambda^3_{\Delta A}}{105u_s}(1+\frac{6\lambda_{\Delta A}a_s}{u_s})-\mathcal{C}\mathcal{S}\mathcal{F}^2\kappa^2
[1-2(\mathcal{C}\mathcal{D}_1+\mathcal{S}\mathcal{E}_1)(a-\frac{\lambda a_s}{u_s})]
\nonumber\\
&&-\frac{\mathcal{C}^2\mathcal{S}^2\mathcal{F}^2a_s^2\kappa^2
(\mathcal{F}^2\kappa^2+2\mathcal{D}_1\mathcal{E}_1\lambda^2)}{6u_s^2}
[1-4(\mathcal{C}\mathcal{D}_1
+\mathcal{S}\mathcal{E}_1)(a-\frac{\lambda a_s}{u_s})]
\nonumber\\
&&-\frac{2\mathcal{C}^2\mathcal{S}\mathcal{F}^2\mathcal{D}_1a_s\kappa^2\lambda}{3u_s}
[1-2(2\mathcal{C}\mathcal{D}_1
+\mathcal{S}\mathcal{E}_1)(a-\frac{\lambda a_s}{u_s})]\Bigr\},\label{RG-eqs-type-I-us}\\
\frac{d\lambda}{dl}
&=&\lambda+\frac{1}{2\pi^2}
\Bigl\{\frac{-8\mathcal{S}^3\mathcal{E}_1^2a_s\lambda^2\hat{g}}
{3\mathcal{D}_1u_s}
[1-6\mathcal{S}\mathcal{E}_1(a-\frac{\lambda a_s}{u_s})]
-\frac{8\mathcal{C}^3\mathcal{D}_1a_s\lambda^2[\beta_2\mathcal{D}_1^2
+(\beta_1-\beta_2)\mathcal{D}_2]}{u_s}\nonumber\\
&&\times
[1-6\mathcal{C}\mathcal{D}_1(a-\frac{\lambda a_s}{u_s})]
-4\mathcal{C}\mathcal{D}_1a_s\lambda^2
[1-2(\mathcal{C}\mathcal{D}_1(a-\frac{\lambda a_s}{u_s})-2a_s)]
-\frac{8\mathcal{C}^2\mathcal{D}_1^2a_s\lambda^3}{3u_s}
[1-2(2\mathcal{C}\mathcal{D}_1(a-\frac{\lambda a_s}{u_s})-a_s)]\nonumber\\
&&+3a_su_s\lambda(1+6a_s)
-\frac{\mathcal{S}\mathcal{F}^2\kappa^2}{\mathcal{D}_1}
[1-2(\mathcal{S}\mathcal{E}_1(a-\frac{\lambda a_s}{u_s})-a_s)]
+\frac{\mathcal{S}^2\mathcal{E}_1\lambda
\hat{g}}{\mathcal{D}_1}
[4\mathcal{S}\mathcal{E}_1(a-\frac{\lambda a_s}{u_s})-1]\nonumber\\
&&
-\frac{3u_s(1+4a_s)\lambda}{4}
+3\mathcal{C}^2\lambda[\beta_2\mathcal{D}_1^2
+(\beta_1-\beta_2)\mathcal{D}_2][4\mathcal{C}\mathcal{D}_1(a-\frac{\lambda a_s}{u_s})-1]
+4\mathcal{C}\mathcal{D}_1\lambda^2
[2(\mathcal{C}\mathcal{D}_1(a-\frac{\lambda a_s}{u_s})-a_s)-1]\nonumber\\
&&-\frac{2\mathcal{S}^2\mathcal{F}^2\mathcal{E}_1\lambda a_s\kappa^2}{3\mathcal{D}_1u_s}
[1-2(2\mathcal{S}\mathcal{E}_1(a-\frac{\lambda a_s}{u_s})-a_s)]
-\frac{2\mathcal{C}\mathcal{S}\mathcal{F}^2a_s\lambda\kappa^2}{3u_s}
[1-2(\mathcal{C}\mathcal{D}_1+\mathcal{S}\mathcal{E}_1)
(a-\frac{\lambda a_s}{u_s})+2a_s]\nonumber\\
&&-\frac{\mathcal{C}^2\mathcal{S}^2\mathcal{E}_1\mathcal{F}^2\lambda a_s^2\kappa^2
[\beta_2\mathcal{D}_1^2
+(\beta_1-\beta_2)\mathcal{D}_2]}{2\mathcal{D}_1u_s^2}
[1-4(\mathcal{C}\mathcal{D}_1
(a-\frac{\lambda a_s}{u_s})+\mathcal{S}
\mathcal{E}_1(a-\frac{\lambda a_s}{u_s}))]\nonumber\\
&&-\frac{\mathcal{C}^2\mathcal{S}^2\mathcal{F}^2\lambda a^2_s\kappa^2
\hat{g}}{6u_s^2}
[1-4(\mathcal{C}\mathcal{D}_1+\mathcal{S}\mathcal{E}_1)
(a-\frac{\lambda a_s}{u_s})]\Bigr\},\label{RG-eqs-type-I-lambda}\\
\frac{d\beta_1}{dl}
&=&\frac{[(\mathcal{D}_1^2-\mathcal{D}_2)\mathcal{E}_2
-(\mathcal{E}_1^2-\mathcal{E}_2)\mathcal{D}_2]}
{(\mathcal{D}_1^2\mathcal{E}_2
-\mathcal{E}_1^2\mathcal{D}_2)}\beta_1
+\frac{2(\mathcal{D}_1^2-\mathcal{D}_2)}{2\pi^2(\mathcal{D}_1^2\mathcal{E}_2
-\mathcal{E}_1^2\mathcal{D}_2)}
\Bigl\{\mathcal{C}^2\hat{g}^2
[4\mathcal{C}\mathcal{D}_1(a-\frac{\lambda a_s}{u_s})-1]\nonumber\\
&&-\frac{\mathcal{E}_1^2\lambda^2(1+4a_s)}{4}-9\mathcal{S}^2[\beta_2\mathcal{E}_1^2
+(\beta_1-\beta_2)\mathcal{E}_2]^2[1-4\mathcal{S}\mathcal{E}_1(a-\frac{\lambda a_s}{u_s})]-\frac{4\mathcal{S}^2\mathcal{E}_1^2a_s\lambda^2[\beta_2\mathcal{E}_1^2
+(\beta_1-\beta_2)\mathcal{E}_2]}{u_s}\nonumber\\
&&\times[1-2(2\mathcal{S}\mathcal{E}_1(a-\frac{\lambda a_s}{u_s})-a_s)]-\frac{4\mathcal{S}\mathcal{E}_1^3a_s\lambda^3}{3u_s}
[1-2(\mathcal{S}\mathcal{E}_1(a-\frac{\lambda a_s}{u_s})-2a_s)]
-\frac{2\mathcal{C}^2\mathcal{F}^2a_s\kappa^2\hat{g}}{3u_s}\nonumber\\
&&\times[1-2(2\mathcal{C}\mathcal{D}_1(a-\frac{\lambda a_s}{u_s})-a_s)]\Bigr\}-\frac{2(\mathcal{E}_1^2-\mathcal{E}_2)}
{2\pi^2(\mathcal{D}_1^2\mathcal{E}_2
-\mathcal{E}_1^2\mathcal{D}_2)}
\Bigl\{\frac{-4\mathcal{C}^2\mathcal{D}_1^2a_s\lambda^2[\beta_2\mathcal{D}_1^2
+(\beta_1-\beta_2)\mathcal{D}_2]}{u_s}\nonumber\\
&&\times[1-2(2\mathcal{C}\mathcal{D}_1(a-\frac{\lambda a_s}{u_s})-a_s)]+9\mathcal{C}^2 [\beta_2\mathcal{D}_1^2
+(\beta_1-\beta_2)\mathcal{D}_2]^2
[4\mathcal{C}\mathcal{D}_1(a-\frac{\lambda a_s}{u_s})-1]\nonumber\\
&&
+\mathcal{S}^2 \hat{g}^2
[4\mathcal{S}\mathcal{E}_1(a-\frac{\lambda a_s}{u_s})-1]
-\frac{\mathcal{D}_1^2\lambda^2(1+4a_s)}{4}
-\frac{4\mathcal{C}\mathcal{D}_1^3a_s\lambda^3}{3u_s}
[1-2(\mathcal{C}\mathcal{D}_1(a-\frac{\lambda a_s}{u_s})-2a_s)]\nonumber\\
&&
-\frac{2\mathcal{S}^2\mathcal{F}^2a_s\kappa^2\hat{g}}{3u_s}
[1-2(2\mathcal{S}\mathcal{E}_1(a-\frac{\lambda a_s}{u_s})-a_s)]\Bigr\},\label{RG-eqs-type-I-beta1}\\
\frac{d\beta_2}{dl}
&=&\frac{(\mathcal{E}_2\mathcal{D}_1^2
-\mathcal{D}_2\mathcal{E}_1^2)}{(\mathcal{D}_1^2\mathcal{E}_2
-\mathcal{E}_1^2\mathcal{D}_2)}\beta_2+\frac{2\mathcal{E}_2}
{2\pi^2(\mathcal{D}_1^2\mathcal{E}_2
-\mathcal{E}_1^2\mathcal{D}_2)}
\Bigl\{\mathcal{S}^2 \hat{g}^2
[4\mathcal{S}\mathcal{E}_1(a-\frac{\lambda a_s}{u_s})-1]
-\frac{\mathcal{D}_1^2\lambda^2(1+4a_s)}{4}\nonumber\\
&&+9\mathcal{C}^2 [\beta_2\mathcal{D}_1^2
+(\beta_1-\beta_2)\mathcal{D}_2]^2
[4\mathcal{C}\mathcal{D}_1(a-\frac{\lambda a_s}{u_s})-1]
-\frac{4\mathcal{C}^2\mathcal{D}_1^2a_s\lambda^2[\beta_2\mathcal{D}_1^2
+(\beta_1-\beta_2)\mathcal{D}_2]}{u_s}\nonumber\\
&&\times[1-2\Bigl(2\mathcal{C}\mathcal{D}_1(a-\frac{\lambda a_s}{u_s})-a_s)]-\frac{4\mathcal{C}\mathcal{D}_1^3a_s\lambda^3}{3u_s}
[1-2(\mathcal{C}\mathcal{D}_1(a-\frac{\lambda a_s}{u_s})-2a_s)]
-\frac{2\mathcal{S}^2\mathcal{F}^2a_s\kappa^2\hat{g}}{3u_s}\nonumber\\
&&\times[1-2(2\mathcal{S}\mathcal{E}_1(a-\frac{\lambda a_s}{u_s})-a_s)]\Bigr\}
-\frac{2\mathcal{D}_2}{2\pi^2(\mathcal{D}_1^2\mathcal{E}_2
-\mathcal{E}_1^2\mathcal{D}_2)}
\Bigl\{\frac{-4\mathcal{S}^2\mathcal{E}_1^2a_s\lambda^2[\beta_2\mathcal{E}_1^2
+(\beta_1-\beta_2)\mathcal{E}_2]}{u_s}\nonumber\\
&&\times
[1-2(2\mathcal{S}\mathcal{E}_1(a-\frac{\lambda a_s}{u_s})-a_s)]-9\mathcal{S}^2[\beta_2\mathcal{E}_1^2
+(\beta_1-\beta_2)\mathcal{E}_2]^2[1-4\mathcal{S}\mathcal{E}_1(a-\frac{\lambda a_s}{u_s})]+\mathcal{C}^2\hat{g}^2\nonumber\\
&&\times
[4\mathcal{C}\mathcal{D}_1(a-\frac{\lambda a_s}{u_s})-1]
-\frac{\mathcal{E}_1^2\lambda^2(1+4a_s)}{4}
+\frac{-4\mathcal{S}\mathcal{E}_1^3a_s\lambda^3}{3u_s}
[1-2(\mathcal{S}\mathcal{E}_1(a-\frac{\lambda a_s}{u_s})-2a_s)]\nonumber\\
&&
+\frac{-2\mathcal{C}^2\mathcal{F}^2a_s\kappa^2\hat{g}}{3u_s}
[1-2(2\mathcal{C}\mathcal{D}_1(a-\frac{\lambda a_s}{u_s})-a_s)]\Bigr\},\label{RG-eqs-type-I-beta2}\\
\frac{d\lambda_{\Delta A}}{dl}
&=&\lambda_{\Delta A}+\frac{2}{2\pi^2}
\Bigl\{\frac{-64a_s\lambda^3_{\Delta A}}{9u_s}
[1-2(\mathcal{D}_1(a-\frac{\lambda a_s}{u_s})-a_s)]
-\frac{3a_su_s\lambda_{\Delta A}(1+6a_s)}{2}-\frac{3u_s\lambda_{\Delta A}(4a_s+1)}{8}\nonumber\\
&&-4\lambda_{\Delta A}^2
[1+2a_s(1+\frac{\lambda_{\Delta A}}{u_s})]
-4a_s\lambda_{\Delta A}^2
[1+\Bigl(4a_s+\frac{2\lambda_{\Delta A}a_s}{u_s})]\Bigr\},\label{RG-eqs-type-I-lamb-DA}\\
\frac{d\kappa}{dl}
&=&\kappa+\frac{2}{2\pi^2}
\Bigl\{\frac{-2\mathcal{C}\mathcal{S}\mathcal{D}_1\mathcal{E}_1a_s\lambda^2\kappa}{3u_s}
[1-2(\mathcal{C}\mathcal{D}_1(a-\frac{\lambda a_s}{u_s})+\mathcal{S}\mathcal{E}_1(a-\frac{\lambda a_s}{u_s})-a_s)]-\frac{3a_su_s\kappa(1+6a_s)}{2}\nonumber\\
&&
-\mathcal{C}\mathcal{S}\hat{g}\kappa
[1-2(\mathcal{C}\mathcal{D}_1(a-\frac{\lambda a_s}{u_s})+\mathcal{S}\mathcal{E}_1(a-\frac{\lambda a_s}{u_s}))]
-\frac{3(1+4a_s)u_s\kappa}{8}
-\frac{2\mathcal{C}\mathcal{S}^2\mathcal{E}_1
\hat{g}a_s\kappa\lambda}{3u_s}\nonumber\\
&&\times
[1-2(\mathcal{C}\mathcal{D}_1(a-\frac{\lambda a_s}{u_s})
+2\mathcal{S}\mathcal{E}_1(a-\frac{\lambda a_s}{u_s}))]
-\frac{\mathcal{C}^2\mathcal{S}^2\mathcal{F}^2
\hat{g}a_s^2\kappa^3}{6u_s^2}
[1-4(\mathcal{C}\mathcal{D}_1(a-\frac{\lambda a_s}{u_s})
+\mathcal{S}\mathcal{E}_1(a-\frac{\lambda a_s}{u_s}))]\nonumber\\
&&-\frac{2\mathcal{C}^2\mathcal{S}\mathcal{D}_1
\hat{g}a_s\kappa\lambda}{3u_s}
[1-2(2\mathcal{C}\mathcal{D}_1(a-\frac{\lambda a_s}{u_s})+\mathcal{S}\mathcal{E}_1(a-\frac{\lambda a_s}{u_s}))]\Bigr\},\label{RG-eqs-type-I-kappa}\\
\frac{d\hat{g}}{dl}
&=&\hat{g}+\frac{1}{2\pi^2}
\Bigl\{\frac{-4\mathcal{C}\mathcal{D}_1^2\mathcal{E}_1a_s\lambda^3}{3u_s}
[1-2(\mathcal{C}\mathcal{D}_1(a-\frac{\lambda a_s}{u_s})-2a_s)]
+\frac{-4\mathcal{S}\mathcal{D}_1\mathcal{E}_1^2a_s\lambda^3}{3u_s}
[1-2(\mathcal{S}\mathcal{E}_1
(a-\frac{\lambda a_s}{u_s})-2a_s)]\nonumber\\
&&
-\frac{32\mathcal{C}^2\mathcal{D}_1^2\hat{g}
a_s\lambda^2}{3u_s}
[1-2(2\mathcal{C}\mathcal{D}_1(a-\frac{\lambda a_s}{u_s})-a_s)]
+\mathcal{S}^3\mathcal{E}_1(a-\frac{\lambda a_s}{u_s})[\beta_2\mathcal{E}_1^2
+(\beta_1-\beta_2)\mathcal{E}_2]]\nonumber\\
&&
-\frac{32\mathcal{S}^2\mathcal{E}_1^2\hat{g}
a_s\lambda^2}{3u_s}
[1-2(2\mathcal{S}\mathcal{E}_1(a-\frac{\lambda a_s}{u_s})-a_s)]
-\frac{\mathcal{F}^2(1+4a_s)\kappa^2}{4}-
\frac{\mathcal{D}_1\mathcal{E}_1(4a_s+1)\lambda^2}{4}\nonumber\\
&&
+12\hat{g}
[\mathcal{C}^3\mathcal{D}_1(a-\frac{\lambda a_s}{u_s})
[\beta_2\mathcal{D}_1^2
+(\beta_1-\beta_2)\mathcal{D}_2]
+8\mathcal{C}\mathcal{S}
\hat{g}^2
[2(\mathcal{C}\mathcal{D}_1(a-\frac{\lambda a_s}{u_s})\nonumber\\
&&
+\mathcal{S}\mathcal{E}_1(a-\frac{\lambda a_s}{u_s}))-1]
-3(\hat{g}-\beta_2\mathcal{D}_1\mathcal{E}_1)[\mathcal{C}^2
[\beta_2\mathcal{D}_1^2
+(\beta_1-\beta_2)\mathcal{D}_2]
+\mathcal{S}^2[\beta_2\mathcal{E}_1^2
+(\beta_1-\beta_2)\mathcal{E}_2]]\nonumber\\
&&
-\frac{2\mathcal{C}^2\mathcal{F}^2[\beta_2\mathcal{D}_1^2
+(\beta_1-\beta_2)\mathcal{D}_2]a_s\kappa^2}{u_s}
[1-2(2\mathcal{C}\mathcal{D}_1(a-\frac{\lambda a_s}{u_s})-a_s)]
-\frac{2\mathcal{S}^2\mathcal{F}^2[\beta_2\mathcal{E}_1^2
+(\beta_1-\beta_2)\mathcal{E}_2]a_s\kappa^2}{u_s}\nonumber\\
&&\times
[1-2(2\mathcal{S}\mathcal{E}_1(a-\frac{\lambda a_s}{u_s})-a_s)]
-\frac{4\mathcal{C}^2\mathcal{S}^2\mathcal{F}^2\hat{g}^2a_s^2\kappa^2}
{3u_s^2}
[1-4(\mathcal{C}\mathcal{D}_1(a-\frac{\lambda a_s}{u_s})
+\mathcal{S}\mathcal{E}_1(a-\frac{\lambda a_s}{u_s}))]\nonumber\\
&&
-\frac{3\mathcal{C}^2\mathcal{S}^2\mathcal{F}^2[\beta_2\mathcal{D}_1^2
+(\beta_1-\beta_2)\mathcal{D}_2]
[\beta_2\mathcal{E}_1^2
+(\beta_1-\beta_2)\mathcal{E}_2]a_s^2\kappa^2}{2u_s^2}
[1-4(\mathcal{C}\mathcal{D}_1(a-\frac{\lambda a_s}{u_s})
+\mathcal{S}\mathcal{E}_1(a-\frac{\lambda a_s}{u_s}))]\nonumber\\
&&
-\frac{\mathcal{C}^2\mathcal{S}^2\mathcal{F}^2\hat{g}^2a_s^2\kappa^2}
{3u_s^2}
[1-4(\mathcal{C}\mathcal{D}_1(a-\frac{\lambda a_s}{u_s})
+\mathcal{S}\mathcal{E}_1(a-\frac{\lambda a_s}{u_s}))]\Bigr\},
\label{RG-eqs-type-I-g-hat}
\end{eqnarray}
\end{small}
where the variable functions are designated as
\begin{eqnarray}
\mathcal{D}_1&\equiv&|\mathbf{n}_X|^2\cos^2\theta,\hspace{0.5cm}
\mathcal{D}_2\equiv|\mathbf{n}^2_X|^2\cos^4\theta,\hspace{0.5cm}
\mathcal{E}_1\equiv|\mathbf{n}_Y|^2\sin^2\theta,\hspace{0.5cm}
\mathcal{E}_2\equiv|\mathbf{n}^2_Y|^2\sin^4\theta,\\
\mathcal{F}&\equiv&|\cos\theta\sin\theta\mathbf{n}_X\cdot \mathbf{n}_Y|
+|\cos\theta\sin\theta\mathbf{n}_X\cdot \mathbf{n}^*_Y|,
\hspace{0.5cm}\mathcal{C}\equiv1/|\mathbf{n}_X\cos\theta|^2, \hspace{0.5cm}
\mathcal{S}\equiv1/|\mathbf{n}_Y\sin\theta|^2.
\end{eqnarray}
Here, we would like to stress that $\theta\in[0,\pi/2]$, and
$\theta=0,\pi/2$ serve as single magnetic order parameter
with $\mathbf{Q}_X$ or $\mathbf{Q}_Y$, respectively.

\begin{figure*}
\centering
\includegraphics[width=3.5in]{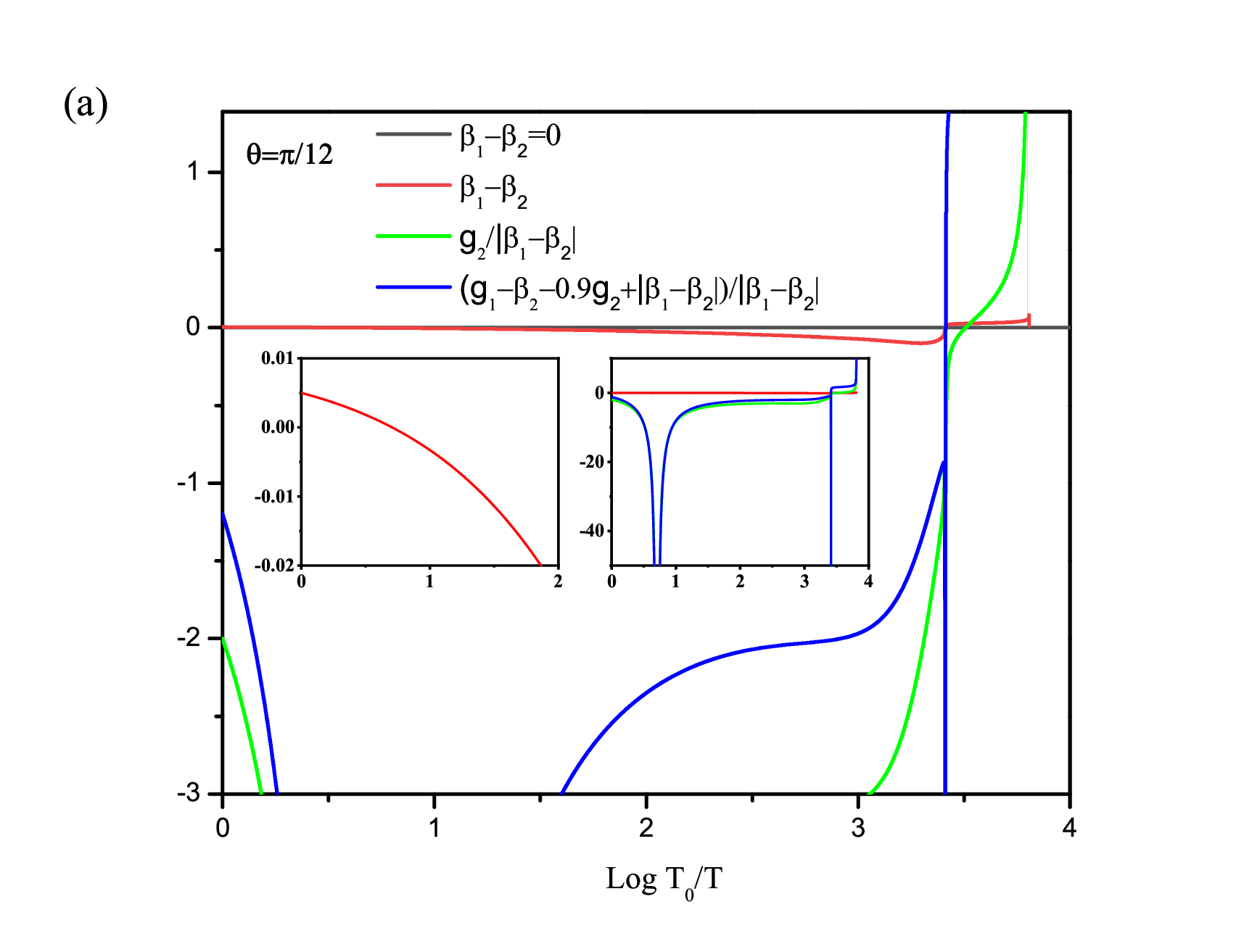}
\includegraphics[width=3.5in]{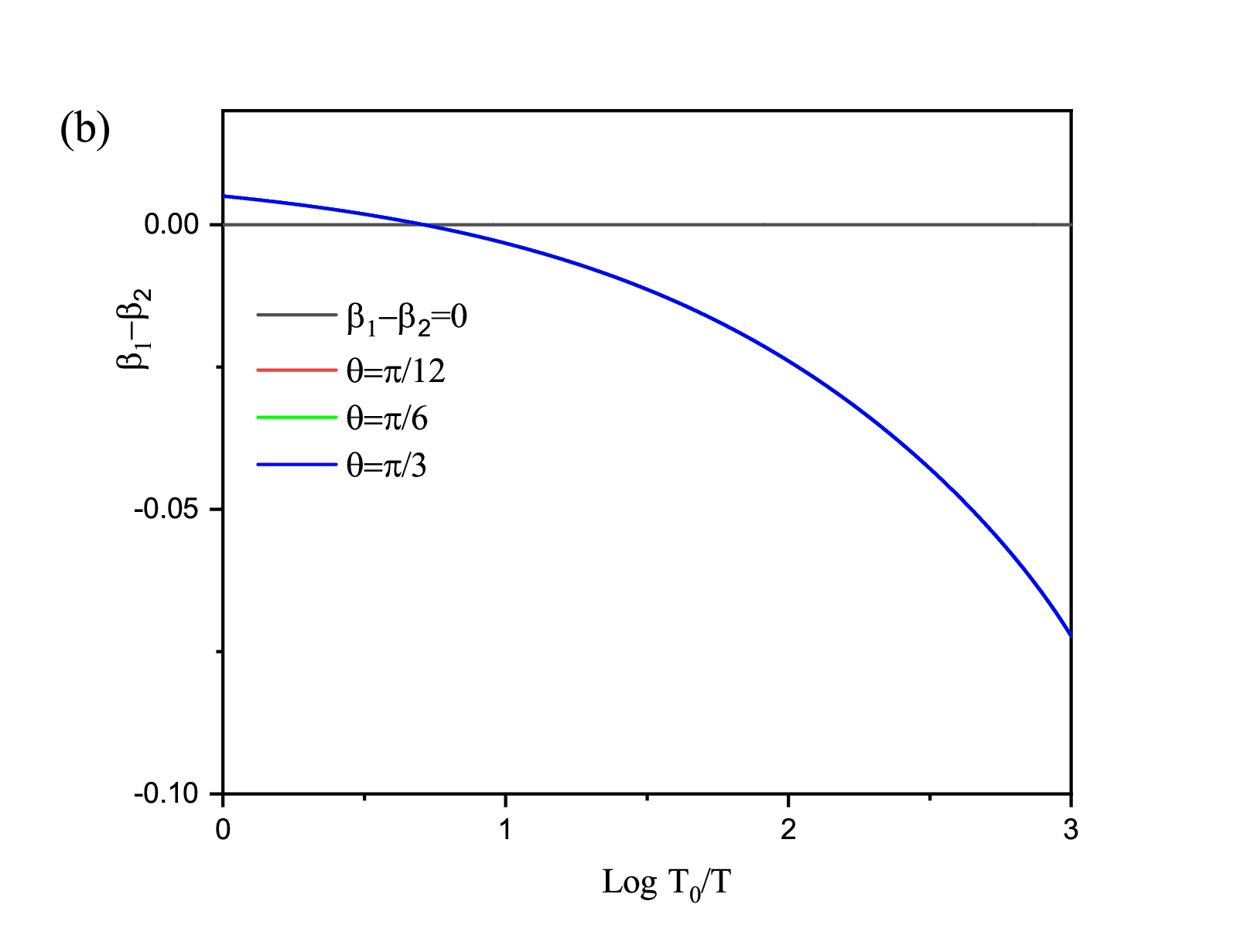}
\vspace{-0.6cm}
\caption{(Color online) (a) Temperature-dependent stable
constraints of the $C_2$-symmetry DPMH state
under the representative starting values of interaction parameters
chosen as  $g_1=-0.015$, $g_2=-0.01$, $u_s=0.05$, $\lambda=0.01$,
$\beta_1=0.005$, $\beta_2=0.01$ (the qualitative results are
insensitive to the initial values). Hereby, the angle $\theta$ is designated
in Sec.~\ref{Sec_S_eff} to specify the direction of magnetic order
in the spin space. Insets: the sign-change region of $\beta_1-\beta_2$ (left panel) and enlarged-region for $\frac{g-\tilde{\beta}}{|\beta-\tilde{\beta}|}-\frac{9}{10}
\frac{\tilde{g}}{|\beta-\tilde{\beta}|}+1$ (right panel).
(b) Sign-change regions of $\beta_1-\beta_2$ under different values of
$\theta$.}\label{Fig_C2-DPMH}
\end{figure*}

For type-II case, at which $|\mathbf{n}^2_i|^2\neq|\mathbf{n}_i|^4$ with $i=X,Y$,
only one of $\beta_1$ and $\beta_2$ flows independently. In this circumstance,
the flows of $a_s$, $a$, $u_s$, $\lambda$, $\lambda_{\Delta A}$ and $\kappa$
share the same evolutions with their type-I counterparts. Nevertheless, the parameter
$\hat{g}$ evolves under the following way
\begin{small}
\begin{eqnarray}
\frac{d\hat{g}}{dl}
&=&\hat{g}+\frac{1}{2\pi^2}
\Bigl\{\frac{-4\mathcal{C}\mathcal{D}_1^2\mathcal{E}_1a_s\lambda^3}{3u_s}
[1-2(\mathcal{C}\mathcal{D}_1(a-\frac{\lambda a_s}{u_s})-2a_s)]
-\frac{4\mathcal{S}\mathcal{D}_1\mathcal{E}_1^2a_s\lambda^3}{3u_s}
[1-2(\mathcal{S}\mathcal{E}_1(a-\frac{\lambda a_s}{u_s})-2a_s)]\nonumber\\
&&
-\frac{32\mathcal{C}^2\mathcal{D}_1^2\hat{g}
a_s\lambda^2}{3u_s}
[1-2(2\mathcal{C}\mathcal{D}_1(a-\frac{\lambda a_s}{u_s})-a_s)]
-\frac{32\mathcal{S}^2\mathcal{E}_1^2\hat{g}
a_s\lambda^2}{3u_s}
[1-2(2\mathcal{S}\mathcal{E}_1(a-\frac{\lambda a_s}{u_s})-a_s)]\nonumber\\
&&
+12\hat{g}
[\mathcal{C}^3\mathcal{D}_1(a-\frac{\lambda a_s}{u_s})
[\beta_2\mathcal{D}_1^2
+(\beta_1-\beta_2)\mathcal{D}_2]
+\mathcal{S}^3\mathcal{E}_1(a-\frac{\lambda a_s}{u_s})[\beta_2\mathcal{E}_1^2
+(\beta_1-\beta_2)\mathcal{E}_2]]-
\frac{\mathcal{D}_1\mathcal{E}_1(4a_s+1)\lambda^2}{4}\nonumber\\
&&
+8\mathcal{C}\mathcal{S}
\hat{g}^2
[2(\mathcal{C}\mathcal{D}_1(a-\frac{\lambda a_s}{u_s})
+\mathcal{S}\mathcal{E}_1(a-\frac{\lambda a_s}{u_s}))-1]
-\frac{\mathcal{F}^2(1+4a_s)\kappa^2}{4}
-3\hat{g}[\mathcal{C}^2
[\beta_2\mathcal{D}_1^2
+(\beta_1-\beta_2)\mathcal{D}_2]\nonumber\\
&&
+\mathcal{S}^2[\beta_2\mathcal{E}_1^2
+(\beta_1-\beta_2)\mathcal{E}_2]]
+\frac{-2\mathcal{C}^2\mathcal{F}^2[\beta_2\mathcal{D}_1^2
+(\beta_1-\beta_2)\mathcal{D}_2]a_s\kappa^2}{u_s}
[1-2(2\mathcal{C}\mathcal{D}_1(a-\frac{\lambda a_s}{u_s})-a_s)]\nonumber\\
&&
+\frac{-2\mathcal{S}^2\mathcal{F}^2[\beta_2\mathcal{E}_1^2
+(\beta_1-\beta_2)\mathcal{E}_2]a_s\kappa^2}{u_s}
[1-2(2\mathcal{S}\mathcal{E}_1(a-\frac{\lambda a_s}{u_s})-a_s)]
-\frac{4\mathcal{C}^2\mathcal{S}^2\mathcal{F}^2\hat{g}^2a_s^2\kappa^2}
{3u_s^2}\nonumber\\
&&\times
[1-4(\mathcal{C}\mathcal{D}_1(a-\frac{\lambda a_s}{u_s})
+\mathcal{S}\mathcal{E}_1(a-\frac{\lambda a_s}{u_s}))]
-\frac{3\mathcal{C}^2\mathcal{S}^2\mathcal{F}^2[\beta_2\mathcal{D}_1^2
+(\beta_1-\beta_2)\mathcal{D}_2]
[\beta_2\mathcal{E}_1^2
+(\beta_1-\beta_2)\mathcal{E}_2]a_s^2\kappa^2}{2u_s^2}\nonumber\\
&&\times
[1-4(\mathcal{C}\mathcal{D}_1(a-\frac{\lambda a_s}{u_s})
+\mathcal{S}\mathcal{E}_1(a-\frac{\lambda a_s}{u_s}))]
-\frac{\mathcal{C}^2\mathcal{S}^2\mathcal{F}^2\hat{g}^2a_s^2\kappa^2}
{3u_s^2}
[1-4(\mathcal{C}\mathcal{D}_1(a-\frac{\lambda a_s}{u_s})
+\mathcal{S}\mathcal{E}_1(a-\frac{\lambda a_s}{u_s}))]\Bigr\}.\label{RG-eqs-type-II-g-hat}
\end{eqnarray}
\end{small}

\begin{figure}
\centering
\includegraphics[width=3.5in]{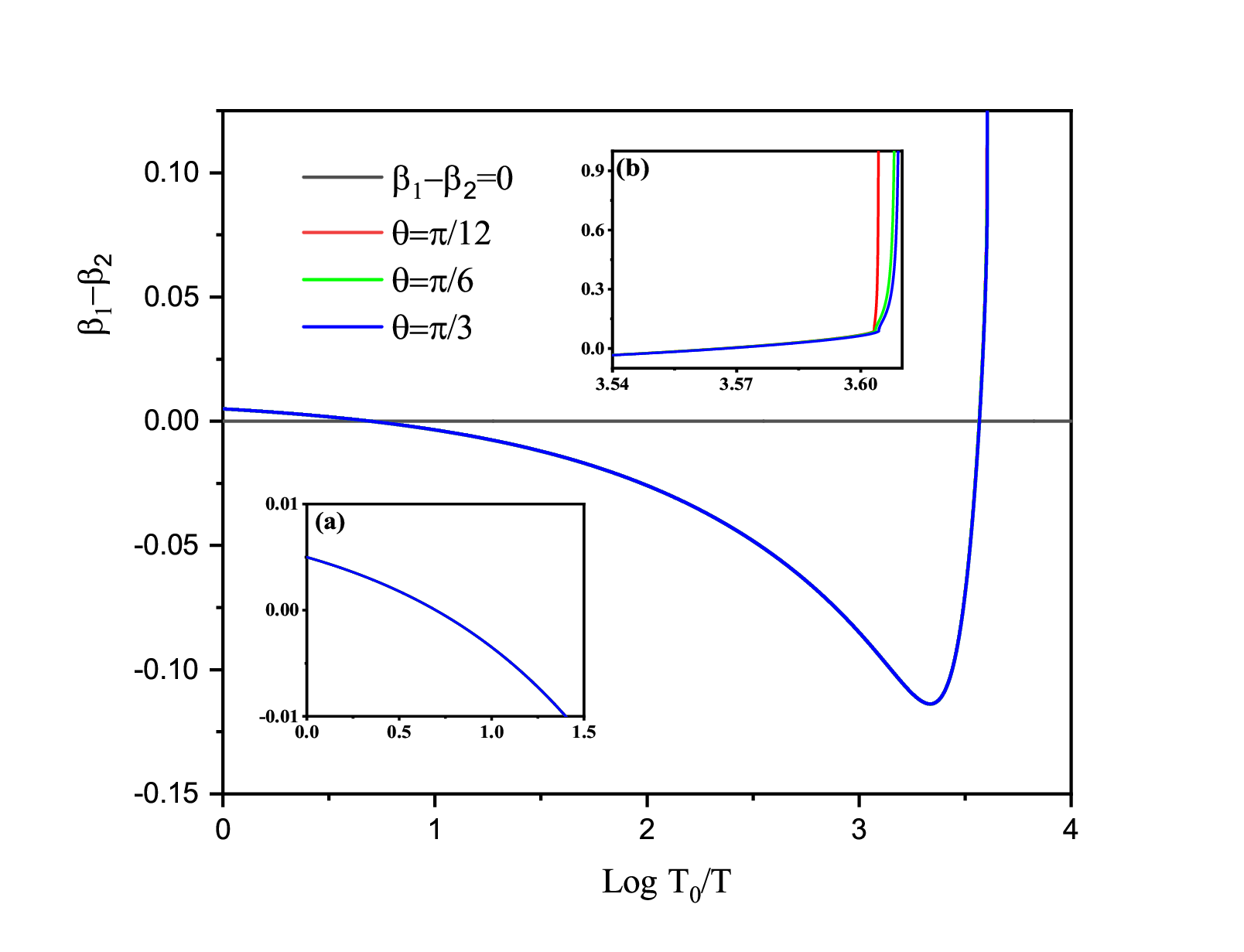}
\vspace{-0.6cm}
\caption{(Color online) Temperature-dependent stable
constraints ($\beta_1-\beta_2$) of the $C_2$-symmetry MH state
under the representative starting values of interaction parameters
chosen as  $g_1=0.01$, $g_2=-0.01$, $u_s=0.05$, $\lambda=0.01$,
$\beta_1=0.01$, $\beta_2=0.005$ with three
representative $\theta=\pi/12,\pi/6,\pi/3$
to satisfy the MH's stable constraint (the qualitative results are
insensitive to the initial values). Hereby, the angle $\theta$ is designated
in Sec.~\ref{Sec_S_eff} to specify the direction of magnetic order
in the spin space. Insets: (a) the sign-change
region of $\beta_1-\beta_2$ and (b) behaviors
around $l_c$.}\label{Fig_mag-helix-beta12}
\end{figure}

\begin{figure*}
\centering
\includegraphics[width=2.6in]{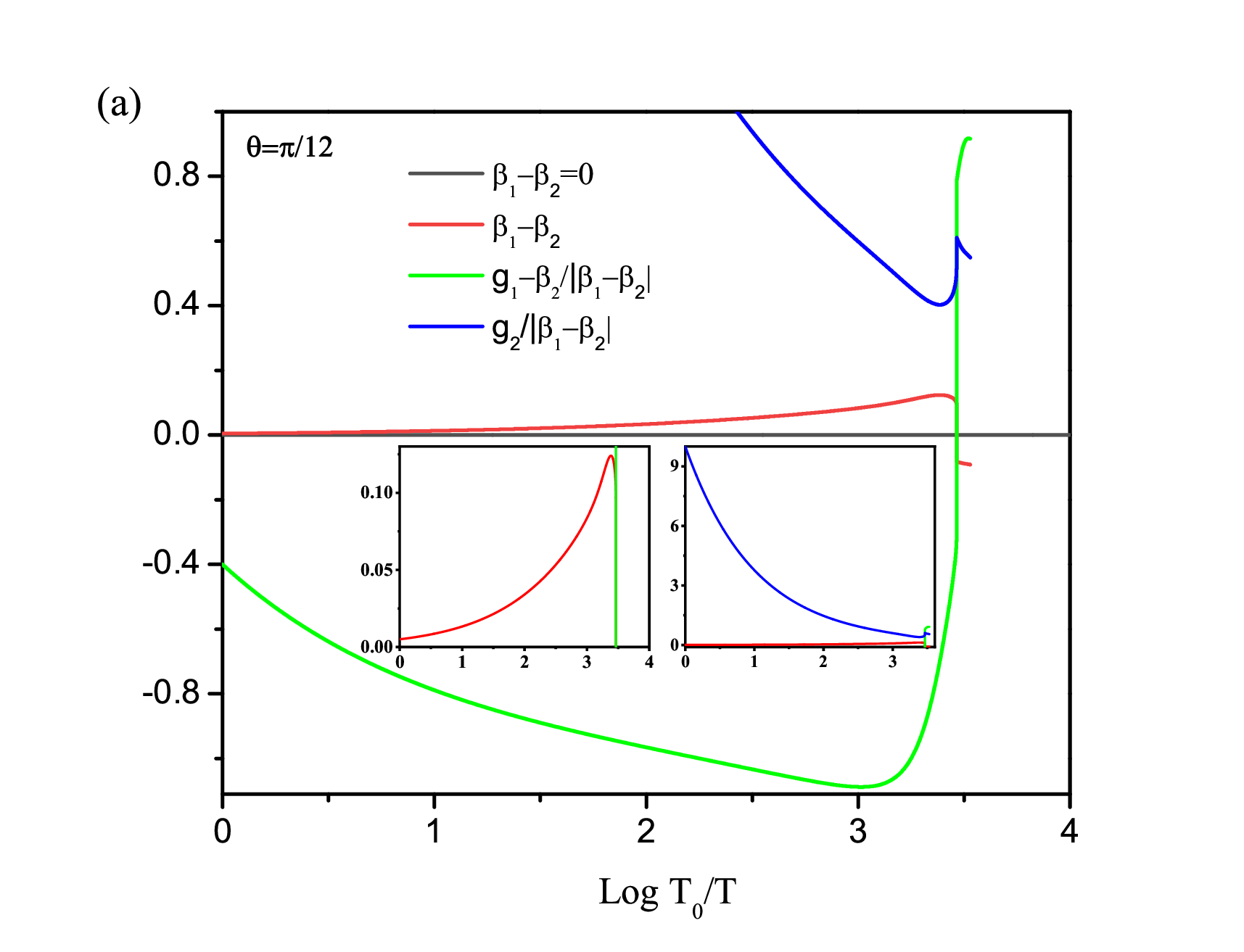}\hspace{-1.1cm}
\includegraphics[width=2.6in]{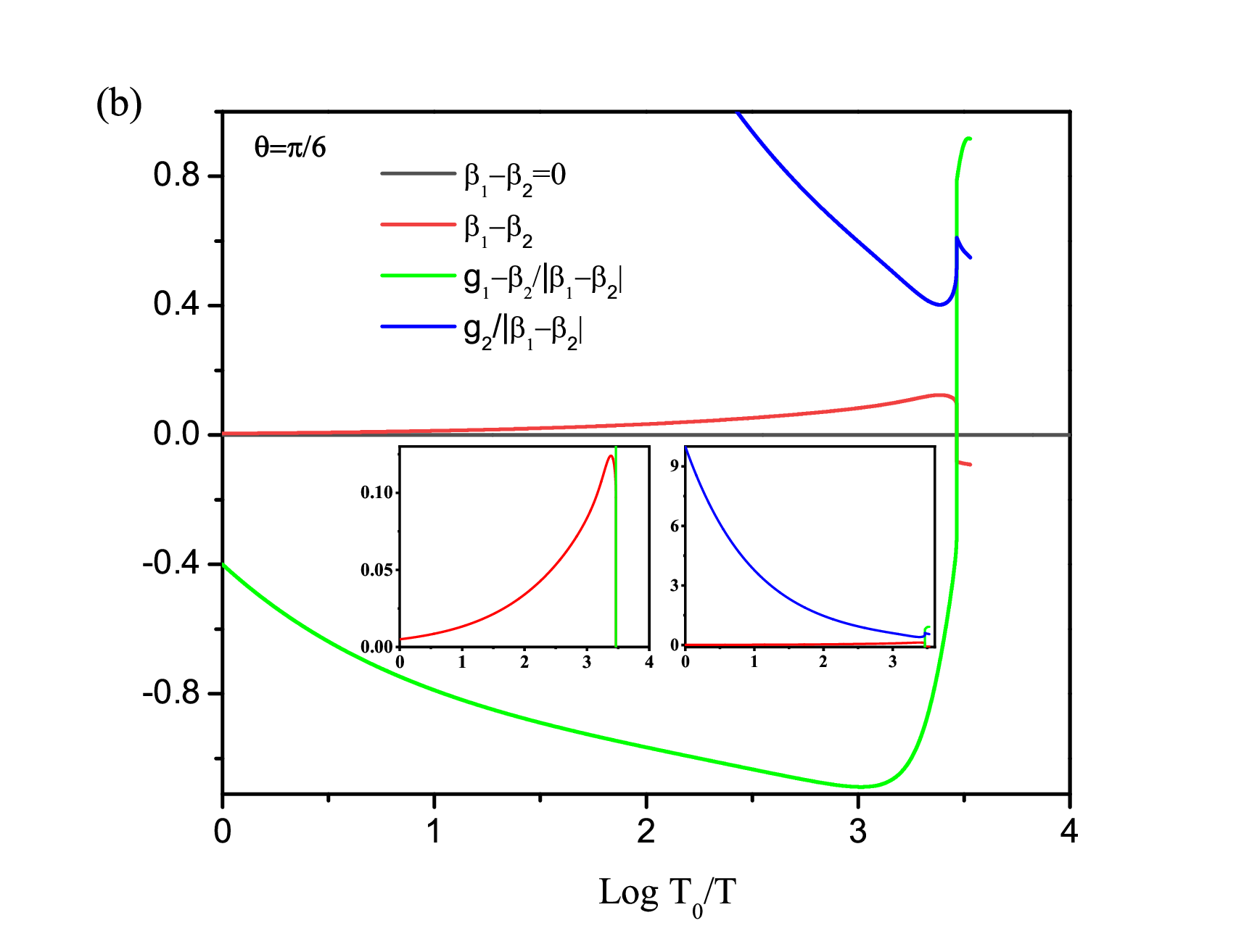}\hspace{-1.1cm}
\includegraphics[width=2.6in]{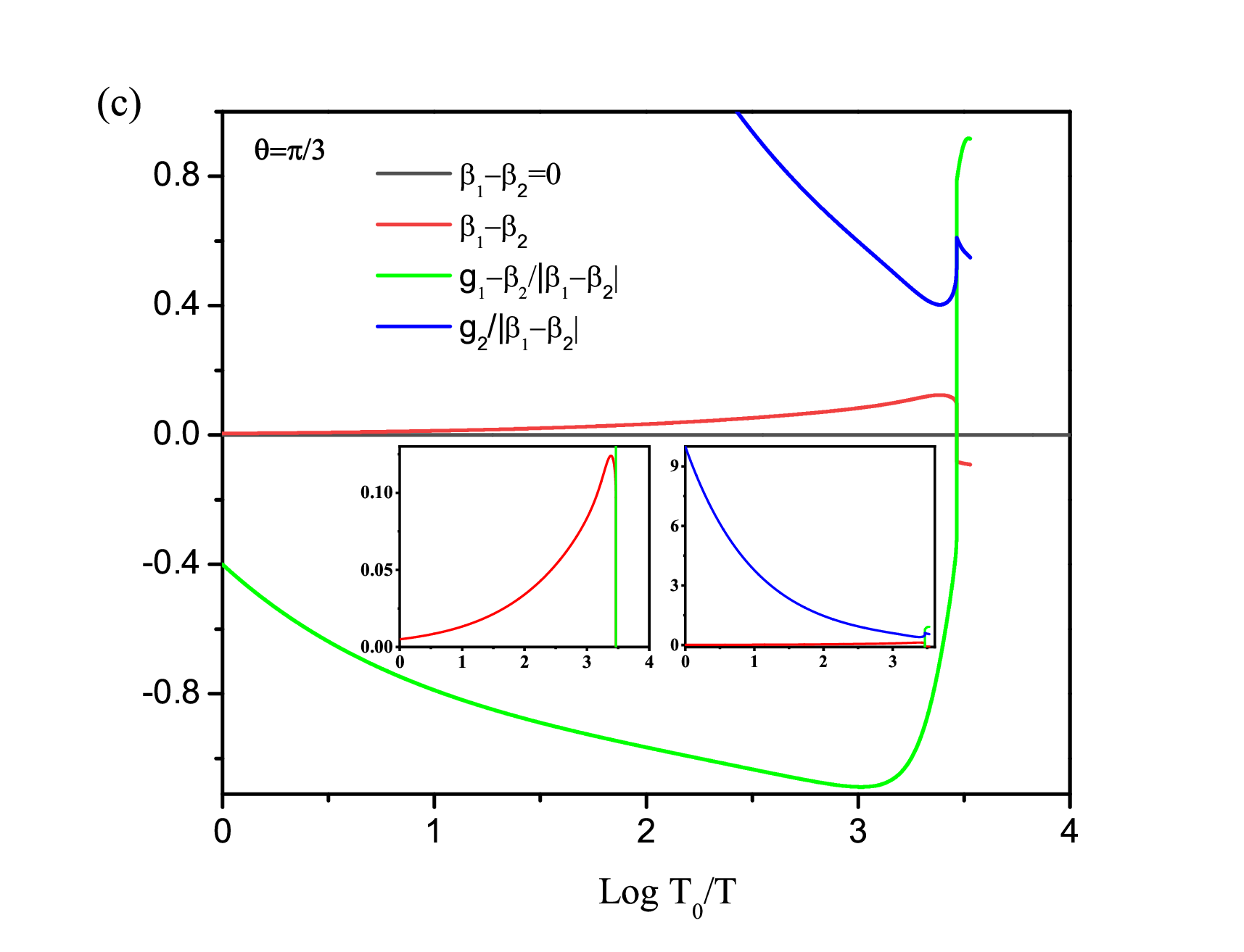}\\
\vspace{-0.26cm}
\caption{(Color online) Temperature-dependent stable
constraints of the $C_2$-symmetry ICS $\perp$ MH state
under the representative starting values of interaction parameters
chosen as  $g_1=0.03$, $g_2=0.05$, $u_s=0.05$, $\lambda=0.01$,
$\beta_1=0.01$, $\beta_2=0.005$ with three
representative $\theta=\pi/12,\pi/6,\pi/3$
to satisfy the ICS $\perp$ MH's stable constraint
(the qualitative results are insensitive to the initial values).
Hereby, the angle $\theta$ is designated
in Sec.~\ref{Sec_S_eff} to specify the direction of magnetic order
in the spin space: (a) $\theta=\pi/12$, (b) $\theta=\pi/6$, and (c) $\theta=\pi/3$.
Insets: the enlarged regions for $\beta_1-\beta_2$ (left panel)
and $g_2/|\beta_1-\beta_2|$ (right panel).}\label{Fig_IC-perp-helix-g-big}
\end{figure*}

Furthermore, the RG equations of parameters $\beta_1$ and $\beta_2$ can be
broken down into six distinct sorts depending on the concrete conditions.

For type-II case-A with $|\mathbf{n}^2_X|^2=|\mathbf{n}_X|^4$, $|\mathbf{n}^2_Y|^2=|\mathbf{n}_Y|^4$,
$|\mathbf{n}^2_Y|^2=0$ and $|\mathbf{n}^2_X|^2\neq0$, $\beta_1$
evolves but $\beta_2$ is an invariant constant,
\begin{small}
\begin{eqnarray}
\frac{d\beta_1}{dl}
&=&\beta_1+\frac{2}{2\pi^2}
\Bigl\{\frac{-4\mathcal{C}^2\mathcal{D}_1^2a_s\lambda^2\beta_1}{u_s}
[1-2(2\mathcal{C}\mathcal{D}_1(a-\frac{\lambda a_s}{u_s})-a_s)]+9\mathcal{C}^2 \mathcal{D}_2\beta^2_1
[4\mathcal{C}\mathcal{D}_1(a-\frac{\lambda a_s}{u_s})-1]-\frac{\lambda^2(1+4a_s)}{4}\nonumber\\
&&
+\frac{\mathcal{S}^2 \hat{g}^2}{\mathcal{D}_2}
[4\mathcal{S}\mathcal{E}_1(a-\frac{\lambda a_s}{u_s})-1]
-\frac{4\mathcal{C}\mathcal{D}_1a_s\lambda^3}{3u_s}
[1-2(\mathcal{C}\mathcal{D}_1(a-\frac{\lambda a_s}{u_s})-2a_s)]\nonumber\\
&&
-\frac{2\mathcal{S}^2\mathcal{F}^2a_s\kappa^2\hat{g}}
{3\mathcal{D}_2u_s}
[1-2(2\mathcal{S}\mathcal{E}_1
(a-\frac{\lambda a_s}{u_s})-a_s)]\Bigr\},\label{RG-eqs-type-II-case-A-beta1}\\
\frac{d\beta_2}{dl}&=&0.\label{RG-eqs-type-II-case-A-beta2}
\end{eqnarray}
\end{small}

For type-II case-B with $|\mathbf{n}^2_X|^2=|\mathbf{n}_X|^4$, $|\mathbf{n}^2_Y|^2=|\mathbf{n}_Y|^4$, $|\mathbf{n}^2_X|^2=0$ and $|\mathbf{n}^2_Y|^2\neq0$, $\beta_1$ evolves whereas $\beta_2$ is an
invariant constant,
\begin{small}
\begin{eqnarray}
\frac{d\beta_1}{dl}
&=&\beta_1+\frac{2}{2\pi^2}
\Bigl\{\frac{-4\mathcal{S}^2\mathcal{E}_1^2a_s\lambda^2\beta_1}{u_s}
[1-2(2\mathcal{S}\mathcal{E}_1(a-\frac{\lambda a_s}{u_s})-a_s)]
-9\mathcal{S}^2\mathcal{E}_2\beta^2_1[1-4\mathcal{S}\mathcal{E}_1(a-\frac{\lambda a_s}{u_s})]-\frac{\lambda^2(1+4a_s)}{4}\nonumber\\
&&+\frac{\mathcal{C}^2\hat{g}^2}{\mathcal{E}_2}
[4\mathcal{C}\mathcal{D}_1(a-\frac{\lambda a_s}{u_s})-1]
-\frac{4\mathcal{S}\mathcal{E}_1a_s\lambda^3}{3u_s}
[1-2(\mathcal{S}\mathcal{E}_1(a-\frac{\lambda a_s}{u_s})-2a_s)]\nonumber\\
&&-\frac{2\mathcal{C}^2\mathcal{F}^2a_s\kappa^2\hat{g}}
{3\mathcal{E}_2u_s}
[1-2(2\mathcal{C}\mathcal{D}_1
(a-\frac{\lambda a_s}{u_s})-a_s)]\Bigr\},\label{RG-eqs-type-II-case-B-beta1}\\
\frac{d\beta_2}{dl}&=&0.\label{RG-eqs-type-II-case-B-beta2}
\end{eqnarray}
\end{small}

For type-II case-C with $|\mathbf{n}^2_X|^2=|\mathbf{n}_X|^4$, $|\mathbf{n}^2_Y|^2\neq|\mathbf{n}_Y|^4$, $|\mathbf{n}^2_Y|^2=0$, and $|\mathbf{n}^2_X|^2=0$, $\beta_2$ evolves but $\beta_1$ is an
invariant constant,
\begin{small}
\begin{eqnarray}
\frac{d\beta_1}{dl}
&=&0,\label{RG-eqs-type-II-case-C-beta1}\\
\frac{d\beta_2}{dl}
&=&\beta_2+\frac{2}{2\pi^2}
\Bigl\{\frac{-4\mathcal{S}^2a_s\lambda^2[\beta_2\mathcal{E}_1^2
+(\beta_1-\beta_2)\mathcal{E}_2]}{u_s}
[1-2(2\mathcal{S}\mathcal{E}_1(a-\frac{\lambda a_s}{u_s})-a_s)]
-9\mathcal{S}^2\beta^2_2\mathcal{E}_1^2[1-4\mathcal{S}\mathcal{E}_1(a-\frac{\lambda a_s}{u_s})]\nonumber\\
&&+\frac{\mathcal{C}^2\hat{g}^2}{\mathcal{E}^2_1}
[4\mathcal{C}\mathcal{D}_1(a-\frac{\lambda a_s}{u_s})-1]-\frac{4\mathcal{S}\mathcal{E}_1a_s\lambda^3}{3u_s}
[1-2(\mathcal{S}\mathcal{E}_1(a-\frac{\lambda a_s}{u_s})-2a_s)]
-\frac{\lambda^2(1+4a_s)}{4}\nonumber\\
&&
-\frac{2\mathcal{C}^2\mathcal{F}^2a_s\kappa^2\hat{g}}
{3\mathcal{E}^2_1u_s}
[1-2(2\mathcal{C}\mathcal{D}_1(a-\frac{\lambda a_s}{u_s})-a_s)]\Bigr\}.\label{RG-eqs-type-II-case-C-beta2}
\end{eqnarray}
\end{small}

\begin{figure*}
\centering
\includegraphics[width=3.5in]{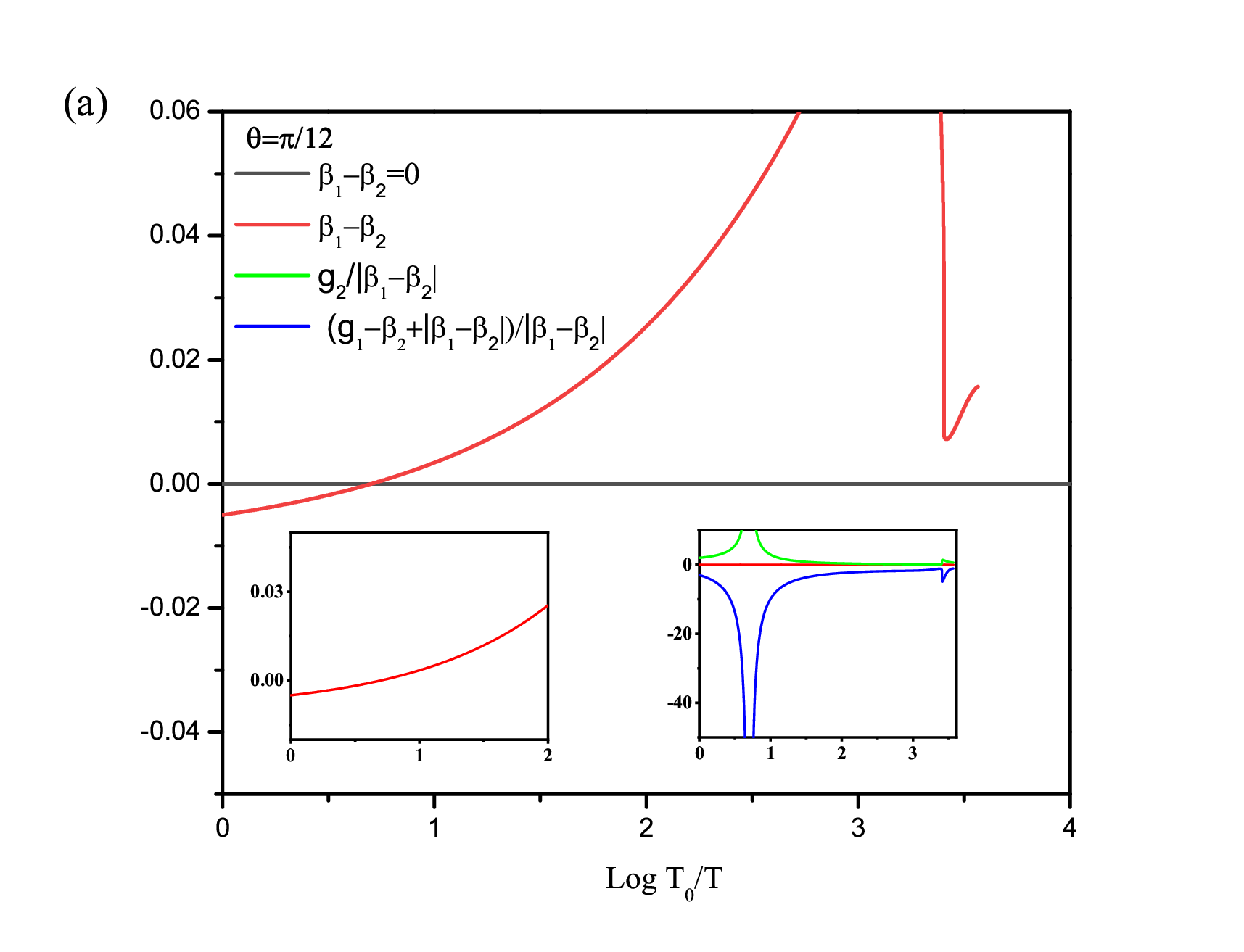}
\includegraphics[width=3.5in]{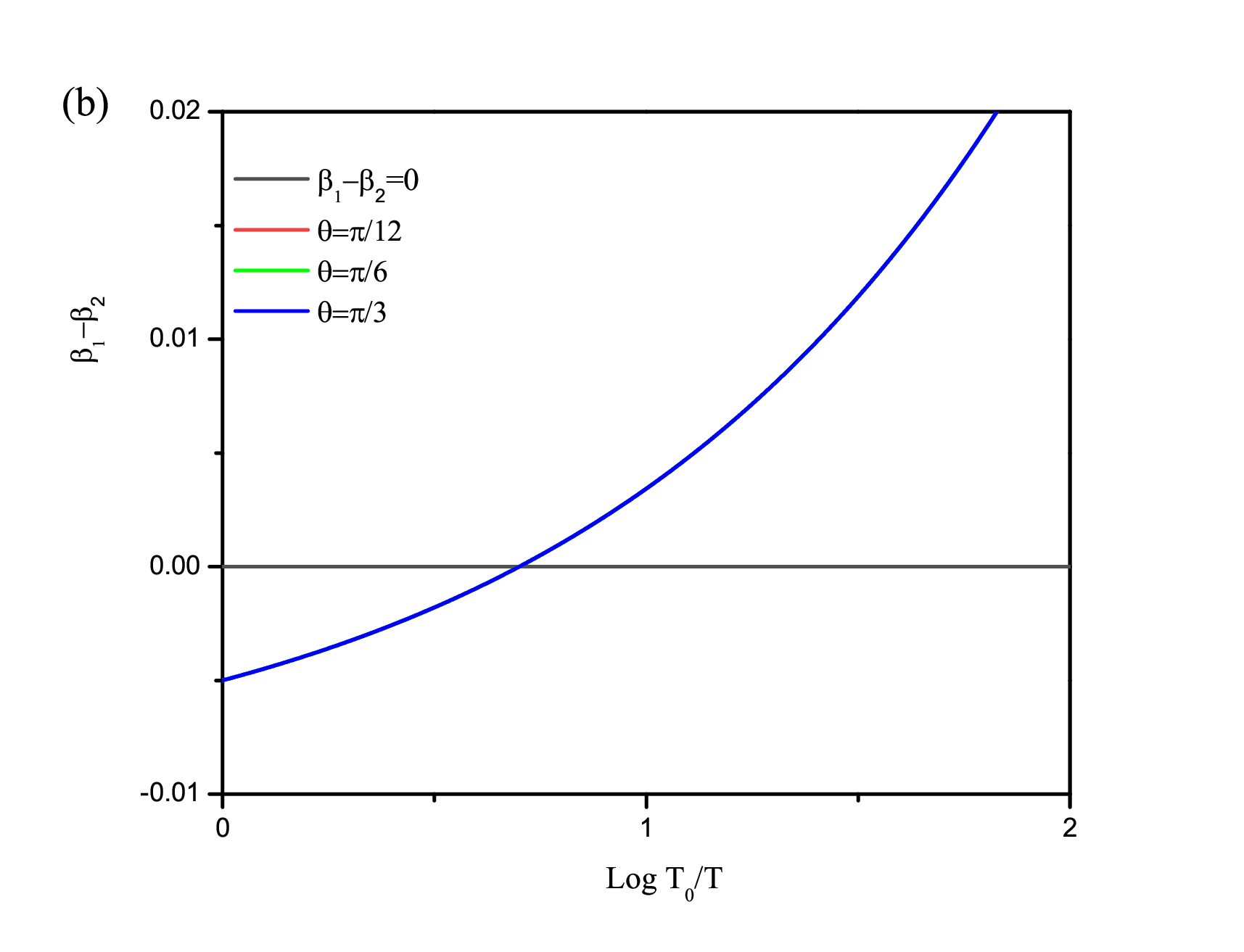}
\vspace{-0.6cm}
\caption{(Color online) (a) Temperature-dependent stable
constraints of $C_4$-symmetry SVC state under the
representative starting values of interaction parameters
chosen as  $g_1=-0.01$, $g_2=0.01$, $u_s=0.05$, $\lambda=0.01$,
$\beta_1=0.005$, $\beta_2=0.01$ with three
representative $\theta=\pi/12,\pi/6,\pi/3$
to satisfy the SVC's stable constraint
(the qualitative results are insensitive to
initial values of parameters). Hereby, the angle
$\theta$ is designated in Sec.~\ref{Sec_S_eff} to
specify the direction of magnetic order
in the spin space. Inset: the enlarge region for $g_2/|\beta_1-\beta_2|$ and
$(g_1+|\beta_1-\beta_2|)/|\beta_1-\beta_2|$.
(b) Sign-change regions at different values of
$\theta$.}\label{Fig_SVC-pi-12}
\end{figure*}

\begin{figure*}
\centering
\includegraphics[width=3.5in]{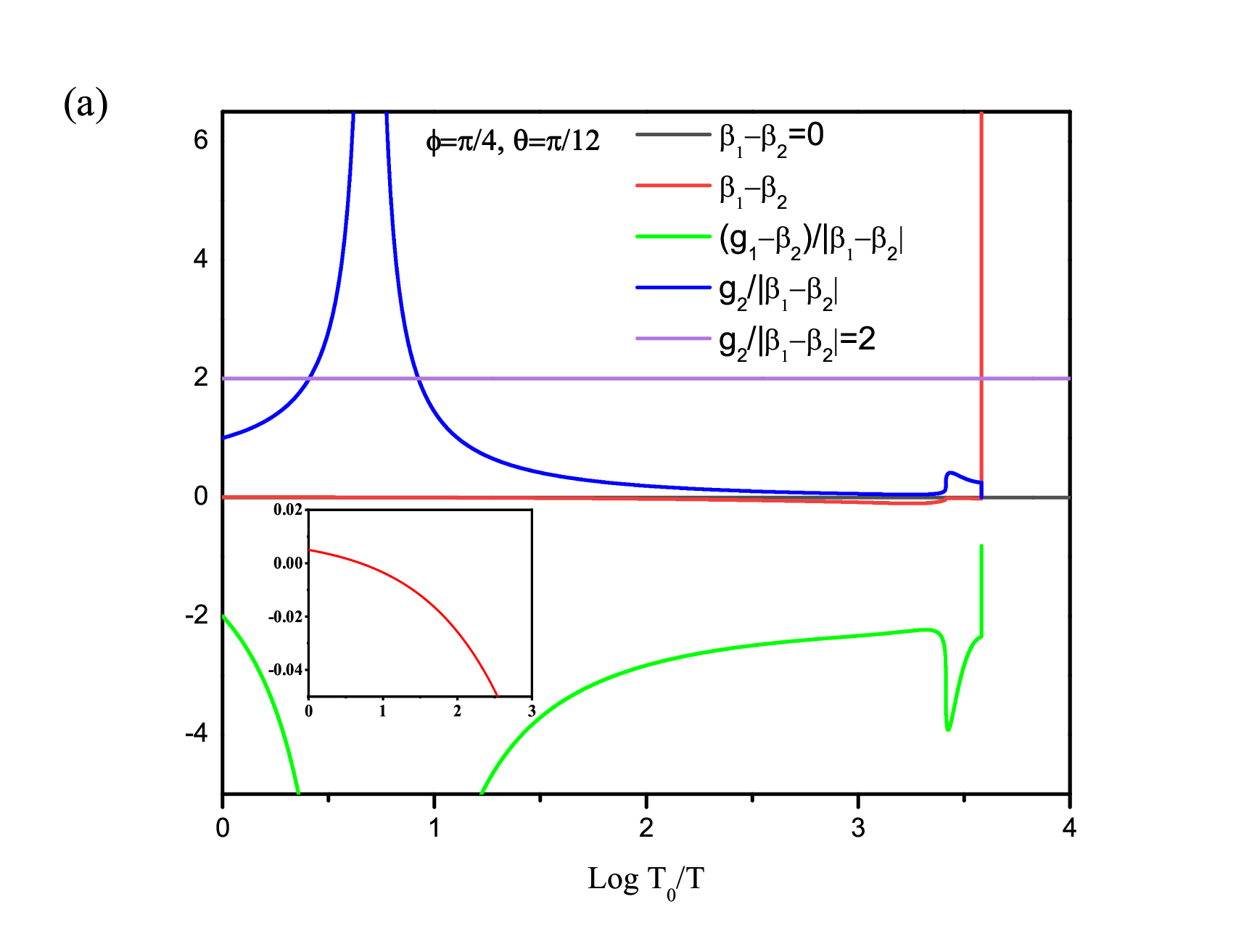}
\includegraphics[width=3.5in]{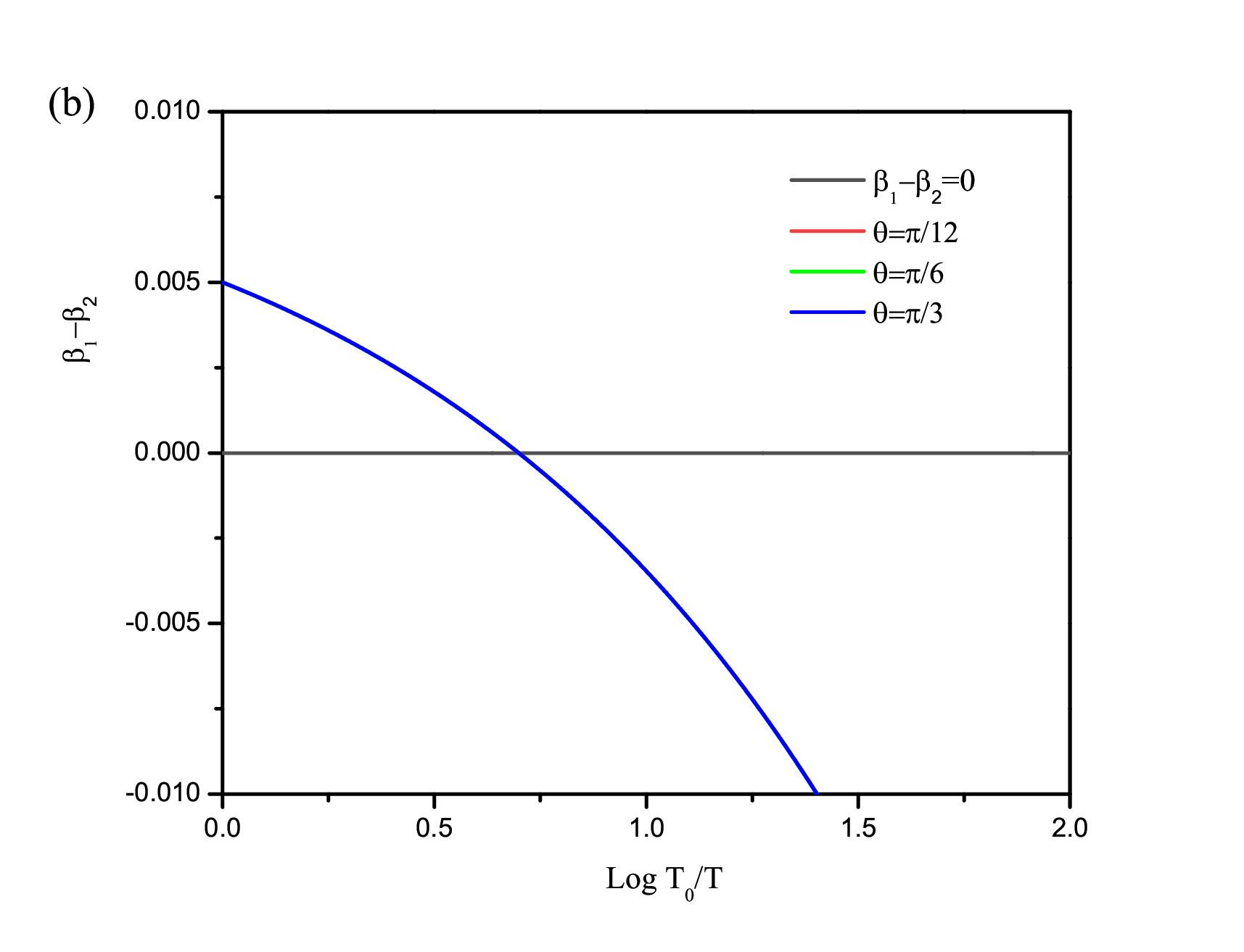}
\vspace{-0.6cm}
\caption{(Color online) (a) Temperature-dependent stable
constraints of symmetric double-$\mathbf{Q}$ noncoplanar SWC state
under the representative starting values of interaction parameters
chosen as  $g_1=-0.005$, $g_2=0.005$, $u_s=0.05$, $\lambda=0.01$,
$\beta_1=0.01$, $\beta_2=0.005$ with three
representative $\theta=\pi/12,\pi/6,\pi/3$
to satisfy the symmetric SWC's stable constraint (the qualitative
results are insensitive to initial values of parameters).
Hereby, the angle $\theta$ is designated
in Sec.~\ref{Sec_S_eff} to specify the direction of magnetic order
in the spin space. In addition, $\phi$ is introduced by
$\mathbf{n}_X\cdot \mathbf{n}_Y=\sin^2\phi$
and $\phi=\pi/4$ corresponds to the
symmetric noncoplanar SWC~\cite{Andersen2018PRX}.
Inset: the enlarge region for $\beta_1-\beta_2$.
(b) Sign-change regions at different values of
$\theta$.}\label{Fig_SWC-symm}
\end{figure*}

For type-II case-D with $|\mathbf{n}^2_Y|^2=|\mathbf{n}_Y|^4$,
$|\mathbf{n}^2_X|^2\neq|\mathbf{n}_X|^4$, $|\mathbf{n}^2_Y|^2=0$,
and $|\mathbf{n}^2_X|^2=0$, $\beta_2$ evolves but $\beta_1$ is an
invariant constant,
\begin{small}
\begin{eqnarray}
\frac{d\beta_1}{dl}
&=&0,\label{RG-eqs-type-II-case-D-beta1}\\
\frac{d\beta_2}{dl}
&=&\beta_2+\frac{2}{2\pi^2}
\Bigl\{\frac{-4\mathcal{C}^2a_s\lambda^2[\beta_2\mathcal{D}_1^2
+(\beta_1-\beta_2)\mathcal{D}_2]}{u_s}
[1-2(2\mathcal{C}\mathcal{D}_1(a-\frac{\lambda a_s}{u_s})-a_s)]
+9\mathcal{C}^2 \beta^2_2\mathcal{D}_1^2
[4\mathcal{C}\mathcal{D}_1(a-\frac{\lambda a_s}{u_s})-1]\nonumber\\
&&
+\frac{\mathcal{S}^2 \hat{g}^2}{\mathcal{D}^2_1}
[4\mathcal{S}\mathcal{E}_1(a-\frac{\lambda a_s}{u_s})-1]
-\frac{\lambda^2(1+4a_s)}{4}-\frac{4\mathcal{C}\mathcal{D}_1a_s\lambda^3}{3u_s}
[1-2(\mathcal{C}\mathcal{D}_1(a-\frac{\lambda a_s}{u_s})-2a_s)]\nonumber\\
&&
-\frac{2\mathcal{S}^2\mathcal{F}^2a_s\kappa^2\hat{g}}
{3\mathcal{D}^2_1u_s}
[1-2(2\mathcal{S}\mathcal{E}_1
(a-\frac{\lambda a_s}{u_s})-a_s)]\Bigr\}.\label{RG-eqs-type-II-case-D-beta2}
\end{eqnarray}
\end{small}

For type-II case-E with $|\mathbf{n}^2_Y|^2\neq|\mathbf{n}_Y|^4$, $|\mathbf{n}^2_X|^2\neq|\mathbf{n}_X|^4$, $|\mathbf{n}^2_Y|^2=0$, and $|\mathbf{n}^2_X|^2=0$, $\beta_2$ evolves but $\beta_1$ is an
invariant constant,
\begin{small}
\begin{eqnarray}
\frac{d\beta_1}{dl}
&=&0,\label{RG-eqs-type-II-case-E-beta1}\\
\frac{d\beta_2}{dl}
&=&\beta_2+\frac{2}{2\pi^2}
\Bigl\{\frac{-4\mathcal{C}^2a_s\lambda^2[\beta_2\mathcal{D}_1^2
+(\beta_1-\beta_2)\mathcal{D}_2]}{u_s}
[1-2(2\mathcal{C}\mathcal{D}_1(a-\frac{\lambda a_s}{u_s})-a_s)]
+9\mathcal{C}^2 \beta^2_2\mathcal{D}_1^2
[4\mathcal{C}\mathcal{D}_1(a-\frac{\lambda a_s}{u_s})-1]\nonumber\\
&&
+\frac{\mathcal{S}^2 \hat{g}^2}{\mathcal{D}^2_1}
[4\mathcal{S}\mathcal{E}_1(a-\frac{\lambda a_s}{u_s})-1]
-\frac{\lambda^2(1+4a_s)}{4}-\frac{4\mathcal{C}\mathcal{D}_1a_s\lambda^3}{3u_s}
[1-2(\mathcal{C}\mathcal{D}_1(a-\frac{\lambda a_s}{u_s})-2a_s)]\nonumber\\
&&
-\frac{2\mathcal{S}^2\mathcal{F}^2a_s\kappa^2\hat{g})}
{3\mathcal{D}^2_1u_s}
[1-2(2\mathcal{S}\mathcal{E}_1(a-\frac{\lambda a_s}{u_s})-a_s)]\Bigr\}.\label{RG-eqs-type-II-case-E-beta2}
\end{eqnarray}
\end{small}

For type-II case-F with $\mathcal{E}_2\mathcal{D}^2_1-\mathcal{D}_2\mathcal{E}^2_1=0$, both
$\beta_1$ and $\beta_2$ are energy-independent constants,
\begin{eqnarray}
\frac{d\beta_1}{dl}=0,\hspace{0.5cm}
\frac{d\beta_2}{dl}=0.\label{RG-eqs-type-II-case-F-beta12}
\end{eqnarray}

\begin{figure*}
\centering
\includegraphics[width=2.6in]{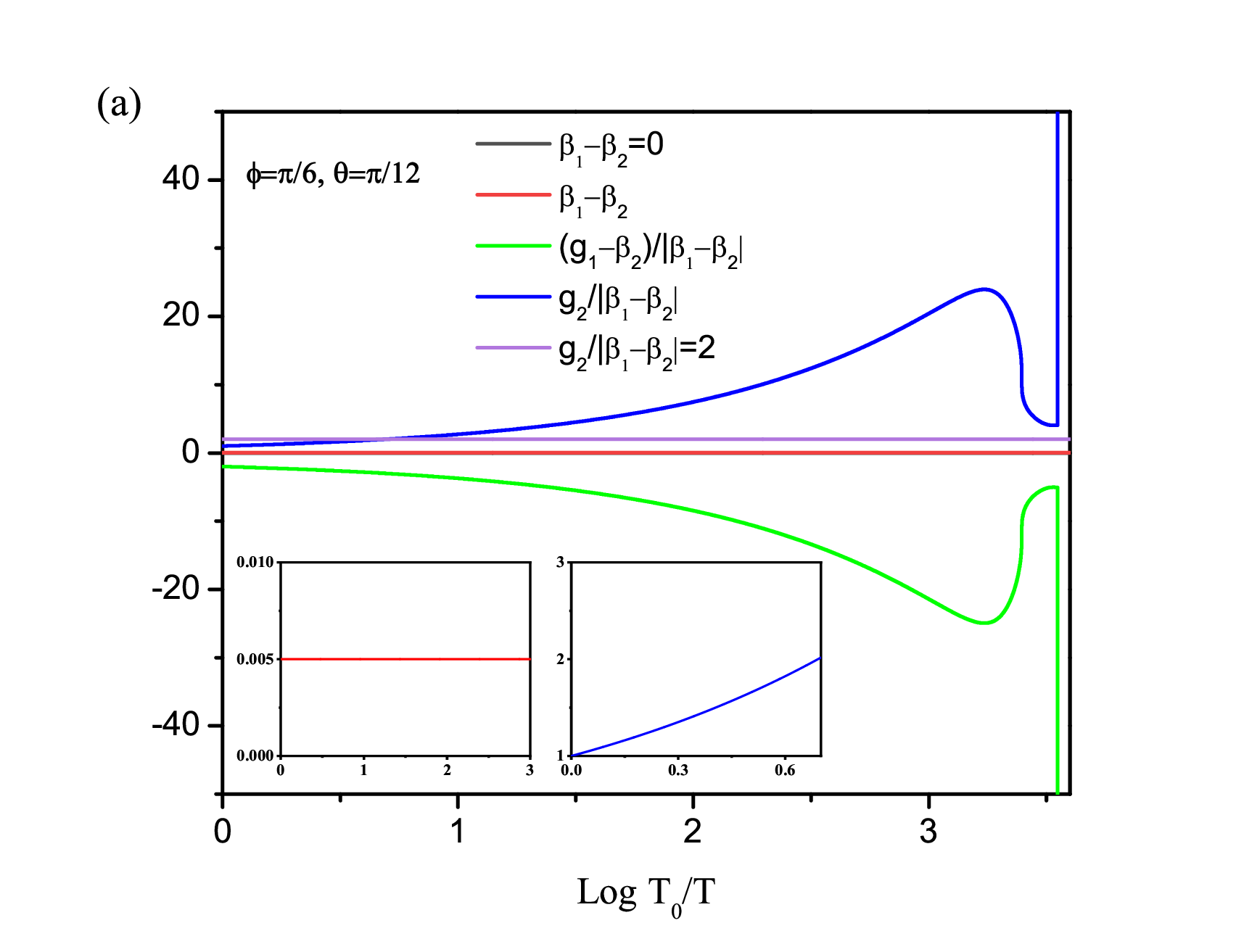}\hspace{-1.1cm}
\includegraphics[width=2.6in]{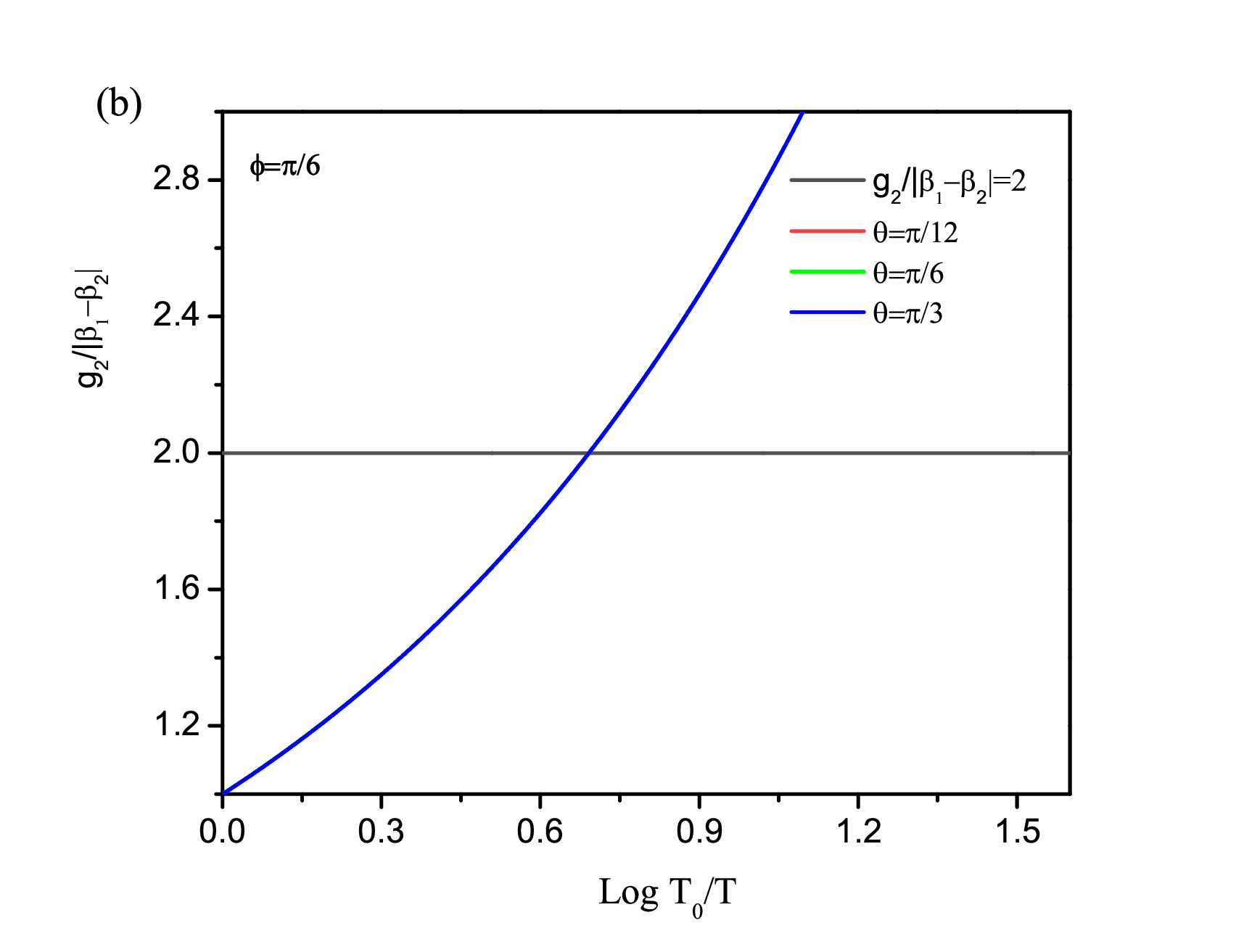}\hspace{-1.1cm}
\includegraphics[width=2.6in]{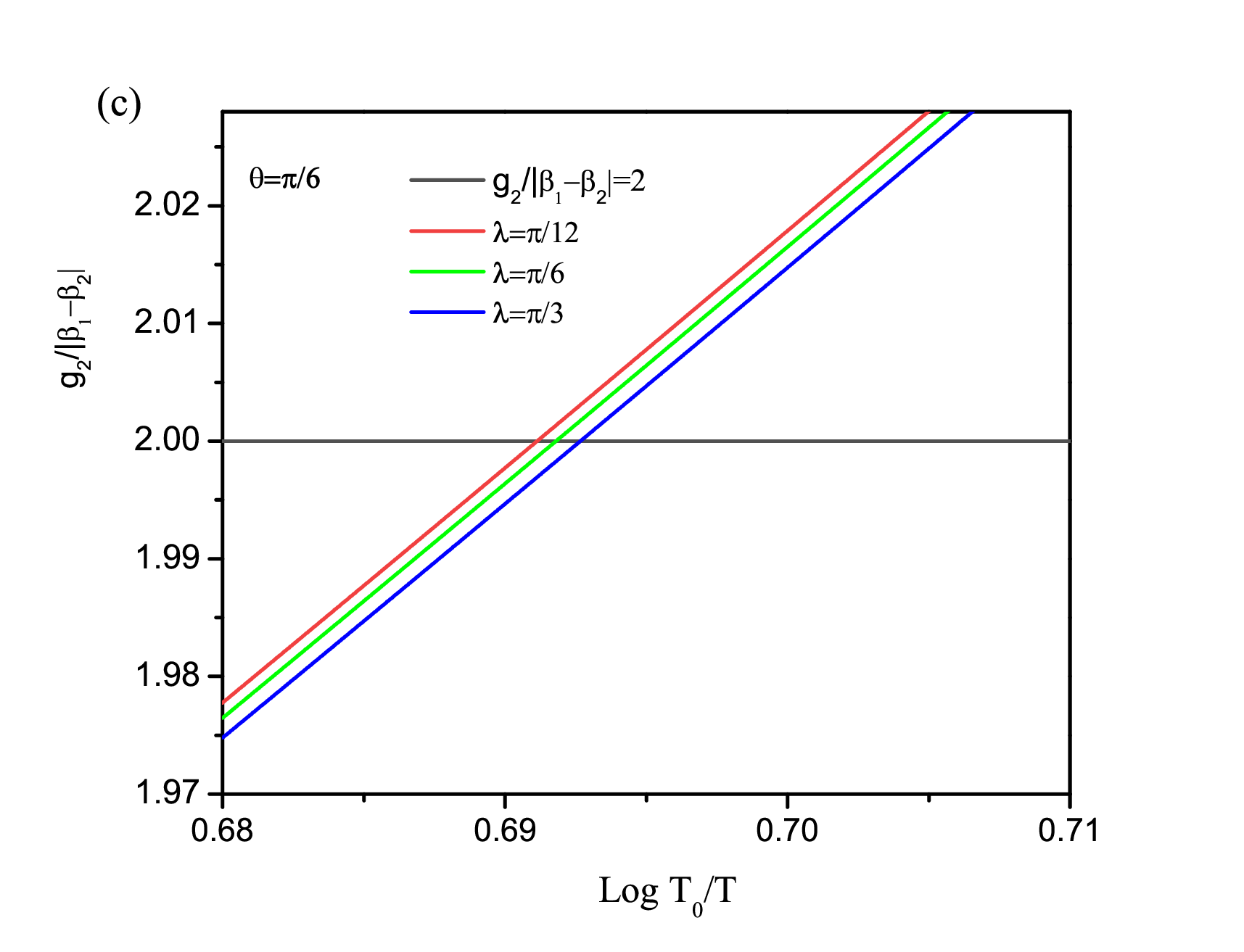}
\vspace{-0.23cm}
\caption{(Color online) (a) Temperature-dependent stable constraints of asymmetric
double-$\mathbf{Q}$ noncoplanar SWC state under the representative starting
values of interaction parameters
chosen as  $g_1=-0.005$, $g_2=0.005$, $u_s=0.05$, $\lambda=0.01$,
$\beta_1=0.01$, $\beta_2=0.005$ with three
representative $\theta=\pi/12,\pi/6,\pi/3$
to satisfy the asymmetric SWC's stable constraint (the qualitative
results are insensitive to initial values of parameters).
Hereby, the angle $\theta$ is designated
in Sec.~\ref{Sec_S_eff} to specify the direction of magnetic order
in the spin space. In addition, $\phi$ is introduced by
$\mathbf{n}_X\cdot \mathbf{n}_Y=\sin^2\phi$
and a concrete value $\phi=\pi/6$ for the asymmetric
case is taken~\cite{Andersen2018PRX}. Inset: the enlarge region for $\beta_1-\beta_2$.
(b) and (c) Sign-change regions at different values of
$\theta$ and $\lambda$.}\label{Fig_SWC-anti}
\end{figure*}

\end{widetext}

\begin{figure*}
\centering
\includegraphics[width=3.5in]{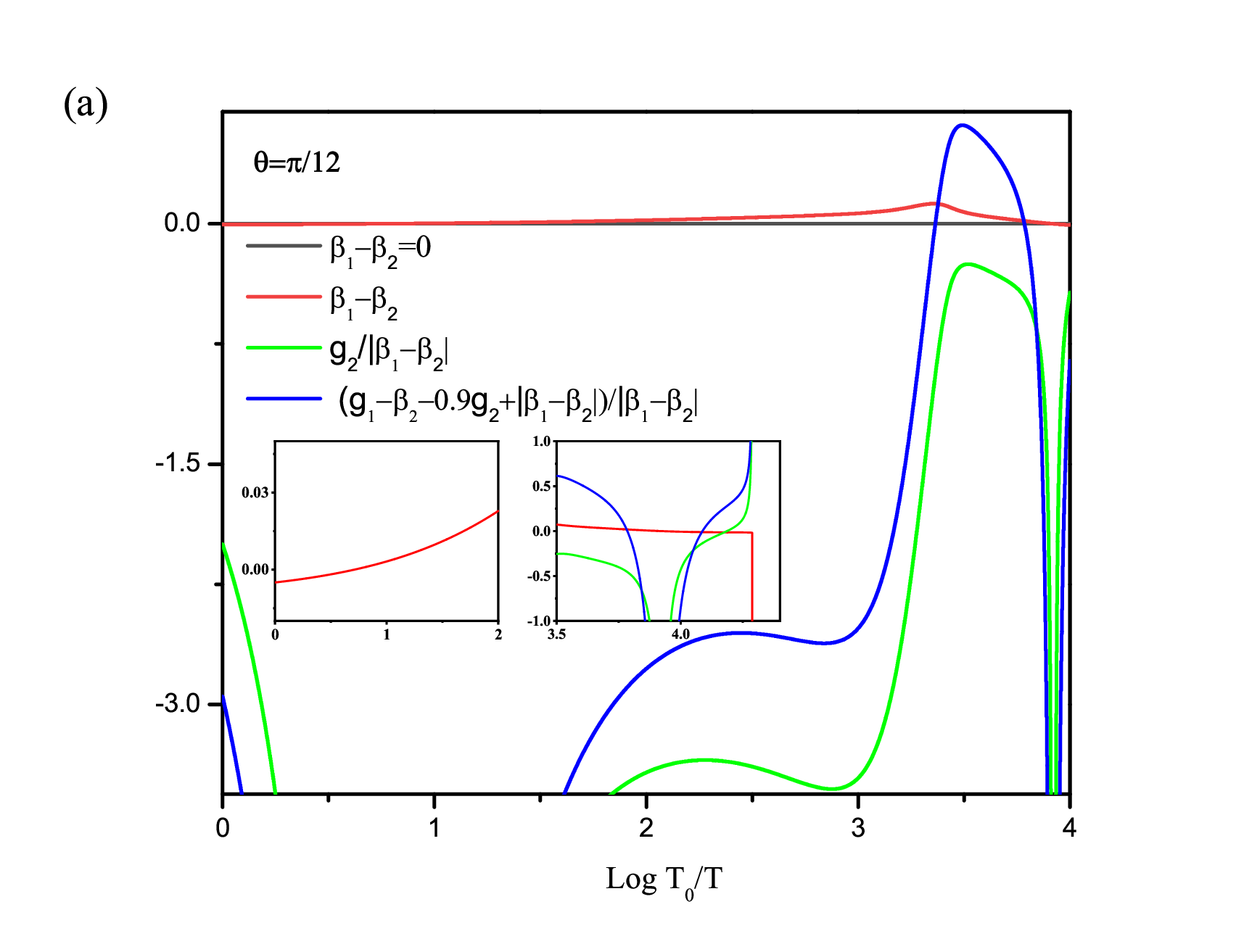}
\includegraphics[width=3.5in]{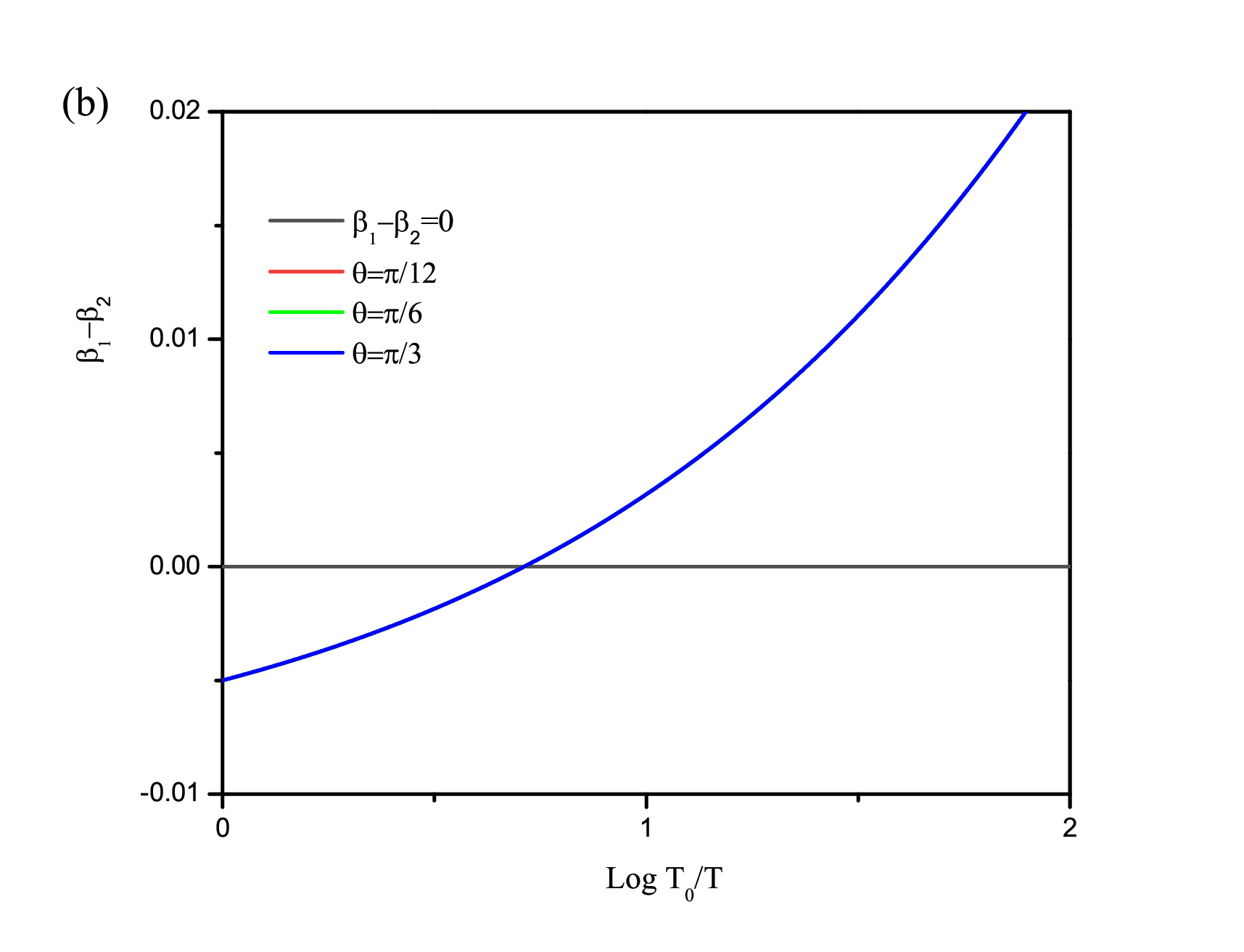}
\vspace{-0.6cm}
\caption{(Color online) (a) Temperature-dependent stable
constraints $C_4$-symmetry IC CSDW state
for case-1 under the representative starting values of interaction
parameters chosen as  $g_1=-0.015$, $g_2=-0.01$, $u_s=0.05$, $\lambda=0.01$,
$\beta_1=0.005$, $\beta_2=0.01$ with three
representative $\theta=\pi/12,\pi/6,\pi/3$
to satisfy the IC CSDW's stable constraint
(the qualitative results are insensitive to
initial values of parameters). Hereby, the angle $\theta$ is designated
in Sec.~\ref{Sec_S_eff} to specify the direction of magnetic order
in the spin space. Inset: the sign-change region of $(g_2+|\beta_1-\beta_2|)/|\beta_1-\beta_2|$.
(b) Sign-change regions at different values of
$\theta$.}\label{Fig_case-1-CSDW-pi-12}
\end{figure*}

\section{Stabilities of incommensurate magnetic states}\label{Appendix_stab-magnetic}

As aforementioned in Sec.~\ref{Sec_intro} of maintext, there are seven different
types of IC magnetic states other than three commensurate ones
including stripe SDW, CSDW, and SVC~\cite{Fernandes2016PRB,
Schmalian2016PRB,Yu1706,Schmalian2018PRB}. To be concrete, these IC magnetic states cover
four different $C_2$ IC cases consisting of
$C_2$ IC ICS, $C_2$ MH, $C_2$ ICS $\perp$ MH, and $C_2$ DPMH,
as well as three distinct $C_4$ IC situations involving $C_4$ IC CSDW,
$C_4$ IC SVC, and $C_4$ IC SWC~\cite{Andersen2018PRX}.
In order to examine whether these IC magnetic states are stable against the
decrease of energy scales, we within this section lean upon the
coupled RG equations~(\ref{RG-eqs-type-I-as})-(\ref{RG-eqs-type-II-case-F-beta12}),
which are completely encoded with the information of ordering competition,
in conjunction with their stable constraints catalogued in Table~\ref{table_criteria-fates}
of the maintext.

In principle, the energy variable of RG evolution is
expressed by $\Lambda=\Lambda_0e^{-l}$ with $l>0$ denoting the running scale.
As our study is concerned with the structure of
schematic phase diagram, it is herein of remarkable convenience to
associate $l$ with temperature via designating $T=T_0e^{-l}$
with $T_0$ being the initial temperature
to measure the evolution of energy scale~\cite{Wang2014PRD,Wang2017PRB,
Fernandes2012PRB,Chubukov2012NP,She2015PRB,
Balents2014PRX,Lee2017PRX,Huh2008PRB,
Xu2008PRB,Foster2008PRB,Chubukov2016PRX,Metlitski2015PRB}.
On the basis of this transformation and RG equations in conjunction
with the strategy addressed in Sec.~\ref{Sec_strategy}, we are now in a proper position to
judge whether these IC magnets are good candidates residing
in the phase diagram of $\mathrm{Ba_{1-x}Na_xFe_2As_2}$ one by one.

We start out by considering the $C_2$ IC magnetic states.
On one hand, the configurations of spin vectors
for $C_2$ ICS magnetic state read $\mathbf{n}_X=(0,0,1)$ and
$\mathbf{n}_Y=(0,0,0)$~\cite{Andersen2018PRX}, which satisfy
the restricted conditions of type-II case-A.
This indicates the interaction parameters obey the RG evolutions of type-II case-A delineated in Eqs.~(\ref{RG-eqs-type-I-as})-(\ref{RG-eqs-type-I-lambda}),
(\ref{RG-eqs-type-I-lamb-DA}), (\ref{RG-eqs-type-I-kappa}),
and (\ref{RG-eqs-type-II-g-hat})-(\ref{RG-eqs-type-II-case-A-beta2}).
As for $C_2$ ICS, its stable constraints can be either $(\beta_1-\beta_2)<0$,
$g_2/|\beta_1-\beta_2|>0$, $(g_1-\beta_2)/|\beta_1-\beta_2|>-1$
or $(\beta_1-\beta_2)<0$, $g_2/|\beta_1-\beta_2|<0$, $(g_1-\beta_2-0.9g_2)/|\beta_1-\beta_2|>-1$~\cite{Andersen2018PRX}.
Based on these, we perform numerical RG
analysis by taking some initial representative values of parameters and obtain the
results shown in Fig.~\ref{Fig_C2-IC-stripe}.
On the other, concerning $C_2$ MH and $C_2$ DPMH, the configurations
of spin vectors are characterized by
$\mathbf{n}_X=\frac{1}{\sqrt{2}}(i,0,1)/$, $\mathbf{n}_Y=(0,0,0)$,
and $\mathbf{n}_X=\frac{1}{\sqrt{2}}(i,0,1)$, $\mathbf{n}_Y=\frac{1}{\sqrt{2}}(i,0,1)$,
respectively~\cite{Andersen2018PRX}. Accordingly, this indicates
that the interaction parameters are dictated by the evolutions for
type-II case-D provided in
Eqs.~(\ref{RG-eqs-type-I-as})-(\ref{RG-eqs-type-I-lambda}),
(\ref{RG-eqs-type-I-lamb-DA}), (\ref{RG-eqs-type-I-kappa}),
(\ref{RG-eqs-type-II-g-hat}), and
(\ref{RG-eqs-type-II-case-D-beta1})-(\ref{RG-eqs-type-II-case-D-beta2}).
To proceed, we parallel the analogous RG numerical analysis
taking advantage of the corresponding constraints~\cite{Andersen2018PRX} $(\beta_1-\beta_2)>0$,
$g_2/|\beta_1-\beta_2|>0$, $(g_1-\beta_2)/|\beta_1-\beta_2|>0$ or $(\beta_1-\beta_2)>0$,
$g_2/|\beta_1-\beta_2|<0$, $(g_1-\beta_2-0.9g_2)/|\beta_1-\beta_2|>-1$
for $C_2$ MH and $(\beta_1-\beta_2)>0$,
$g_2/|\beta_1-\beta_2|<0$, $(g_1-\beta_2-0.9g_2)/|\beta_1-\beta_2|<-1$ for $C_2$ DPMH, respectively.
The conclusions are underscored in Fig.~\ref{Fig_C2-DPMH} and Fig.~\ref{Fig_mag-helix-beta12}
with taking some representative beginning values of parameters.
Learning from Fig.~\ref{Fig_C2-IC-stripe}, Fig.~\ref{Fig_C2-DPMH},
and Fig.~\ref{Fig_mag-helix-beta12}, we apparently figure out that
the sign change of $\beta_1-\beta_2$ is occurred explicitly once
temperature is slightly lowered owing to the effects of ordering competition.
As a consequence, we infer that $C_2$ ICS, $C_2$ MH and $C_2$ DPMH are not
stable states in the low-energy regime and hence not good candidates for
IC magnetic state nearby the QCP in phase diagram of
$\mathrm{Ba_{1-x}Na_xFe_2As_2}$. However,
these three $C_2$ IC states might be suitable states for the quantum critical
region with high temperatures, such as $C_2$ IC SDW illustrated
in Fig.~\ref{Fig_schematic_phase_diagram}.

In a sharp contrast, with respect to $C_2$ ICS $\perp$ MH, whose the
configurations of spin vectors
are related to $\mathbf{n}_X=(0,0,1)$ and $\mathbf{n}_Y=\frac{1}{\sqrt{2}}(i,1,0)$~\cite{Andersen2018PRX},
their interaction parameters are therefore subject to type-I coupled RG
equations~(\ref{RG-eqs-type-I-as})-(\ref{RG-eqs-type-I-g-hat}).
Carrying out the similar numerical analysis gives rise to temperature-dependent
evolutions depicted in Fig.~\ref{Fig_IC-perp-helix-g-big}. It manifestly heralds
that stable constraints of $C_2$ ICS $\perp$ MH, i.e.,
$(\beta_1-\beta_2)>0$, $(g_1-\beta_2)/|\beta_1-\beta_2|<0$, and
$g_2/|\beta_1-\beta_2|>f(g_1-\beta_2)\approx2$ for a finite value of $g_1-\beta_2$
and $\lim_{(g_1-\beta_2)\rightarrow0}f(g_1-\beta_2)\rightarrow 0$~\cite{Andersen2018PRX},
are remarkably robust against ordering competition as the temperature
is decreased. Despite of relative stability, it is very necessary to point out
that $C_2$ ICS $\perp$ MH can be destroyed as long as the magnetic QCP
is closely accessed, at which the ordering competition becomes so ferocious
that any state cannot present solely.

Next, we go to judge $C_4$ IC magnetic states, which include
$C_4$ IC SVC, $C_4$ SWC, and $C_4$ IC CSDW.
In analogy to $C_2$ IC magnetic states, we inspect low-energy fates of
these states by combining their RG equations and stable constraints.
For $C_4$ IC SVC with the configurations of spin vectors being
$\mathbf{n}_X=(0,0,1)$ and $\mathbf{n}_Y=(0,1,0)$~\cite{Andersen2018PRX},
the interaction parameters are governed by the type-II case-A RG
equations~(\ref{RG-eqs-type-I-as})-(\ref{RG-eqs-type-I-lambda}),
(\ref{RG-eqs-type-I-lamb-DA}), (\ref{RG-eqs-type-I-kappa}),
(\ref{RG-eqs-type-II-g-hat})-(\ref{RG-eqs-type-II-case-A-beta2}) and
the stable constraints correspond to $(\beta_1-\beta_2)<0$,
$g_2/|\beta_1-\beta_2|>0$, and $(g_1-\beta_2)/|\beta_1-\beta_2|
<-1$~\cite{Andersen2018PRX}. The numerical results presented in Fig.~\ref{Fig_SVC-pi-12}
reflect that $C_4$ IC SVC cannot be a well stable state in the phase diagram
caused by the influence of ordering competition.

To proceed, we turn to $C_4$ IC SWC, which is well protected by constraints
$(\beta_1-\beta_2)>0$, $(g_1-\beta_2)/|\beta_1-\beta_2|<0$,
and $0<g_2/|\beta_1-\beta_2|<2$~\cite{Andersen2018PRX}. In addition,
the configurations of spin vectors are equivalent to
$\mathbf{n}_X=(i\cos\phi,0,\sin\phi)$ and $\mathbf{n}_Y=(0,i\cos\phi,\sin\phi)$.
Before going further, it is of particular interest to address that
they can be clustered into two sub-situations distinguished by the
parameter $\phi$ which is introduced by $\mathbf{n}_X\cdot \mathbf{n}_Y=\sin^2\phi$
and characterize the symmetric double-$\mathbf{Q}$ noncoplanar SWC with $\phi=\pi/4$ and asymmetric double-$\mathbf{Q}$ noncoplanar with $\phi\neq\pi/4$, respectively~\cite{Andersen2018PRX}.
As a result, the former interaction parameters are dictated by type-II case-A RG equations~(\ref{RG-eqs-type-I-as})-(\ref{RG-eqs-type-I-lambda}),
(\ref{RG-eqs-type-I-lamb-DA}), (\ref{RG-eqs-type-I-kappa}),
and (\ref{RG-eqs-type-II-g-hat})-(\ref{RG-eqs-type-II-case-A-beta2}) but instead the
latter ones evolve under type-II case-F RG equations exhibited in Eqs.~(\ref{RG-eqs-type-I-as})-(\ref{RG-eqs-type-I-lambda}), (\ref{RG-eqs-type-I-lamb-DA}), (\ref{RG-eqs-type-I-kappa}), (\ref{RG-eqs-type-II-g-hat}), and (\ref{RG-eqs-type-II-case-F-beta12}).
Carrying out analogous RG steps yields to Fig.~\ref{Fig_SWC-symm}
and Fig.~\ref{Fig_SWC-anti}, which explicitly signals $C_4$ SWC is
not suitable to be present in the phase diagram.

\begin{figure}
\centering
\includegraphics[width=3.5in]{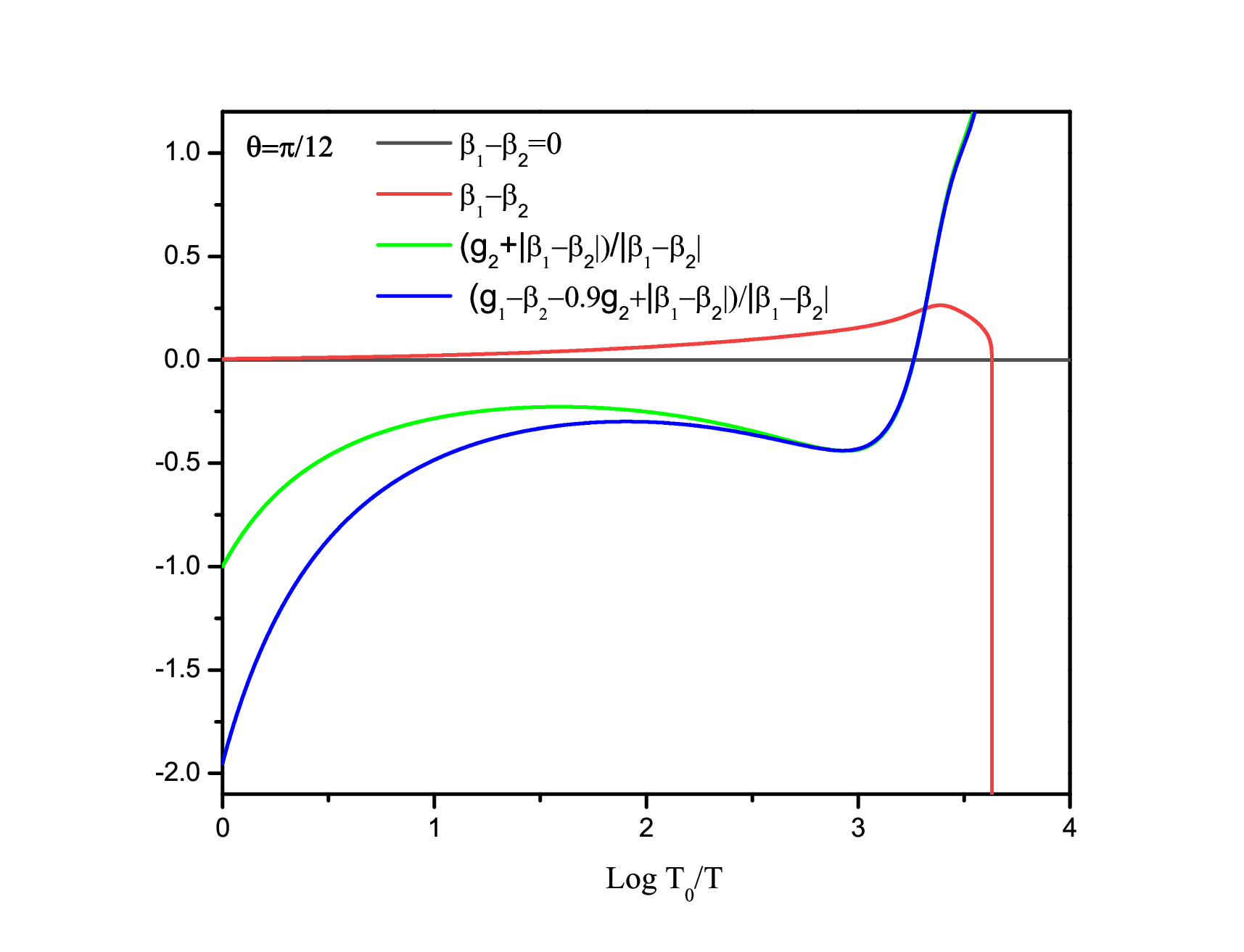}
\vspace{-0.6cm}
\caption{(Color online) Temperature-dependent stable constraints $C_4$-symmetry IC CSDW state
for case-2 under the representative starting values of interaction
parameters chosen as  $g_1=-0.015$, $g_2=-0.01$, $u_s=0.05$, $\lambda=0.01$,
$\beta_1=0.01$, $\beta_2=0.005$ with a
representative $\theta=\pi/12$ to satisfy the IC CSDW's stable constraint
(the qualitative results are insensitive to initial values
of parameters). Hereby, the angle $\theta$ is designated
in Sec.~\ref{Sec_S_eff} to specify the direction of magnetic order
in the spin space.}\label{Fig_case-2-CSDW-pi-12}
\end{figure}

Further, we move to $C_4$ IC CSDW state, at which the configurations of
spin vectors are of the form $\mathbf{n}_X=(0,0,1)$ and
$\mathbf{n}_Y=(0,0,1)$~\cite{Andersen2018PRX}, and thus type-II case-A RG equations~(\ref{RG-eqs-type-I-as})-(\ref{RG-eqs-type-I-lambda}),
(\ref{RG-eqs-type-I-lamb-DA}), (\ref{RG-eqs-type-I-kappa}),
and (\ref{RG-eqs-type-II-g-hat})-(\ref{RG-eqs-type-II-case-A-beta2}) are
in charge of the low-energy fates of interaction parameters.
Hereby, it is necessary to highlight that $C_4$-symmetry IC CSDW~\cite{Andersen2018PRX}
can be stabilized by either $(\beta_1-\beta_2)<0, g_2/|\beta_1-\beta_2|<0,(g_1-\beta_2-0.9g_2)/|\beta_1-\beta_2|
<-1$ (case-1) or $(\beta_1-\beta_2)>0, g_2/|\beta_1-\beta_2|<-1,
(g_1-\beta_2-0.9g_2)/|\beta_1-\beta_1|<-1$ (case-2).
Fig.~\ref{Fig_case-1-CSDW-pi-12} and
Fig.~\ref{Fig_case-2-CSDW-pi-12}
collect the central results stemming from RG analysis,
which manifestly exhibit the temperature (energy) dependence of associated
parameters for $C_4$ IC CSDW. %that govern its low-energy fates.
In the light of these figures, we are informed that stable constraints
for both case-1 and case-2 are considerably robust with
the decrease of temperature, which of course can be sabotaged due to
sufficiently strong fluctuations so long as the
magnetic QCP is closely approached. Consequently, $C_4$ IC CSDW,
like its $C_2$ ICS $\perp$ MH counterpart, is of fair robustness
against ordering competition and an appropriate candidate for $C_4$
magnetic state in phase diagram of $\mathrm{Ba_{1-x}Na_xFe_2As_2}$.

%%%%%%%%%%%%%%%%%%%%%%%%%%%%%%%%%%%%%%%%%%%%%%%%%%%%%%%%%%%%%%%%%%%%%%%%%%%%%%%%%%%%%%%%%
%%%%%%%%%%%%%%%%%%%%%%%%%%%%%%%%%%%%%%%%%%%%%%%%%%%%%%%%%%%%%%%%%%%%%%%%%%%%%%%%%%%%%%%%%


\begin{thebibliography}{10}
\bibitem{notice}
Besides $a$, $a_s$, $u_s$, $\lambda$, $\beta_1$, $\beta_2$,
$g_1$, and $g_2$ appearing in the free energy~(\ref{Eq_free-energy}),
$\kappa$ and $\lambda_{\Delta A}$ are two additional fundamental parameters.
\end{thebibliography}
\end{document}